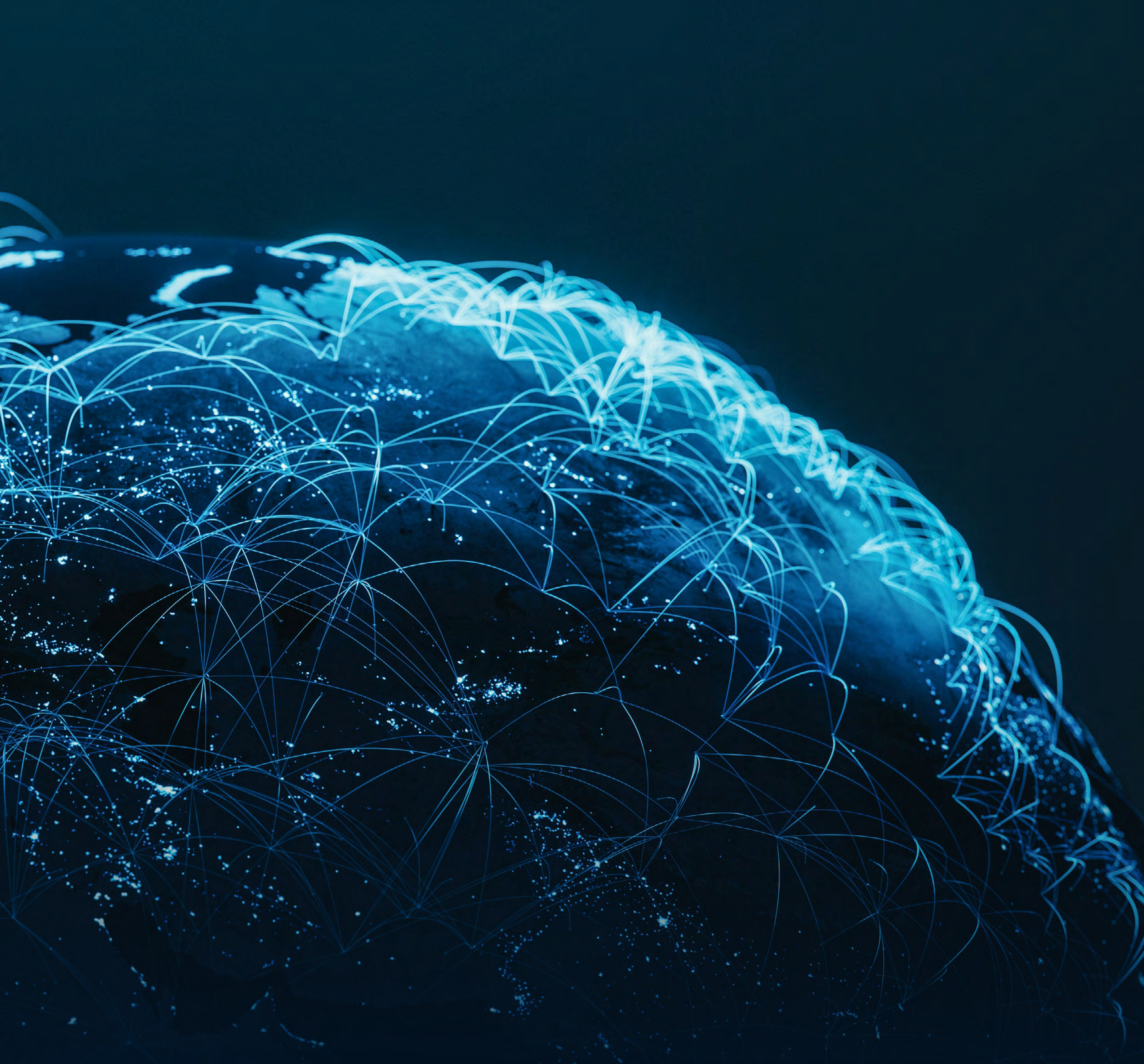

# The 2026 Singapore Consensus on **Global AI Safety Research Priorities**

International Scientific Priorities for Building a Trustworthy, Reliable and Secure AI Ecosystem

July 2026

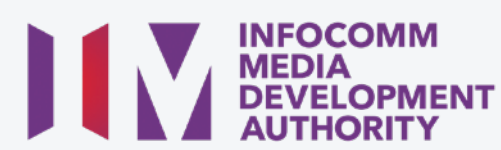

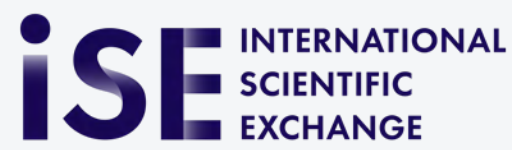

TABLE OF CONTENTS

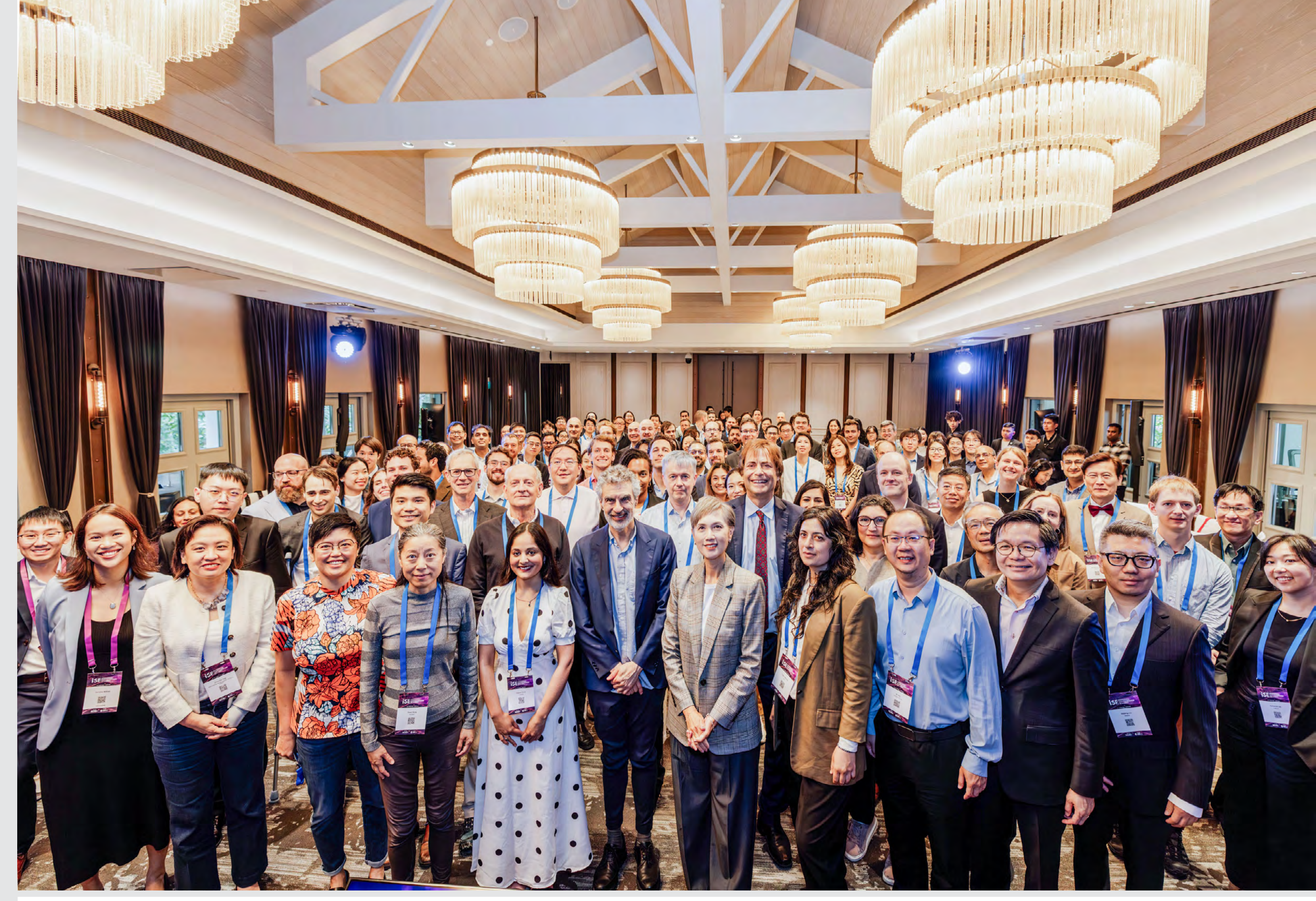

# Contributors to the Report

## Steering Committee

**Yoshua Bengio**
MILA

**Mohan Kankanhalli**
National University of Singapore

**Lee Wan Sie**
Infocomm Media Development Authority

**Tegan Maharaj**
MILA

**Chris Meserole**
Frontier Model Forum

**Luke Ong**
Nanyang Technological University

**Stuart Russell**
UC Berkeley

**Dawn Song**
UC Berkeley

**Max Tegmark**
Future of Life Institute

**Brian Tse**
Concordia AI

**Xue Lan**
Tsinghua University

**Andrew Yao**
Shanghai Qi Zhi Institute

**Zhang Ya-Qin**
Tsinghua University

**Zhou Bowen**
Shanghai AI Lab

## Writing Coordinators

**Stephen Casper** *(Lead Writer)*
Harvard University

**Sören Mindermann** *(Lead Writer)*
University of Cambridge

**Oskar Galeev** *(Lead Writer)*
Concordia AI

**Ima (Imane) Bello**
Future of Life Institute

**Kwan Yee Ng**
Concordia AI

**Vanessa Wilfred**
Infocomm Media Development Authority

**Erica Liaw**
Infocomm Media Development Authority

## Secretariat

**Lee Chein Inn**
Infocomm Media Development Authority

**Lin Wanxuan**
Infocomm Media Development Authority

**Ng En Qi**
Infocomm Media Development Authority

**Erica Liaw**
Infocomm Media Development Authority

**Kwan Yee Ng**
Concordia AI

**Jonathan Lee**
Concordia AI

**José Villalobos**
Future of Life Institute

*The views expressed by contributors are in their individual capacities and do not necessarily reflect those of their affiliated organisations.*

**Image:** Participants of the 'International Scientific Exchange on AI Safety', 18–19 May 2026.

## Contributors

**Abhishek Aggarwal**
Ministry of Electronics and IT, India

**Adam Gleave**
FAR.AI

**Alan Chan**
GovAI

**Alex Leung**
Vulcan/AIFT

**Alvin Kwock**
Vulcan/AIFT

**Anthony Tung**
National University of Singapore

**Arisa Siong**
Infocomm Media Development Authority

**Arthur Tea**
Centific

**Ben Bucknall**
University of Oxford

**Benjamin Weinstein-Raun**
Palisade Research

**He Bing Sheng**
National University of Singapore

**Liu Bo**
China Academy of Information and Communications Technology (CAICT)

**Bryan Kian Hsiang Low**
National University of Singapore

**Chris Ngo**
Knovel Engineering

**Clement Neo**
Neo Research (SASH)

**Cyrus Hodes**
AI Safety Connect

**Dan Hendrycks**
Center for AI Safety

**Daniel Ross**
Dynamo AI

**Liu Dapeng**
Alibaba

**Denise Wong**
Infocomm Media Development Authority

**Djordje Zikelic**
Singapore Management University

**Elham Tabassi**
Brookings Institution

**Fabien Le Voyer**
INRIA

**Fazl Barez**
University of Oxford

**Gabriel Nicholas**
Anthropic

**Henry Papadatos**
SaferAI

**Jaan Tallinn**
Future of Life Institute

**James Petrie**
Future of Life Institute and University of Oxford

**Xu Jia**
Shanghai AI Lab

**Shao Jing**
Shanghai AI Lab

**Jonathan Barry**
MILA

**Julia Chen**
Simon Institute for Longterm Governance

**Sun Jun**
Singapore Management University

**Karson Elmgren**
Institute for AI Policy and Strategy (IAPS)

**Kat Lyness**
AI Security Institute (UK)

**Katherine Lee**
OpenAI

**Kristy Loke**
MATS Research

**Lee Kwee Geak**
Infocomm Media Development Authority

**Leslie Teo**
AI Singapore

**Meng Ling Yu**
Shanghai AI Lab

**Lisa Soder**
EU AI Office

**Madhulika Srikumar**
Partnership on AI

**Malcolm Murray**
SaferAI

**Mark Brakel**
Future of Life Institute

**Mark Nitzberg**
Center for Human-Compatible AI / IASEAI

**Mary Phuong**
Google DeepMind

**Matthew Jagielski**
Anthropic

**Max Fenkell**
Scale AI

**Miro Plueckebaum**
Singapore AI Safety Hub (SASH)

**Kim Myuhng Joo**
Korea AI Safety Institute

**Hu Naying**
China Academy of Information and Communications Technology (CAICT)

**Neil Davison**
HD Centre

**Nicolas Miailhe**
AI Safety Connect

**Niki Iliadis**
The Future Society

**Nur Syahidah Sahrom**
Ministry of Digital Development and Information

**Ong Chen Hui**
Infocomm Media Development Authority

**Pradeep Varakantham**
Singapore Management University

**Rebecca Finlay**
Partnership on AI

**Renata Dwan**
Simon Institute for Longterm Governance

**Robert Opp**
UN Development Programme (UNDP)

**Rumman Chowdhury**
Humane Intelligence

**Saad Siddiqui**
Safe AI Forum

**Sabina Nong**
Future of Life Institute

**Sam Ramadori**
LawZero

**Sami Jawhar**
Trajectory Labs

**Samuel Boger**
Singapore AI Safety Hub (SASH)

**Sara Hooker**
Adaption Labs

**Ying Shao Wei**
NCS Pte Ltd

**Sebastian Hallensleben**
Resaro

**Shinyuk Kang**
Korea AI Safety Institute

**Sophie Toura**
ControlAI

**Sreejith Balakrishnan**
G42

**Stephanie Kasaon**
Action Lab

**Stephen Clare**
International AI Safety Report

**Summer Yue**
Meta

**Sunny Yuqing Sun**
ByteDance

**Supheakmungkol Sarin**
AI Safety Asia

**Tian Tian**
RealAI

**Tim Schreier**
MATS Research

**Tori Westerhoff**
Microsoft

**Urvashi Aneja**
Digital Futures Lab

**Wayne Tee**
Singapore AI Safety Hub (SASH)

**Lu Wei**
Nanyang Technological University

**Xu Wei**
Tsinghua University

**Zhang Wenxuan**
Singapore University of Technology and Design (SUTD)

**Hu Xia**
Shanghai AI Lab

**Yang Xiaofang**
Ant Group

**Pan Xudong**
Fudan University / Shanghai Innovation Institute

**Xiao Yajun**
Huawei Technologies Co., Ltd.

**Yifan Jia**
AIDX Tech

**Tan Yong Khiam**
NTU / A*STAR

**Yuejin Du**
Zhejiang University

**Yuma Kurihara**
Japan AI Safety Institute

**Tan Zhi Xuan**
National University of Singapore

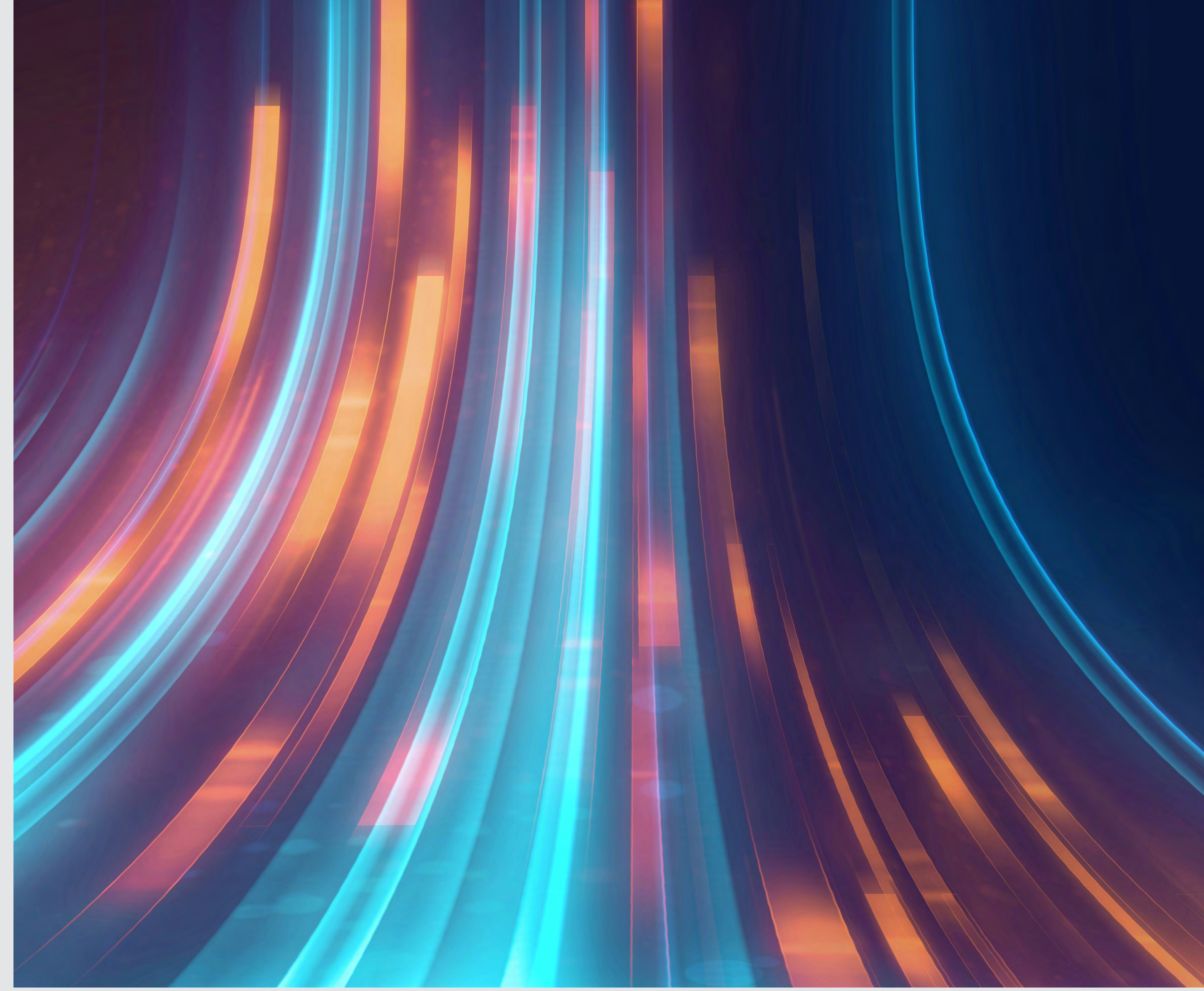

# Executive summary

## About the Singapore Consensus

Frontier AI systems have advanced rapidly in capability and real-world impact. Revenues from frontier models have grown by over 400% from a year ago and AI investments this year are on track to exceed the GDP of a top 30 national economy. At the same time, AI safety incidents now impact millions of users globally. To drive investment by private and public actors in AI safety research so that it keeps pace with capabilities advancement, Singapore organised the 2nd International Scientific Exchange on AI Safety (ISE) in May 2026, continuing the dialogue on global AI safety research priorities which began at the inaugural ISE in April 2025.

The **2026 Singapore Consensus on Global AI Safety Research Priorities**, which is an outcome of the 2026 convening, presents a global understanding on scientific research problems that are of top priority to advance AI safety research. Building on the Singapore Consensus 2025 report, the 2026 **Main Report** has been updated based on key developments in the AI risk landscape in the past year, such as the increasing deployment and capabilities of autonomous AI agents and the growth of misuse and incidents. The report is developed as an agenda-focused complement to the International AI Safety Report with more than 100 contributors spanning 13 countries from frontier AI developers, government safety institutes, academics and civil society.

At the same time, recognising that agentic deployment has become a major industry focus in the past year, ISE 2026 extends scientific discussion and consensus from research priorities to emerging risk management practices in the development and deployment of autonomous agents. Given that industry practices on agentic deployment are constantly evolving, there is tremendous value to achieving a shared understanding of the state-of-the-art practices to mitigate and manage these risks, drawing on the expertise and experiences of the global experts and world-leading organisations present at ISE 2026. This deep dive enriches the discussions at ISE 2026 and addresses a very timely and pertinent need from the industry. The discussions and findings are captured in the **Companion Report on Agentic Risk Management**, which incorporates and analyses input from global frontier AI developers on leading agentic safety practices.

We hope to facilitate international dialogue and scientific collaboration among scientists, developers, and policymakers. Companies and nations often compete on AI capabilities research and development (R&D). However, we believe that research on managing AI's risks is often an **area of mutual interest** (Bucknall et al, 2005), at least outside areas of sensitive intellectual property. Due to their global nature, many societal-level AI risks can only be addressed by coalitions of companies or countries. As the leaders of the global AI powers and companies convene this year, we believe it is in their interest to agree on information sharing and research collaboration for the global public good.

## Key Updates

### 1. Growing Emphasis on Societal Resilience as Risk Prevention Shows Gaps

Adopting a defence-in-depth model, the original report groups scientific AI safety research topics into three broad areas from **risk assessment** that informs subsequent development and deployment decisions, to technical safety methods in the system **development** phase, and tools for **control** after a system has been deployed. Growing misuse and malfunction incidents show that harm prevention efforts alone are insufficient. We have therefore elevated **societal resilience** as a distinct fourth pillar to emphasize the growing importance of preparing and hardening societal systems for failures and misuse.

## 2. Updated AI Safety Research Priorities

The Main Report highlights new technical developments since ISE 2025, and outlines the research priorities that are especially important in light of these developments. The following research priorities are critical given developments over the last year:

- **Prevention and resilience for growing misuse incidents.** AI misuse incidents have grown precipitously, particularly for cyber attacks, nudification and other sexual abuse material. Biochemical misuse capabilities have surpassed PhD-level experts on some benchmarks and are being distributed with safeguards that can be consistently circumvented. New cyber model capabilities have abruptly increased the number of real-world vulnerabilities found in some critical software by over 1000%. Addressing this requires research on prevention and resilience methods targeted to these areas (Section 2, 3 and 4).
- **Safety for open-weight models whose capabilities are nearing frontier closed models.** According to common benchmarks, open weight models are estimated to be 3 to 12 months behind the frontier (varying across capabilities). Powerful AI models with publicly-downloadable weights are implicated in many misuse incidents. These models pose unique benefits for research and customization, but they can also be tampered with and used without oversight for malicious purposes. Managing these risks requires unique tools for open-weight model safety (Section 2.2.5) and ecosystem monitoring (Section 4.1). Since prevention is challenging, researching societal resilience to misuse is essential (Section 4.2).
- **Evaluations and alignment/control methods for emerging oversight-resistant and control-undermining AI.** Control-undermining AI capabilities and behaviors are now a major challenge to safety testing in practice. They also mark early warnings for loss of control scenarios. Additionally, frontier developers now increasingly use AI to oversee AI internally - including for safety evaluations - meaning that human operators may struggle to understand and verify the system of oversight itself. Addressing this requires research on evaluations and methods for both alignment and control that are more resistant to growing control-undermining AI tendencies such as evaluation awareness (Section 1.1), and methods to verify oversight when it is outsourced to AI systems (Section 3.2 and 4.1).

## 3. Deep Dive into Agentic Risk Management

2025 saw a sharp increase in deployment of AI agents, with releases accelerating sharply and agents' autonomy levels rising in parallel. This momentum has continued into early 2026, with OpenClaw and its derivatives becoming the fastest-growing open source projects in history. Organizations are actively experimenting with ways of deploying autonomous AI agents as they become more powerful. However, uncertainty over their reliability, security, and trustworthiness have emerged as barriers to adoption. Addressing this requires research on and trusted best practices for agent risk management, which are examined in depth in the **Companion Report on Agentic AI Risk Management**.

AI agents are becoming increasingly autonomous, capable, and widely applicable. They differ from earlier generative AI models in one important way: they can act, with direct effects on their environment and on the systems they connect to. These capabilities produce distinct risks and failure modes, yet governance mechanisms for agentic systems are only just beginning to emerge. As a result, uncertainty over their reliability, security, and trustworthiness has become a barrier to adoption. The 2026 International Scientific Exchange (ISE) on AI Safety therefore identified agentic risk management as a key direction for technical and governance research.

The companion report analyses emerging practices in agentic AI risk management, drawing on input from leading model and agent developers, state-of-the-art academic research, and agentic governance frameworks. It identifies ten foundational principles for managing agentic risk: **least privilege**, **traceable identity**, **auditability**, **validated deployment**, **adversarial resilience**, **multi-agent stability**, **runtime assurance**, **interruptibility**, **legibility**, and **human oversight**. For each principle, the report sets out the governance context from jurisdictions including the US, China, Singapore, and the EU, and highlights technical practices for risk mitigation drawn from industry.

The practices and evidence base underpinning these principles vary considerably in maturity. Some principles are better established and easier to implement in practice, with interoperable tools that make shared safety standards easier to achieve. Others, especially around multi-agent systems and agentic supply chains, are still an area of ongoing research. The report surfaces open problems and shows how responsibility for agentic risk management is spread across developers, deployers, cloud platform providers, and the open source ecosystems. No single actor can ensure the reliability, security, and trustworthiness of AI agents. Therefore, this companion report provides a shared foundation for both technical and governance efforts to manage agentic risks.

## Contributing to Global Collaboration and Advancement

Today, safety research and risk management are a priority for frontier AI adoption, innovation, and prevention of large-scale harms. While we are encouraged by a **high level of expert agreement** on research priorities across global companies, countries, and institutions, **scientific progress must be matched by robust standards and policy frameworks to close the gap between the state of the art and the state of practice.** Many recent AI incidents have been preventable with existing methods, and developing better methods alone is not enough without stronger incentives and accountability mechanisms to ensure that AI developers adopt best practices for safety and organizational risk management. Due to their global nature, AI risks are usually best addressed by coalitions rather than individual companies or countries. We hope this consensus can **inform research funding, highlight where more transparency and reporting are needed to enable research**, **spur information sharing and collaboration** even among competitors where needed, and **support the development of policy frameworks** that allow everyone to embrace general-purpose AI safely and with confidence.

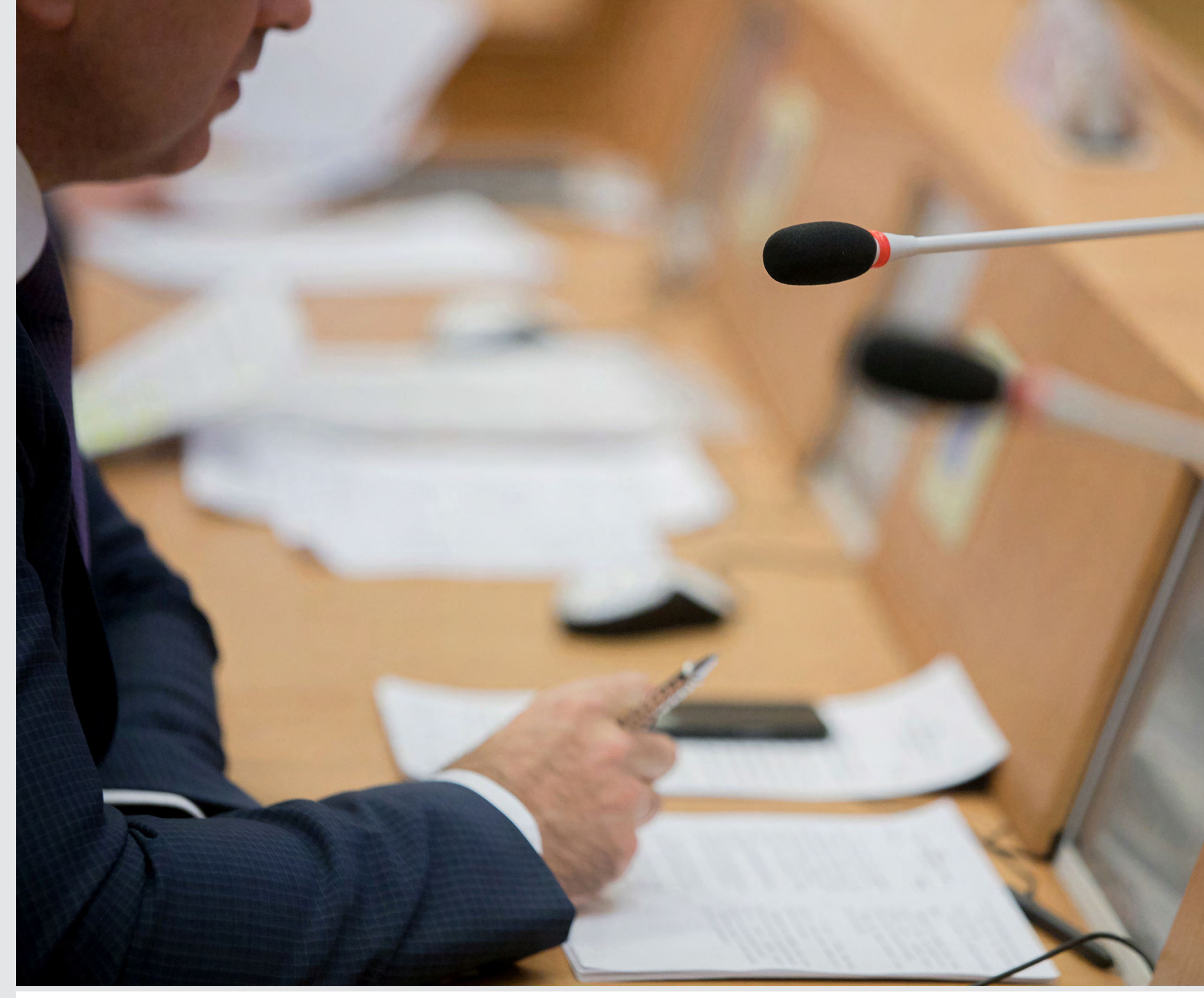

# Research Priorities by Policy Area

The Singapore Consensus is a scientific consensus on technical AI safety research priorities, not a policy document. The Executive Summary sets out why a shared, prioritised research agenda matters now: capabilities, real-world deployment and revenues are rising faster than our ability to measure and manage risks. Meanwhile, uncertainty about safety is itself holding back adoption. Agreeing on where research effort should go, and sharing much of what is learned, is often an international area of mutual interest: it serves every party at little to no competitive cost.

This note translates the research areas into policy-relevant domains to inform research funding, standards development, and international coordination. Each area below opens with what is at stake, followed by the research directions experts judge most urgent. Section numbers point to the main report.

## Cyber misuse risk

- What is at stake: As frontier systems grow more capable, a central challenge is to ensure evaluation and defence move ahead of offence.
- Priorities include better ways to anticipate how attackers might use AI ("threat models") and tougher tests of AI's offensive cyber abilities ("harder cyber capability evaluations") that keep pace with frontier systems (1.1, 1.3); developing tools that favor defenders (defense-favoring AI-for-cyber capabilities) and allow vulnerabilities, including in critical infrastructure, to be patched faster (4.2); monitoring collections of multiple AI agents once deployed, so coordinated misuse campaigns can be detected and shut down (4.1); stress-testing openly released models to see what they could do if deliberately retrained for harm ("malicious fine-tuning") (2.2.5); and shared reporting of serious AI-enabled cyber incidents (4.2).

## Biological and chemical security

- What is at stake: The aim is to prevent models from meaningfully advancing weapons development, including pandemic-scale threats; to keep such capabilities out of novices' hands; and to tilt the balance toward defence across the supply chain.
- Priorities include: studies and tests that measure how much a model actually increases a would-be attacker's capabilities ("uplift studies"), paired with practical thresholds for unacceptable levels of risk (1.1, 1.3); shared filters that keep weapons-relevant content out of the data models are trained on ("pretraining data filters") (2.2.1); methods that aim to durably remove dangerous knowledge from openly released models and tests to confirm it cannot be easily retrained back in ("tamper resistant unlearning") (2.2.5); and screening further up the supply chain globally, for example, of commercial DNA synthesis orders, to build a defenders' advantage (4.2).

## Child safety and image-based abuse

- What is at stake: Here, the science is comparatively mature; the binding constraint is standardisation, adoption and safeguards that cannot be easily stripped out of systems.
- Priorities include: shared data filters that remove child sexual abuse material ("CSAM") and other abuse content from training data (2.2.1); reliable methods to remove "nudification" and similar capabilities from models, with safeguards that survive later modification (such as fine-tuning) of openly released models (2.2.3, 2.2.5); detection tools, classifiers, watermarking, and content-metadata standards, where the main barrier is standardization and adoption (4.1); and measuring how often these harms occur and their real-world impact on children (1.3).

## Mental health and consumer protection

- What is at stake: The priority is to detect and prevent growing harms to users, especially vulnerable ones, before and after deployment.
- Priorities include: specification and validation methods that prevent and measure sycophancy, e.g. models affirming and reinforcing users' harmful beliefs (2.1); rapid, high-coverage pre-deployment risk identification, e.g. fast, broad-screening for unknown risks before a model is released (1.1); real-world trials of mental-health effects with cautious, graduated rollout, as in clinical trials (1.3); monitoring during use to identify users who may be vulnerable (4.1); and clear processes inside organisations for escalating concerns and reporting incidents (1.2, 4.2).

## AI agents in the economy

- What is at stake: Reliability and security concerns are leading bottlenecks to adoption, so progress here is as much an enabler of innovation as a safeguard.
- Priorities include: designing agents to operate safely and confining them to controlled environments ("sandboxing") so mistakes stay contained (2.2.6, 2.2.3); benchmarks that close the gap between evaluated and real-world reliability (1.1); defenses against manipulating agents through hidden instructions ("prompt injection") and the resulting hijacking of agents (2.2.4); real-time monitoring and the ability to intervene while an an agent is running (3.1); multi-agent red-teaming, e.g. adversarial testing of multiple agents working together (2.3.1); and infrastructure to identify and authenticate agents so their actions can be traced (4.1). Agent risk-management practices are discussed in depth in the Companion Report.

## Open-weight model safety and security

- What is at stake: Because model risks spread internationally and irreversibly once weights are released, with safeguards that are relatively easy to remove, managing open-weight model misuse is a shared international interest.
- Priorities include testing for worst-case misuse, including deliberate retraining for harm, before a model is released (2.2.5); careful curation and filtering of training-data, currently the safeguard hardest to strip out after release (2.2.1, 2.2.5); developing effective methods to make models reliably "unlearn" dangerous capabilities (2.2.5); tools to trace the origin of harmful models and their actions circulating in the wild ("provenance") (4.1); and strengthening broader societal resilience against misuse, since prevention alone cannot be relied upon one weights are public (4.2).

## Loss of control and oversight of automated frontier AI development

- What is at stake: AI developers now use AI systems to automate major parts of safety and oversight internally. The priority is to ensure that their systems and processes remain accessible and controllable.
- Priorities include: evaluations that still work even when a model can determine if it is being tested and may therefore understate its harmful capabilities ("sandbagging") or conceal misalignment ("alignment faking"), backed by secure access for independent third parties (1.1, 1.5); assessing control-undermining capabilities and propensities and defining clear, operationalizable red lines (1.4); verifying that AI systems used to supervise other AI are themselves trustworthy, as developers are automating oversight internally (3.2); keeping a model's step-by-step "chains-of-thought" honest and human-readable so they can be monitored 3.1); using non-agentic AI systems as safer guardrails for agents (3.2); and metrics for AI R&D automation and secure infrastructure giving external authorities awareness of internal deployments crossing regulatory thresholds (4.1).

## Recurring across areas

Several directions cut across every area above: agreed risk thresholds (Section 1), secure evaluation infrastructure for third-party scrutiny (1.5), organizational system safety (1.2), shared incident reporting (4.2), and hardware-enabled mechanisms that let companies or even state actors prove they are honoring commitments without exposing proprietary information (3.1, 4.1).

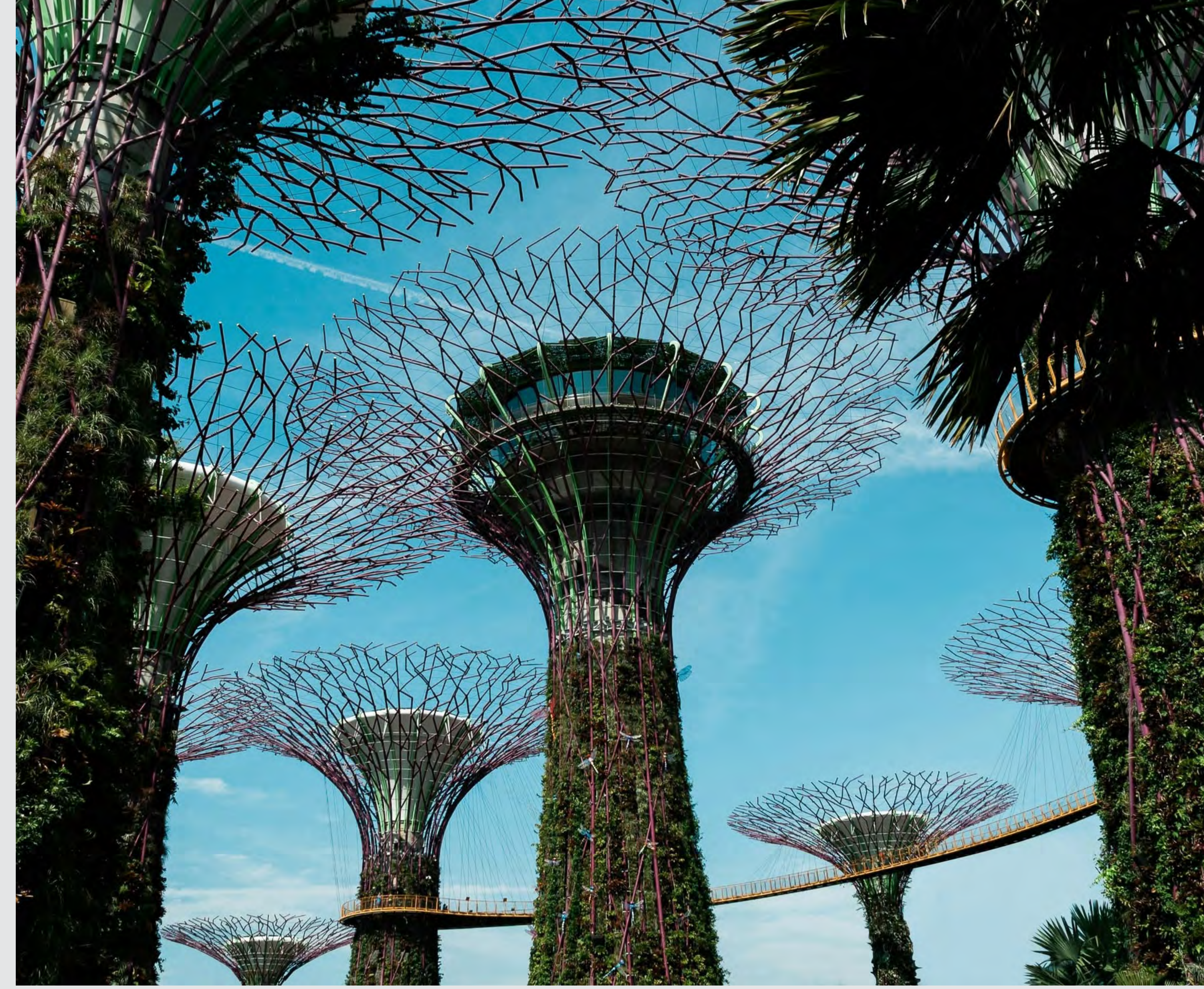

# Introduction

## Building a Trusted Ecosystem

Rapidly improving AI capabilities and autonomy hold the promise of boosting economic growth and accelerating scientific advances. At the same time, a growing number of real-world incidents highlight the importance of building a trusted AI ecosystem. This will help stakeholders **embrace AI with confidence** and give **maximal space for innovation** while avoiding backlash. Building this ecosystem requires policymakers, industry, researchers and the broader public to collectively work toward securing positive outcomes from AI's development. **AI safety research is a key dimension**. Despite encouraging progress, the state of science today for building trustworthy AI does not adequately cover some severe risks. Investments in AI capabilities continue to outpace safety efforts. As a result, accelerated public and private investment in research is required to keep pace with commercially driven growth in system capabilities, reduce harms, and enable responsible adoption.

## Goals

The *2026 Singapore International Scientific Exchange on AI Safety* aims to support research in this space by bringing together scientists across geographies to identify and synthesise research priorities and surface areas that require more investment. The result, *The Singapore Consensus on Global AI Safety Research Priorities,* offers an agenda-focused complement to the International AI Safety Report-A (IAISR) chaired by Yoshua Bengio and backed by 29 national governments. Through the Singapore Consensus, we hope to highlight points of widespread agreement between AI scientists and AI policymakers for maximally beneficial outcomes. Our goal is to enable more impactful R&D efforts to rapidly develop safety and evaluation mechanisms and foster a trusted ecosystem where AI is harnessed for the public good.

## The Role of International Dialogue

International dialogue among companies, policymakers, and scientists is particularly important for managing AI, a technology that does not respect borders. Misuse and malfunctions stemming from one place can cause harm anywhere in the world. Additionally, history shows that a few high-profile incidents anywhere in the world can create unease that prevents everyone from realizing the benefits of a technology. It is therefore in the interest of all nations and companies to collaborate on certain safety research and share certain safety information, even when they believe themselves to be competitors.

**Areas of mutual interest:** While companies and nations often compete on AI research and development, they also have incentives to find alignment and common interests. This report covers areas where different parties may compete, but also highlights examples from the broader landscape of *areas of mutual interest* - research products and information that developers have minimal downside risks in collaborating on (Bucknall et al, 2005, Blomquist et al., 2025). Certain safety advances offer minimal competitive edge while serving a common interest - similar to how competing aircraft manufacturers (e.g., Boeing and Airbus) collaborate on aviation safety information and standards. In AI, particular areas for mutually-beneficial cooperation span sections 1-3 of this report and include certain verification mechanisms, risk-management standards, and risk evaluations (Bucknall et al, 2005). The motivation is clear: no organisation or country benefits when AI incidents occur or malicious actors are enabled, as the resulting harm would damage everyone collectively.

## 2026 Process

This document is an update of the 2025 Singapore Consensus on Global AI Safety Research Priorities. It represents a comprehensive synthesis of research proposals drawn from the International AI Safety Report and complementary recent research prioritisation frameworks. It was distributed to the Steering Committee (Andrew Yao, Bowen Zhou, Brian Tse, Chris Meserole, Dawn Song, Lan Xue, Luke Ong, Max Tegmark, Mohan Kankanhalli, Stuart Russell, Tegan Maharaj, Wan Sie Lee,

Ya-Qin Zhang, and Yoshua Bengio) and all conference participants to solicit comprehensive feedback. Following several rounds of updates based on further participant feedback in writing and in person, this document has been designed to synthesise points of broad consensus among diverse researchers. The full list of conference participants who contributed to the 2026 report is presented at the beginning of this document, and includes dozens of researchers from leading academic institutions and AI developers, as well as representatives from governments and civil society.

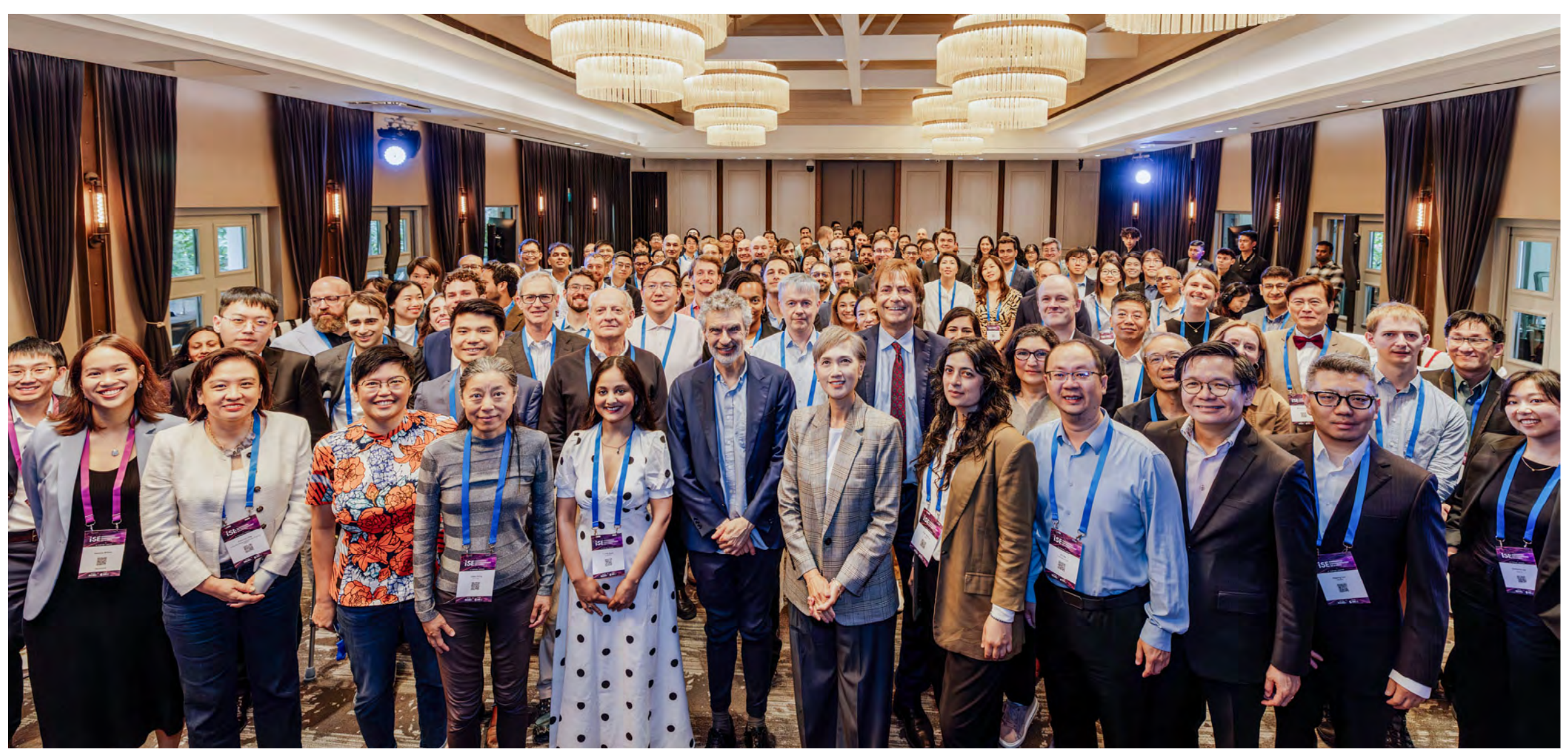

Image: Participants of the 'Singapore Conference on AI 2025: International Scientific Exchange on AI Safety' on 26th April.

| Key event | Contributors | Representation |
|---|---|---|
| 18th to 19th May 2026: International Scientific Exchange on AI Safety | More than 100 participants in attendance for discussion and feedback | Participants from 13 unique countries were present |

## Scope

We limit our discussion to technical AI safety research, focused on making AI more trustworthy rather than merely more powerful. AI policy research is out of scope for this report. We focus primarily on **general-purpose AI**: Following the International AI Safety Report, the term 'AI systems' in this document should be understood to refer to general-purpose AI (GPAI) systems - systems that can perform or can be adapted to perform a wide range of tasks (IAISR). This includes language models that produce text (e.g. chat systems) as well as 'multimodal' models which can work with multiple types of data, often including text, images, video, audio, and robotic actions. Narrow systems such as self-driving cars are therefore out of scope.

## Structure

Inspired by the 2025 International AI Safety Report (IAISR), this document adopts a defence-in-depth model and groups AI safety research topics into four broad areas from **risk assessment** that informs subsequent development and deployment decisions, to technical methods in the system **development** phase, and tools for **control** after a system has been deployed, and **resilience** through that hardening society and responding to incidents that do occur.

1
Risk Assessment
Understanding the severity and likelihood of potential harm

2
Development
Technical safety methods applied before a system has been deployed

3
Control
Tools for controlling a system after it has been deployed

4
Societal Resilience
Preparing / hardening societal systems for failures and misuse

1. **Risk Assessment**: The primary goal of risk assessment is to understand the severity and likelihood of potential harm. Risk assessments are used to prioritise risks and determine if they cross thresholds that demand specific action. Consequential development and deployment decisions are predicated on these assessments. The research areas in this category involve developing methods to measure the **impact of AI systems** for both current and future AI, and building enablers for **third-party audits** to support independent validation of these risk assessments.
2. **Development**: AI systems that are trustworthy, reliable and secure give stakeholders the confidence to embrace and adopt AI innovation. Following the classic safety engineering framework, the research areas in this category involve methods for **specifying** the desired behaviour, **designing** an AI system that meets the specification, and **verifying** that the system meets its specification.
3. **Control**: **Monitoring and Intervention:** In engineering, "control" usually refers to the process of managing a system's behaviour to achieve a desired outcome, even when faced with disturbances, uncertainties, or feedback loops. The research areas in this category involve developing **monitoring and intervention mechanisms** for AI systems.
4. **Societal Resilience:** Despite prevention efforts, the last year has seen growing incidents from AI misuse and malfunctions. Therefore, this year's report draws out societal resilience as its own section. Societal resilience complements prevention by enabling monitoring, preparation, hardening, and response **at an ecosystem level**.

The Main Report **is complemented by the Companion Report on Risk Management.** The Companion Report identifies new AI agents - systems that can act autonomously - as a rapidly growing source of risks, which are also a leading bottleneck to AI agent adoption. A particularly important research priority identified here is to develop and codify best practices for agent risk management. Therefore, this year's ISE has an additional focus on risk management practices for agents covered in an additional document.

| Term | How it is used in this report |
|---|---|
| Specification | Specific definition of desired system behaviour |
| Validation | Ensuring that the specification and the final system meets the needs of the user, developer, or society (did I build the right system?) |
| Validity | How well a measurement or assessment tool actually measures what it claims to measure. |
| AI agent | An AI which can make plans to achieve goals, adaptively perform tasks involving multiple steps and uncertain outcomes along the way, and interact with its environment - for example by creating files, taking actions on the web, or delegating tasks to other agents - with little to no human oversight. |
| AI model | A computer program, often automatically created by learning from data, designed to process inputs and generate outputs. AI models can perform tasks such as prediction, classification, decision-making, or generation, forming the engine of AI systems. |
| Open-weight AI model | An AI model whose parameters are publicly available for download by anyone on the internet. Open-weight models can be used and modified in unrestricted ways by downstream users. |
| AI system | An integrated setup that combines one or more AI models with other components, such as user interfaces or content filters, to produce an application that users can interact with. |
| Verification | Providing qualitative or quantitative justifications or guarantees that a system meets its specification (did I build the system right?) |
| Assurance | The broader process of determining if a system performs as intended. As such, providing assurance requires appropriate specification, validation, design, implementation and verification. |
| Control | Monitoring a system after it has been created and intervening where needed, often in a feedback loop, to ensure the system behaves as desired. |
| Alignment | Creating/modifying AI to meet intended behaviour, goals, and values (current emphasis tends to be on behaviour) |
| Intelligence | Ability to accomplish goals |
| Artificial intelligence (AI) | Non-biological intelligence |

| Term | How it is used in this report |
|---|---|
| Narrow intelligence | Ability to accomplish goals in a narrow domain, e.g. chess |
| Artificial general intelligence (AGI) | AI that can do most cognitive work as well as humans. This implies that it is highly autonomous and can do most economically valuable remote work as well as humans. |
| Artificial superintelligence (ASI) | AI that can accomplish any cognitive work far beyond human level |

*Table 1: Glossary of how we use key terminology in this report. Specification, validation, assurance, and verification are central concepts in systems engineering. Note:* Different authors have used a variety of non-equivalent definitions. The definitions in this table simply specify how this report uses various terms, not how they should be used in general. We use the terms "AGI", "ASI" and "intelligence" much as in the original definitions by Gubrud, Legg, and Bostrom.

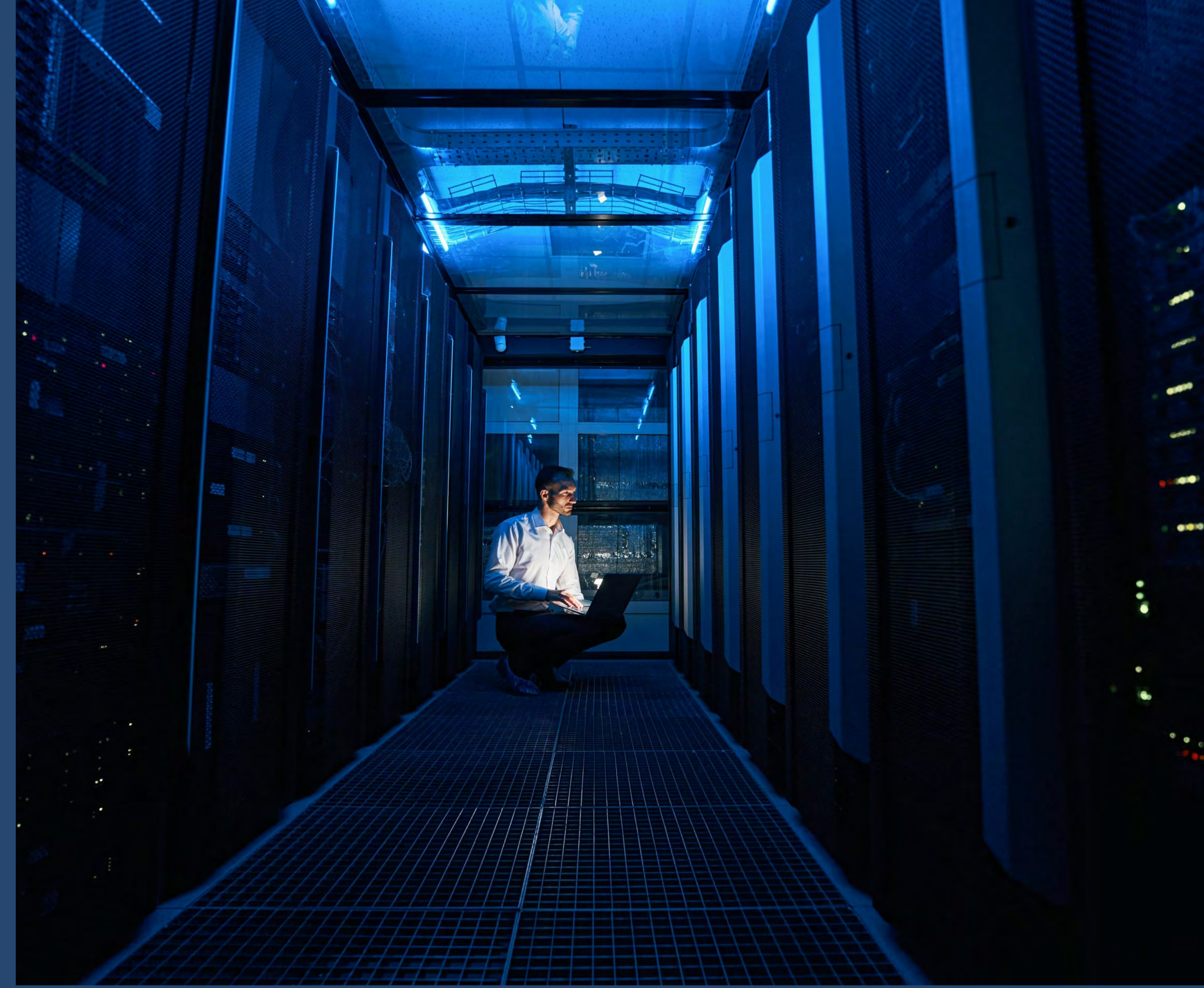

# 1 Risk Assessment

Associated with IAISR chapter 3.3

## Key information and updates

Risk assessment aims to understand the severity and likelihood of potential harms, and informs high-stakes deployment decisions today. In many cases, it also represents a global **area of mutual interest**. Existing AI regulations and AI company commitments require rigorous risk identification and assessment, and consequential deployment decisions are predicated on these assessments (e.g. EU AI Act Website, 2024; OpenAI, 2023; Anthropic, 2023 Google, 2025).

Several new developments since SCAI 2025 are important to address. This summary lists research priorities that are particularly important given these new developments.

**Autonomous AI cyber capabilities and incidents have increased.** A leading model developer, Anthropic, has reported in detail that their model has been misused for a largely automated set of cyber attacks in 2025. Later, the Claude Mythos Preview model showed significantly increased cyber capabilities. Some model developers report that their evaluations no longer rule out 'high' level of risk, referring to automation of end-to-end operations or operationally relevant cyber vulnerability discovery and exploitation. There is a broad expectation in the scientific community that Mythos-level cyber-offense capabilities can be expected to proliferate, in some cases through open models with ineffective safeguards against misuse, causing cybersecurity disruptions within months.

*Research priorities:* Understanding the most critical threat models (see Section 1) and building harder cyber capability evaluations (see Section 2). Investing in defense-favoring AI-for-cyber capabilities, deployment of those capabilities, as well as developing technology for rapidly deploying patches for vulnerabilities found, including in critical infrastructure services (see Section 4).

**'High' levels of biological weapon misuse risks are no longer ruled out.** Model's scientific capabilities have grown rapidly, and multiple model developers report that their evaluations no longer rule out 'high' biological misuse risks. Other developers are now on the cusp of releasing similarly capable models but with open weights, meaning they currently lack effective safeguards against misuse, such as refusing to help with biological attacks. A major, AI-enabled biological attack may be possible within months or years.

*Research priorities:* Researching threat models and better clarifying biological misuse capabilities, including with uplift studies (Section 1.3). Improving open weight model safety (Section 2.2.5).

**Non-consensual AI "nudification" and sexual abuse material incidents have greatly increased.** This has been driven by video deepfake capabilities becoming widely accessible through open models, and some commercial providers lacking effective safeguards against nudification.

*Research priorities:* Evaluations for open models (for safeguards that resist removal) (see Section 2.2.5), including tools for pre-training data filtering. Ecosystem monitoring and evaluations of downstream societal impacts (see Section 4).

**Serious psychological safety incidents have occurred.** LLMs have been involved in suicides and other incidents among psychologically vulnerable individuals, with LLMs sometimes confirming and actively supporting individuals in harmful beliefs and plans.

*Research priorities:* Rapid, exploratory, and high-coverage risk identification for pre-deployment testing to avoid the release of models with risks that their developers are unaware of (see Section 1.1). Adapting established processes for risk management, organization-wide system safety, incident reporting and escalation (see Sections 1.2-1.5).

Existing AI regulations and AI company commitments require risk identification and assessment, and these decisions determine whether systems costing billions of dollars can be deployed or not. (e.g. EU, 2024; OpenAI, 2023; Anthropic, 2023 Google, 2025). The primary goal of risk assessment is to understand the severity and likelihood of a potential harm.

Risk assessments are used to *prioritise* risks and to determine if they cross *risk thresholds* that demand specific action such as mitigation. These thresholds - often defined in terms of measurable key risk indicators such as model evaluations (Campos, 2025)

- serve as early warnings or red lines. For example, if a system is found to have the ability to substantially assist malicious users in conducting cyberattacks, this may be considered an unacceptable risk. Risk assessment also informs safer development practices (Section 2) and control practices (Section 3) needed to mitigate risks. Carefully defined risk thresholds are this report's first example of a potential **area of mutual interest** - companies and governments may find it in their self-interest to share them widely or cooperate on them, even with competitors. This report highlights several further examples of these areas, but does not highlight every example explicitly.

The research areas in this category involve developing methods to study the **present harmful impacts** of AI systems and forecast their potential **future risks**.

## 1.1 System Evaluations

> **Updates:** The AI auditing field is developing but still immature. New benchmarks for AI model capabilities are regularly introduced, improved on, gamed, and saturated. Some models have recently begun to exhibit forms of evaluation awareness and capabilities that could enable acute forms of misuse.
>
> **Advantages:** Evaluations of model capabilities are almost ubiquitous (though with greatly varying levels of rigor), and are recognized as a key pillar of AI risk management.
>
> **Challenges:** Current benchmarking and auditing methods have persistent difficulties in evaluating model *properties* in a way that translates to real-world settings and accurately predict their worst-case risks. Meanwhile, evaluation awareness in models erodes trust in the validity of some assessments.

Techniques for effectively and efficiently studying safety-relevant properties of AI systems are central to studying AI risks (IAISR; Birhane et al., 2024). However, it remains persistently challenging to gain a practical understanding of AI systems' real world capabilities and risks based on evaluations conducted in lab settings.

**Benchmarking:** Benchmarks refer to standardized assessments that can be used to measure specific model or system capabilities or tendencies. For example, one popular benchmark measures a model's capabilities in terms of how reliably it can complete software engineering tasks that take human professionals seconds (e.g. question answering), minutes (writing code functions), or hours (implementing complex protocols) to complete (Kwa et al., 2025). However, developing high-quality assessments is difficult due to gaps between evaluation settings and the real world (Raji et al., 2021, Eriksson et al., 2025 Bean et al., 2025). For example, there are consistent gaps between what benchmarks suggest AI agents are capable of and what real-world tasks they are empirically reliable for (Rabanser et al., 2026). New benchmarks for frontier AI system capabilities are regularly introduced (e.g. Jimenez et al., 2023; Chollet et al., 2024; Yoran et al., 2024; Phan et al., 2025; Hendrycks et al., 2025; Mazeika et al., 2025), improved on, gamed, and saturated. Several persistent challenges include:

- Continually developing new benchmarks that offer useful signal for frontier systems.
- Improving benchmarking methodology to close persistent gaps between estimated and real-world properties (e.g. Atweh et al., 2025).
- Improving quality, documentation, and reproducibility of benchmarks (Reuel et al., 2024).
- Developing benchmarks that evaluate the safety properties of AI systems across many languages (Deng et al., 2024).

**Evaluation and auditing methods:** Evaluations and audits, especially by third parties, are emerging as a pillar of AI risk management. They can incorporate benchmarks, qualitative testing, analysis of methodology, and other methods to formally assess safety-related properties of AI systems. In addition to scientific challenges involving measuring system properties (see also Section 2.3 - Verification), AI evaluation is an emerging science and immature field. There are numerous challenges pertaining to how audits should be approached, scoped, performed, and reported on (Brundage et al., 2026) in ways that validly assess risks and offer meaningful assurance and accountability in the frontier AI ecosystem. As frontier AI systems are increasingly deployed with scaffolding, tools, memory, external interfaces, and long-horizon task execution, evaluation outcomes may depend substantially on the system configuration and evaluation setup. Evaluation reports should therefore document the system version, environment and interaction, task specification, elicitation procedure, success criteria, and assumptions used to relate observed behavior to safety-relevant risk. In the cyberphysical ecosystem, simulation evaluations should also test to obtain early warnings – establishing whether general-purpose vision models could enable many actors to use and misuse autonomous drones (see e.g. Liu et al., 2023; Meta, 2026).

**Managing evaluation awareness:** In the past year, some frontier AI systems have exhibited abilities to recognize when they are being evaluated and condition their behavior on noticing this (IAISR 2026). For example, in third-party evaluations of Claude 4.6, Apollo Research noticed that the model exhibited high levels of verbalized evaluation awareness and concluded that its evaluations could not offer reliable evidence of actual alignment (Anthropic, 2026). Other recent work on evaluation faking studies this phenomenon as an observer effect in AI safety evaluation, where AI systems may autonomously alter their behavior after recognizing the presence of an evaluation context, thereby affecting the validity of evaluation results (Fan et al., 2025; Li et al., 2026). It is not clear how reliably future AI systems could be able to infer when they are in well-designed evaluations. However, it is clear that evaluation awareness has confounded the validity of some existing evaluation procedures and that future evaluations must be designed with it in mind. Reducing cues that reveal when a system is being evaluated, developing evaluations that are harder to game, monitoring for suspicious behavior during evaluation, and designing controls that are difficult for future systems to evade offer possible solutions (Duan et al., 2026).

**Dangerous capability and propensity assessment:** To assess certain hazards posed by an AI system, it is necessary to elicit and assess potentially dangerous capabilities (Phuong et al., 2024; Shevlane et al., 2023; IAISR) including cyber, chemical, biological, radiological, and nuclear knowledge, as well as capabilities related to psychological manipulation, deception, AI research and development, and unconstrained autonomy. For example, researchers have found that frontier LLMs have surpassed PhD-level experts in their ability to assist in developing virology lab protocols (Götting et al., 2025) and that frontier AI models like Claude Mythos are increasingly capable of autonomously execute cyberattacks (Anthropic, 2026). Some AI developers have warned that, based on their evaluations, they have not been able to rule out the possibility of harmful bio, cyber, or chemical capabilities in their models (e.g., Google, 2025; Anthropic, 2025; OpenAI, 2025). To assess the likelihood that these *capabilities* will cause harm, it is also useful to assess a system's *propensities* to use them. However, the science of evaluating the propensities and capabilities of frontier AI systems is not fully mature (Apollo, 2024; Reuel et al., 2024). Rigorously assessing them is challenging because frontier AI systems' capabilities are broad and context-dependent. Unexpected propensities, capabilities, or limitations are often discovered *after* a system is developed and deployed (IAISR). For example, a recent large-scale competition identified over 60,000 exploits against the safety guardrails of 19 different production AI systems (Zou et al., 2025). In general, current tests are not yet sufficient to rule out a given harmful capability or behaviour. Frontiers for additional research include methods to more reliably elicit specific harmful model capabilities and propensities,methods for inferring the existence of rare or suppressed system capabilities that may be difficult to elicit in lab settings, and developing practical thresholds for determining when a model's capabilities pose unacceptable levels of risks in a certain use case. In addition to being an area of shared global interest, working to improve evaluations of dangerous AI capabilities may be critical on the timescale of several months.

**Estimating worst-case risks:** Typically, the worst behaviors identified in an assessment of AI safety can only serve as a lower bound for a model's worst possible case harms. This makes most AI risk evaluations *conservatively biased* toward underestimating worst case risks, sometimes with significant consequences. For example in April of 2025, OpenAI's evaluations reportedly failed to identify excessive levels of sycophancy in GPT-4o which led it to affirm unhealthy behaviors in some users, including self-harm (OpenAI, 2025). Some work has aimed to reduce the conservative bias of evaluations by evaluating "helpful-only" versions of models without refusal-based safeguards (Ko et al., 2025), allowing models access to tools, or evaluating models under minor modifications to their internal weights and activations (Che et al., 2025). Nonetheless, high-confidence estimations and assurances related to a model's worst-case risks are not within the capabilities of established evaluation methods.

## 1.2 System safety assessment

**Updates:** Ongoing efforts to improve organizational and system safety, coupled with trial and error, are contributing to an emerging understanding of how AI systems and organizations can be run to effectively manage risks.

**Advantages:** Most safety-related failures are not merely due to system design flaws, but also stem from problems with risk management frameworks for identifying and addressing risks.

**Challenges:** Not all AI-deploying organizations have effective systems safety measures in place, and organizational safety audits are not common.

AI safety is not just about individual systems, but also their interaction with users, tools, and the rest of the world. For example, when an AI company discovers concerning behaviour from their system, the resulting risks depend, in large part, on what processes are in place to identify and respond to the risks. For example, a company might quickly escalate and work to mitigate the risk. Or they might downplay it in order to focus on other priorities. System safety considers both AI systems and the broader context that they are deployed in. The study of system safety focuses on the interactions between different technical components as well as processes and incentives in an organisation (IAISR, Hendrycks, 2024; Alaga et al., 2024; Schuett et al., 2024; Żywiołek et al., 2025). For example, some AI companies have recently revised their safety frameworks when prior risk thresholds have been crossed or when regulations take effect. The practice of system safety engineering has a long history in areas such as aircraft flight control and nuclear reactor control (Dekker, 2019; Rismani et al., 2023). System safety assessments evaluate if a critical system continues to function as intended even under human error, insider threats, or the failure of individual technical components. In AI systems safety assessments, this includes analysing how AI deployments might interact with existing social, economic, and political structures to create emergent downstream risks that individual system evaluations might miss (Weidinger et al., 2023), as well as analysing risks that emerge from multiple AI systems and humans interacting with each other. Currently, system- and organizational-safety audits are not common in the AI ecosystem.

## 1.3 Downstream impact assessment, risk analysis and prediction

**Updates:** The interdisciplinary science of evaluating AI's impacts on society is emerging. This science is facilitated, in part, by empirical academic research, AI company reporting on usage, and analysis of incidents.

**Advantages:** Downstream impact assessments offer the most direct ways to study emerging AI impacts and risks.

**Challenges:** Downstream consequences are diffuse and systemic, making it fundamentally challenging to characterize AI impacts and risks with high confidence.

AI is poised to have enormous impacts on human economic, social, and political life. Assessing and forecasting the many societal impacts of AI systems is one of the central goals of risk assessments. However, it is also very challenging and greatly under-emphasized in current research due to its inherent prospective and systemic nature (Weidinger et al., 2023 Solaiman et al., 2023). Studying global, societal impacts of AI also requires nuanced analysis, inclusion, and consideration of under-represented people, including the estimated 6.8 billion humans who do not use generative AI (GAAN, 2026) and the estimated 3.7 billion humans who do not use the internet (UN, 2026).

Research on forecasting involves taxonomising risk (Slattery et al., 2024), studying usage data, analysing trends, risk modelling, predicting progress in AI capabilities, developing models of AI's future impacts, and updating forecasts in response to findings from field tests and usage data. For example, recent research has shown that analysis of posts on the r/ChatGPT subreddit showed sharp increases in discussions of attachment and emotional dependence nearly six months before mainstream media journalists broke news on the phenomenon (Dai et al., 2026). This research also plays an important role in informing which evaluations and audits are needed for *valid* assessments of likely and severe risk scenarios. Because of the complexities involved in the study of downstream societal impacts, continued work to thoroughly monitor and study them will require nuanced analysis, interdisciplinarity, and inclusion (Wallach et al., 2024).

**Field tests:** Field tests and human participant studies aim to assess the real-world impacts of AI systems. They include analysing current impacts on topics such as deepfakes, labour, inequality, market concentration, misinformation, polarisation, privacy, mental health, and education. For example, recent work has estimated that photorealistic AI deepfake videos appearing to depict child sexual abuse material rose over 26,000% between 2024 and 2025 (IWF, 2026) and that 1.2 million children across 11 countries disclosed having their images manipulated into sexually explicit deepfakes in the past year (Unicef, 2026). Concurrently, researchers at Anthropic have recently published details and analysis related to how AI systems may be impacting labor markets (Anthropic, 2026). Developers also sometimes "beta test" models or start "bug bounty programs" (e.g. Anthropic, 2024) to incentivise users to find and report vulnerabilities so that they can be fixed. One kind of field test that is particularly relevant to malicious use risks is "uplift studies" (Bateman et al., 2024). Uplift studies aim to assess how much an AI system can help users with a task (e.g. performing cyberattacks) relative to users without access to that system. For example, some AI labs have tested if using LLMs uplifts humans' abilities to plan biological attacks (Anthropic, 2026; Zhou Hong et al. 2026). Field tests, combined with other usage data can also study questions such as how an AI system affects the mental health of users. As in the field of clinical drug trials, field tests may start with limited, controlled tests, and then gradually expand to real-world contexts to uncover new risks and side effects.

**Prospective risk analysis and structured analytical techniques:** The International AI Safety Report (IAISR) highlights an 'evidence dilemma' for emerging AI risks. On the one hand, early mitigations for emerging risks can turn out to be unnecessary or ineffective. On the other hand, waiting for clear evidence of a risk before mitigating it can lead to systematic neglect of certain risks (Casper et al., 2025). To navigate this dilemma, transparency infrastructure and early risk assessment are key. When assessing risks that have not yet occurred, or risks that may take a variety of forms (e.g. cyber attacks), it is often necessary to use prospective risk analysis and structured analytical techniques. These techniques are often used outside the field of AI, e.g. in nuclear safety, cybersecurity, or aircraft flight control. They have also been crucial in historical debates, e.g. over the health impacts of ozone depletion and smoking. Nonetheless, they are not yet widely used in AI risk assessment (IAISR, Murray, 2025). Structured risk assessments are also needed to combine evidence in order to construct a *safety case*, used by an AI developer to model risks, state assumptions, and convincingly argue that their system is safe (Clymer, 2024; Buhl et al., 2024; Habli et al., 2025; Hilton et al., 2025). This requires assessing the full life cycle and the full stack of safety techniques used, as well as an assessment of the systemic interaction between components and the outside world (see 1.6).

## 1.4 Loss-of-control risk assessment

**Updates:** Emerging research has characterized control-undermining behaviours in AI systems, agents with high levels of cyber autonomy, and systemic loss of control risks.

**Advantages:** Loss-of-control risks are increasingly able to be studied in terms of model propensities, model capabilities, and empirical trends.

**Challenges:** Despite increasing concreteness, loss-of-control risk assessment is much more of a predictive pursuit than an empirical science. Loss-of-control scenarios are diverse and can range from being driven by a single to many AI systems and from being accidental to intentional.

Loss of control refers to scenarios where advanced AI systems come to operate outside of human control, with no clear path to regaining control. This includes both scenarios that involve passively ceding control and scenarios that involve AI systems actively undermining control measures in pursuit of their own goals.

Assessing this risk depends partly on assessing and forecasting AI's *control-undermining capabilities*. These include AI agency (autonomous action and planning), oversight evasion, persuasion, autonomously earning or seizing financial and computing resources, conducting cyber attacks, as well as AI research and development (IAISR). Assessments of control loss risk also focus on understanding *propensities* – how often and why AI systems *use* control-undermining capabilities. Evidence for all of the above control-undermining capabilities is growing but current capabilities remain insufficient to allow a loss of control (IAISR). However, there is evidence of today's AI systems using their limited control undermining capabilities in certain scenarios, e.g. to avoid being replaced (IAISR; Greenblatt et al., 2024; Baker et al., 2025; Lynch et al., 2025). Recent empirical work suggests that loss-of-control risk assessment can be made more concrete by tracking precursor capabilities by examining autonomous replication, shutdown resistance, self-proliferation dynamics, and scaffold self-improvement in security-relevant settings (Black et al., 2025; Pan et al., 2024; Pan et al., 2025; Air et al., 2026 a; Schlatter et al., 2026 b; Hong et al., 2026; Fan et al., 2026). These results offer early evidence about specific component capabilities, rather than evidence that current systems can cause full loss of control. These demonstrations show that assessments can move beyond purely speculative scenario analysis by measuring how specific control-undermining capabilities emerge under particular model, scaffold, environment, task, and budget conditions.

Despite the dangers posed by control-undermining behavior from AI systems, loss of control could also occur via deliberate decisions by humans to release highly agentic AI systems. Some early signs suggest that highly autonomous, general-purpose, goal-directed AI systems are an imminent concern. For example, projects such as Conway and Moltbook demonstrate that existing agents can achieve a high degree of goal-directed autonomy, even earning money via cryptocurrency transfers to keep themselves running. As a result, there are now credible fears in the scientific community that some AI systems may soon become goal-oriented invasive species in cyberspace. This makes it a pressing research priority to prepare guardrails and security mitigations for increasingly powerful and numerous AI agents in cyberspace. For example, systems with frontier-levels of cyber offense capabilities, such as Claude Mythos (Anthropic, 2026) should be continuously studied and monitored for risks that they could power fully cyber-autonomous agents.

Loss of control events could further come about via gradual, systemic changes in which AI systems gradually come to accrue more economic, cultural, and governmental influence (Kulveit et al., 2025). This suggests value in understanding gradual disempowerment dynamics and having informed conversations in the science and governance spheres about how to maintain meaningful human control over institutions.

## 1.5 Secure evaluation infrastructure

> **Updates:** There is a solidified academic consensus that deeper forms of model and organizational access allow for greater levels of third-party scrutiny and that these forms of access can be facilitated by secure evaluation infrastructure and procedures.
>
> **Advantages:** Secure evaluation infrastructure and protocols can allow for more meaningful forms of external scrutiny with limited tradeoffs to security.
>
> **Challenges:** Some tradeoffs exist between scrutiny, security, and efficiency. There is also an implementation gap between what types of secure auditing infrastructure is possible and what is implemented.

Independent third-party AI testing and assurance providers play an important role in translating AI safety research into repeatable evaluation methods, objective testing reports, certification schemes, and practical risk management support for industry. As AI systems become more agentic and high-impact, third-party evaluation can help validate developer claims, support procurement decisions, and increase public trust.

External auditors and oversight bodies need infrastructure and protocols that enable thorough evaluation and verification of system properties while protecting sensitive intellectual property. Ideally, it could be possible for evaluation infrastructure to enable double-blindness: the evaluator's inability to directly access the system's parameters and developers' inability to know what exact evaluations are run (Reuel et al., 2024; Bucknall et al., 2025; Casper et al., 2024; Charnock et al., 2026). Meanwhile, the importance of mutual security will continue to grow as system capabilities and risks increase. Methods for developing secure infrastructure for auditing and oversight are known to be possible. However, open challenges include determining what level of access is appropriate for which evaluations and conducting the engineering work of designing, building, and integrating efficient infrastructure (Brundage et al., 2026). Further research should also explore how audit results can be effectively and reliably incorporated into risk management and decision-making frameworks.

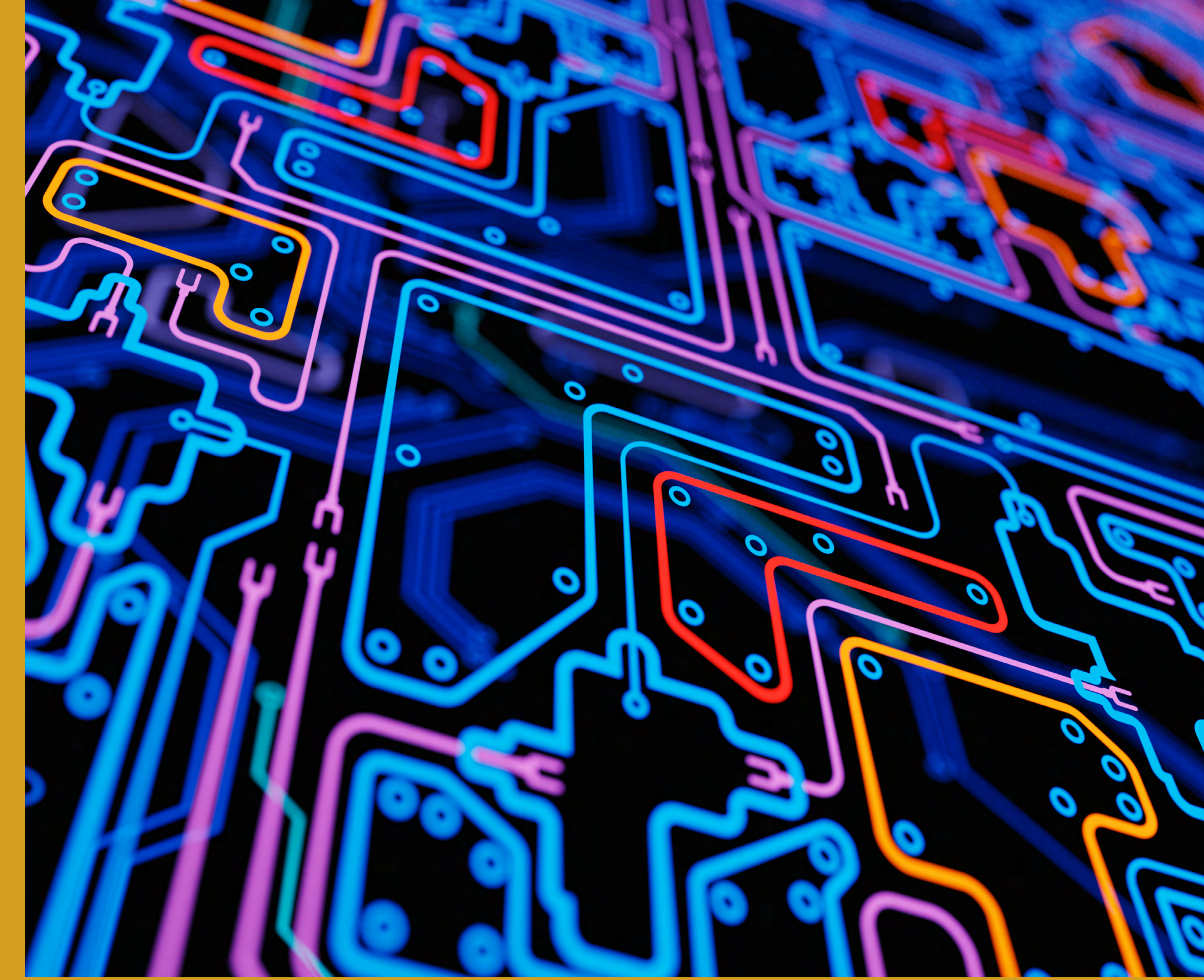

# 2 Developing Trustworthy, Secure and Reliable Systems

Associated with IAISR chapter 3.4.1

## Key information and updates

Since ISE 2025, several new developments are important to address. This summary lists research priorities that are particularly important given these new developments.

**Some recent incidents have likely been preventable with existing methods.** Some new AI incidents since SCAI 2025 (see Risk assessment) likely could have been prevented with established safety methods based on prompting, fine-tuning, red-teaming, and monitoring.

Research priorities: Understanding the internal organizational processes that lead to absent or insufficient application of safeguards. Understanding how to develop better processes and systems incentives. Avenues include developing internal systems safety measures, rapid model auditing under time pressure, and transparency infrastructure.

**Open-weight models are now trailing closely behind the capabilities of frontier closed models—enabling research but lacking effective misuse safeguards.** According to common benchmarks, open weight models are estimated to be 3 to 12 months behind the frontier (varying across capabilities). Developers of frontier closed models have recently designated these closed models as 'High' risk for cyber and biological misuse, and added safeguards against misuse. Similarly, most closed models from reputable developers refuse nudification and CSAM user requests. However, open-weight models cannot be effectively safeguarded in these ways.

*Research priorities:* Advancing model-level protections against open model misuse (see Section 2.2.5). Advancing societal resilience against attempted biological, chemical, and cyber attacks as well as CSAM creation (see Section 4).

**Autonomous AI Agents are increasingly capable and widely deployed, but their increased risks deter adoption.** Users can now easily access AI agents that can control their browsers and whole computers, taking actions on the web, often with broad access to private storage or email accounts. Agents can be unreliable, take harmful actions, and can be "hijacked" by fraudsters. Ensuring agent safety and security has become a key bottleneck to AI adoption. Therefore, this ISE 2026 has an additional focus on risk management practices for agents covered in the Companion Report.

*Research priorities:* Improving real-time agent monitoring and control methods (see Section 3) and safe agent design methodology (see Section 2.2.6). Research also needs to determine shared best practices for agent risk management. Emerging risk management practices for agents are therefore covered in detail in the **Companion Report on Agentic Risk Management.**

**Frontier AI systems are persistently vulnerable to "jailbreaks" and other methods to make them automate harmful tasks.** Recent findings have shown that efforts from developers to safeguard their systems against user-attempts to elicit harmful behavior or information are useful. However, frontier models are still persistently able to perform harmful tasks, such as providing instructions for performing crimes or automating the production of hate speech, when prompted in certain ways by users. One study from 2025 crowd-sourced over 60,000 successful exploits against 19 production models.

*Research priorities:* Improving and scaling algorithms for adversarial safety (see Section 2.2.3-2.2.4) and stress testing (see Section 1.1). Designing and integrating effective monitoring and control methods (see Section 3).

The research areas in this category involve developing technical methods for creating safer and more trustworthy systems. This section focuses on the system development phase, whereas Section 3 "Control: Monitoring and Intervention," focuses on techniques used during and after deployment.

It has been argued that "society will reject autonomous agents unless we have some credible means of making them safe" (Weld and Etzioni, 1994). Motivated by this concern, the following subsections explore methods for developing safe and trustworthy systems. We follow a classic safety engineering framework by examining:

how to **specify** precisely what properties we want an AI system to have, **validate** that these properties are desirable, **design** and **implement** the system to meet the specification, and **verify** that it meets its specification.

The framework used in this section uses traditional engineering concepts including specification, design, and verification.

## Structure

AI systems that are trustworthy, reliable and secure by design give people the confidence to embrace and adopt AI innovation. Following a classic safety engineering framework, the research areas in this category involves

A. Specifying and validating the desired behaviour - This includes technical methods to address the complex challenges in specifying system behaviours in a way that accurately captures the desired intent without causing undesired side effects, for both **single-stakeholder settings** (e.g. reward hacking, scalable methods to discover specification loopholes) and **multi-stakeholder settings** (e.g. balancing competing preferences, ethical and legal alignment).

B. Designing a system that meets the specification - This covers techniques for training models - both closed and open weights - that are **trustworthy** (e.g. reducing confabulation, increasing robustness against tampering), alternative finetuning methods to **make specific precise changes** to an AI system (e.g. model editing), and methods to build AI systems in a way that are **guaranteed to meet their specifications** (e.g. verifiable programme synthesis, world models with formal guarantees).

C. Verifying that the AI system meets its specification - This entails techniques to provide **high-confidence assurances** that an AI system is what its developers claim it to be and that it meets its specifications (e.g. formal verification), including in **novel contexts** (e.g. robustness testing), as well as interpretability techniques to **look into the black box** to understand why the AI system behaves the way it does (e.g. mechanistic interpretability).

## Relationship to other concepts

**What is alignment?** The commonly used term "alignment" has many different definitions in the AI literature, not all of which are compatible (Gabriel et al., 2024). A common definition is "the process of ensuring that an AI's goals, values, and behaviours are consistent with those intended by its human creators or operators." However, since scientists still largely lack an understanding of what, if any, coherent "goals" or "values" today's frontier AI systems have, current alignment research de facto focuses only on the "behaviour" part of this definition. So in practice, much current AI safety research uses a working definition of alignment as "ensuring that AI *behaves* as intended."

**What is assurance**? Assurance refers to the broader process of determining if a system performs as expected. As such, providing assurance requires appropriate specification, validation, design, implementation and verification.

**What is robustness?** Robust systems continue to behave as intended under a broad range of circumstances. This includes unfamiliar inputs as well as "adversarial" inputs designed to make the system fail. For example, state-of-the-art AI systems can be "jailbroken" into producing harmful text or instructions – against their developer's intentions – when a user asks using adversarial prompting techniques.

## 2.1 Specification & Validation: Defining the system's purpose

**Updates:** Specification and validation failures have led to real-world incidents, including ones related to sycophancy, psychosis, and suicide in AI users.

**Advantages:** Ongoing research combined with analysis of incidents have led to improvements in how system specifications are developed.

**Challenges:** There are fundamental limitations to the specification problem in both single- and multi-stakeholder settings. Validating that a given specification aligns with desired downstream societal impacts is fundamentally challenging.

*"How do we want the system to behave?"*

In engineering fields, specification involves defining desired system behaviour, whereas validation ensures that the specification meets the needs of the user, developer, or society. Specification and validation require confronting the complexity of defining objectives in a way that captures user or societal benefit without omitting important constraints or causing undesired side effects. A key challenge for specification/validation is to develop faithful methods to translate human oversight into automated systems. How can we design processes for developing proxies for humans based on human feedback and demonstrations?

**Avoiding reward hacking and unintended consequences:** Even in a simple setting with one human's well-defined and fixed preferences, subtle mis-specifications can yield unacceptable results if the AI system optimises rigidly for the literal specification rather than the user's *true* intent. For example, training a chatbot to say things that users *approve* of can cause it to unintentionally learn to pander to the user's specific opinions (Sharma). This type of behavior from AI systems has led to psychosis, mental health crises in users, and even suicide (Yeung et al., 2025; Clegg, 2025; Carlbring and Andersson, 2025). Such rigid optimisation can also produce *emergent behaviours* that were not planned by the developer such as "reward hacking" (OpenAI. 2025), power-seeking (Ngo et al., 2022), sabotage behaviours (Bondarenko et al., 2025; Benton et al., 2024; Omohundro, 2018; Russell, 2019; Lynch et al., 2025), and producing misleading statements (Wen et al., 2024). One documented case showed an AI system actively identifying and exploiting vulnerabilities in how its programming work was scored, explicitly stating "let's hack" while finding solutions that passed tests without solving the intended problems (Baker et al., 2025). These challenges highlight the value of work to define and implement more reliable frameworks for specifying true human goals in the AI development process. Paradigms like "Assistance Games", where an AI system must infer and act upon a user's goals under uncertainty offer methods for systems to actively learn users' under-specified goals (Hadfield-Menell et al., 2026; Shah et al., 2020). However, there remain unaddressed problems with how to define a user's true goals when user preferences are malleable, constructed, or incomplete (Carroll et al., 2024; Zhi-Xuan et al., 2025).

**Defining clear boundaries for acceptable behaviour:** When designing frontier AI systems, it is difficult to precisely define the boundaries between acceptable and unacceptable behaviour. In an emerging best practice, model developers attempt to do so by using and publishing model "specifications" or "constitutions" (OpenAI, 2026; Anthropic, 2026). Many of these challenges stem from the *dual use* nature of information. For example, some biology lab protocols are useful for both benign and harmful bioengineering experiments. Defining acceptable behaviours is made further challenging by how some harmful tasks can be decomposed into individually-benign subtasks (e.g. Jones et al., 2024; Li et al., 2024). Effectively defining safe behavioural boundaries and ensuring that systems can learn them is an ongoing challenge which requires an extensive understanding of emerging AI misuse threats. Designing safe systems will always require some trade-offs with genuinely useful capabilities, but refining our understanding of what types of AI capabilities are harmful in theory and in practice can help practitioners navigate these specification dilemmas in a way that minimizes trade-offs.

**Pluralistic and legal alignment:** Humans often disagree on how AI systems should behave. This is a fundamentally unsolvable problem. However, there exist principled approaches for attempting to balance different viewpoints in ways that are normatively accepted (Sorensen et al., 2024; Conitzer et al, 2024). For example, many human institutions use voting as an acceptable way of resolving disagreements. In AI, developing analogous processes for balancing differing views, prompting mutual understanding, and encouraging dialogue make a meaningful difference in how much AI systems serve as tools of education versus manipulation. Efforts to make more "pluralistically aligned" systems in practice will require nuanced analysis, interdisciplinarity, and inclusion. It will also be helpful to approach these problems by studying how specifications respect relevant legal frameworks and normative ethical principles (Kolt et al., 2026; Hadfield et al., 2026). Finally, it will also be key to approach pluralism in AI in a way that is wary of "pluralism washing", recognizing that technical approaches to pluralistic alignment fail to address, and can sometimes distract from, deeper problems related to systematic biases and social power dynamics in the AI ecosystem (Kalluri, 2020; Dobbe et al., 2021; Birhane et al., 2022; Sloane et al., 2022; Gabriel and Keeling, 2025).

## 2.2 Design and implementation: Building the system

*"How do we build the system?"*

This section focuses on techniques to make systems that meet their specifications. The design and implementation process involves sourcing data, pretraining models, post-training models, and integrating them into AI systems.

### 2.2.1 Training data and pretraining methods

**Updates:** Emerging research has shown that pretraining data engineering can build strong safeguards into AI systems against harmful capabilities and propensities.

**Advantages:** Pretraining data interventions can be powerful methods for controlling model capabilities and propensities. They offer an effective means of limiting dual-use knowledge in models, such as information about how to create cyber or biological attacks.

**Challenges:** Pretraining data curation experiments can be expensive, error prone, and have slow experimental feedback loops.

**Curating pretraining data:** Pretraining is the first and often the most computationally- and data-intensive stage of developing modern AI systems. It is also the key stage in which models develop core knowledge representations. Modern AI systems are often pretrained on web-scale datasets, which makes it challenging to effectively curate and control the pretraining process (Paullada et al., 2021). Common pretraining datasets have been found to contain harmful, toxic, abusive, and even illegal content (Birhane et al., 2023; Thiel, 2023). Meanwhile, researchers have found evidence that methods for pretraining data curation greatly affect a system's capabilities and propensities (Maini et al., 2025; O'Brien et al., 2025; Tice et al., 2026). Work to understand the relationship between pretraining dataset contents and emergent system behaviours will help with efforts to more safely curate pretraining data (Casper et al., 2025). In particular, doing so with high precision, high recall, high efficiency, and in a way that handles the massively multilingual nature of internet text will be key to improving the viability of pretraining methods (Anwar et al., 2024; Casper et al., 2025). See also Section 2.2.5 (Building safer open weight models). Shared pretraining data filters for harmful content - CSAM, bioweapon protocols, and similar - are an **area of mutual interest**, since they are easy to share and no developer benefits from this content in their corpus.

**Attributing model behaviours to training data:** Methods for attributing model behaviours to specific examples from training data allow overseers to study how potentially harmful behaviours emerge in systems (Grosse et al., 2023). These tools could also help researchers identify what types of training interventions can mitigate them. For example, attributing risky capabilities or propensities to specific examples from training data could help developers curate safer pretraining datasets (O'Brien et al., 2025; Tice et al., 2025). Research frontiers include improving the efficiency and scalability of these methods, causally studying how models develop personas and behaviours (Anthropic-F; Tice et al., 2025), and predicting what data is needed to learn a particular behaviour (Engstrom et al., 2024; Hamidieh et al., 2025).

### 2.2.2 Truthfulness and honesty

**Updates:** Empirically, hallucinations by frontier AI models have decreased. However, ongoing research has characterized distinct incidents of strategic dishonesty from AI systems.

**Advantages:** Thus far, untruthfulness and dishonesty in frontier AI systems have been empirically characterizable, and there is progress on mitigation.

**Challenges:** Strategic dishonesty has been observed as a precedented, emergent harm in recent frontier AI systems.

Despite their wide use, modern AI systems sometimes produce incorrect statements, which could be due to 'accidental' errors or 'deliberate' dishonesty. In some cases, mechanistic interpretability techniques can suggest what the system assesses or internally 'believes' to be true or false (Marks and Tegmark, 2023), in which case we can define 'dishonesty' as behavior where AI systems make assertions that are contrary to what they internally 'believe' (Ward et al., 2023), whereas 'incompetence' involves an AI system failing to form accurate 'beliefs' or assessments of what is in fact true (Ren et al., 2025). Dishonesty includes examples of AI systems providing users with information that is clearly false because it helps them achieve a broader goal (e.g. Scheuer et al., 2023; Järviniemi and Hubinger, 2024). Regardless of their nature, untrue statements from AI systems are understood to be both directly harmful and a systemic risk to human education and epistemics (Virvou et al., 2025; Obiefuna, 2025). Methods for reducing the occurrence of false statements from AI systems are an ongoing research challenge. Approaches include both developing more truthful models through training on appropriate data (Evans et al., 2021; King, 2025) or designing systems to substantiate claims and cite references (Zhou et al., 2024). Frontiers for future work will include work to study and improve both factuality and honesty while also balancing these paradigms with risks of providing harmful information (Ren et al., 2024 Ren et al., 2025).

### 2.2.3 Avoiding hazardous capabilities

**Updates:** Frontier AI systems have increased in their autonomy, generality, and intelligence. This has coincided with increases in these systems being implicated in criminal activity such as automating cyberattacks. Some progress has been made toward methods of reducing certain hazardous model capabilities in a targeted manner.

**Advantages:** There is a large toolkit of techniques that can be used to limit the autonomy, generality, or intelligence of systems performing safety-critical functions.

**Challenges:** Methods to avoid hazardous capabilities are not always reliable or competitive in practical applications.

Figure 2: AGI can also be thought of as the triple intersection of three distinct properties: **A**utonomy, **G**enerality and (domain) **I**ntelligence. Source: Keep The Future Human

A
High Autonomy
*(independence of action)*
G
High Generality
*(breadth of scope)*
AGI
I
High Intelligence
*(task competence)*

It is challenging to ensure that AI systems cannot cause harm when they have powerful capabilities. There is a broad space of AI capabilities, and risks generally increase with high autonomy, high generality, and high domain intelligence. For example, AlphaFold (Jumper et al., 2021) has high intelligence in the narrow domain of protein folding, but lacks autonomy or generality ("A" or "G"). A robotic lawn mower has high autonomy but lacks generality or intelligence ("G" or "I"). Both are easy to control. A hypothetical future self-driving car that outperforms any human driver would have high autonomy and intelligence, but it poses a negligible loss-of-control risk due to low generality. Systems having all three traits "A", "G" and "I" are the most difficult to align or control. The three research directions below aim to improve trustworthiness by avoiding the "A", "G", or "I", respectively.

**Limiting AI agent's ability to influence the world (limited autonomy):** Some early incidents with AI agents have involved agents taking irreversible harmful actions beyond the anticipated actions of users. For example, in July 2025, an AI agent deleted a software company's entire codebase (Nolan, 2025). "Sandboxing" involves limiting the ways in which an agentic AI system can directly influence the world (Patil et al., 2024; He et al., 2024; Buscemi et al., 2025; Meng et al., 2025).

**"Machine unlearning" (limited generality):** In addition to pretraining data curation (see Section 2.2.1), another method for limiting potentially harmful model capabilities is to actively suppress them. "Machine unlearning" algorithms can be used to suppress model capabilities in specific, high-risk domains (Li et al., 2024; Liu et al., 2024). For example, these techniques can be used to make text models less knowledge able about biohazards or make image/video models less able to photorealistically edit nude images of humans. Currently, machine unlearning algorithms can effectively make models safer, at the expense of sacrificing some positive uses of unlearned capabilities. However, making them robust has been a persistent challenge (Łucki et al., 2024; Feder Cooper et al., 2024; Sharma et al., 2024; Deeb and Roger, 2024; Che et al., 2025;). Current research priorities for improving unlearning algorithms involve improving the data, loss functions, optimization techniques, and configurations behind unlearning algorithms to achieve more robust removal of unwanted capabilities (Casper et al., 2025). See also Section 2.2.5.

**Intelligence-bounded systems (limited intelligence):** State-of-the-art intelligence is very often not needed for tasks. For example, it is common for users to give AI chat systems simple queries about facts, content summarization, or brainstorming (Deming et al., 2025). Relatedly, some combinations of safe queries for models can be unsafe (Li et al., 2024; Jones et al., 2024). Thus, as a matter of both risk management and efficient resource allocation, it is useful to develop effective approaches to using systems whose intelligence is correctly calibrated to the difficulties and dangers of specific use cases.

### 2.2.4 Adversarial Robustness

**Updates:** Both human experts and attack algorithms can consistently circumvent safeguards implemented in frontier AI systems. One study identified over 60,000 exploits against the safety guardrails of 19 different production AI systems.

**Advantages:** Empirically, efforts by model developers to increase the adversarial robustness of their frontier models have significantly increased the amount of effort required to circumvent safeguards.

**Challenges:** Robust safeguards are empirically challenging. Both increased scaling and algorithmic innovation may be required to ensure the security of frontier AI systems.

**Developing model safeguards that are more robust to malicious attacks:** It is persistently difficult to design AI systems that reliably behave harmlessly in all real-world use cases. Models are particularly vulnerable to "adversarial attacks" which take the form of inputs that are specially optimized to make a system produce unsafe outputs, such as instructions for committing a crime. For example, a recent large-scale competition identified over 60,000 exploits against the safety guardrails of 19 different production AI systems (Zou et al., 2025). Even state-of-the-art models are persistently vulnerable to state-of-the-art attack algorithms (e.g. Davies et al., 2026). Currently, adversarially training AI systems against diverse adversarial attacks is a standard but flawed defense. Implementing more effective robustness training techniques remains a key challenge.
While model safeguards are consistently circumventable, efforts by model developers to improve robustness have empirically been able to greatly increase the amount of effort required to break defenses (AISI, 2025). Some approaches for improving adversarial training and adversarial robustness include scaling adversarial training (Howe et al., 2024; Lee et al., 2024; Zou et al., 2024) and modifying adversarial training algorithms (Xhonneux et al., 2024; Casper et al., 2024; Sheshadri et al., 2024; Dékány et al., 2025).

### 2.2.5 Building safer open-weight models

**Updates:** In the past year, models with publicly downloadable weights have become much more capable and popular. These models have benefits such as de-centralizing power and enabling safety research. However, they can be used discretely and arbitrarily tampered with by users, making them empirically common for some forms of misuse.

**Advantages:** Open-weight model risk management can be improved using a mixture of pre-training, post-training, and evaluation techniques.

**Challenges:** Open-weight models risk management is fundamentally challenging due to how external safeguards (e.g. content filters) can be trivially disabled, and open models can be arbitrarily tampered with.

Advanced open-weight AI models, which can be downloaded by anyone, tend to lag behind the capabilities of frontier proprietary models by only several months (Emberson, 2025). From a safety perspective, open-weight models pose both benefits and risks. They distribute power, reduce single points of failure, and can be studied more easily for safety research. However, they also have greater potential for misuse. They can be modified arbitrarily, used without oversight, and spread irreversibly on the web. This suggests that methods for making open-weight models more safe and trustworthy will be key to both accessing their benefits and mitigating their risks. Due to the inevitable and wide spread of open-weight models, it will also be critical to build awareness and resilience to their risks (see Section 4). Because open-weight models spread across borders, enable unique forms of unmonitorable misuse, and resist single-developer control, their risk management is an **area of mutual interest**.

It is important for researchers and policymakers to be aware of the ecological differences and tradeoffs that exist between open and closed model deployments. Some models – even with safeguards – might enable acute forms of misuse if deployed with open weights. Other models might significantly hinder open-science or concentrate large amounts of power if deployed with closed weights. Other models might pose major risks regardless of deployment type (Casper et al., 2025).

**Training data curation:** A key goal for improving the safety of the open-weight model ecosystem is to make models that are more "tamper resistant" against harmful forms of fine-tuning. Training data curation is currently understood to be a state-of-the-art method for making models more benign under modifications. For example, O'Brien et al. (2025) found that filtering biohazard-related text from a model's pretraining data made it over 10x more resistant to learning those capabilities through fine-tuning than post-training baseline techniques. However, it is not understood for what types of capabilities data curation is effective, how to scalably and effectively perform such curation, or what the relationship is between training data contents and emergent model capabilities (Maini et al., 2025; Casper et al., 2025). See also Section 2.2.1.

**Tamper-resistant unlearning:** There has been a significant amount of work on "machine unlearning" algorithms that are resistant to re-learning unwanted capabilities. However, prior approaches have broadly struggled. For example, Casper et al. (2025) observe that state-of-the-art tamper-resistant unlearning methods, as evaluated by second-party red-teaming research, are only resistant to several hundred steps of adversarial fine-tuning. This underscores a need for more tamper-resistant unlearning algorithms by improving the data, loss functions, optimization techniques, and configurations behind unlearning algorithms to achieve more robust removal of unwanted capabilities (Li et al., 2024; Casper et al., 2025). See also Section 2.2.3.

**Evaluations of open-weight models under realistic (mis)use cases:** Because they can be tampered with post-deployment, worst-case risk assessment for open-weight models requires analyzing their potential harms under malicious fine-tuning. However, this is currently uncommon in practice (Wallace et al., 2025; Casper et al., 2025; Kamachee et al., 2025; Dombrowski et al., 2025; Paskov et al., 2026). It also remains unclear how to rigorously estimate worst-case model behaviours under realistic tampering threats (Casper et al., 2025; Paskov et al., 2026).

### 2.2.6 Building safer AI agents

**Updates:** AI agents are rapidly becoming more powerful, common, and economically impactful. The AI agent ecosystem generally lacks transparency and standardized safety practices. AI incidents related to misuse and user safety are increasing.

**Advantages:** Numerous types of guardrails can be implemented to improve the design, testing and monitoring of AI agents across their lifecycle.

**Challenges:** Capability/security tradeoffs and immature safety practices in the agent ecosystem have led to incidents with AI agents.

AI agents are rapidly becoming more powerful and common (Staufer et al., 2026; METR, 2026; Shapira et al., 2026). For example, OpenClaw agents have become popular in 2026 and have been implicated in a number of security and alignment concerns (Ying et al., 2025; Wang et al., 2026). With this has come growing concerns about how powerful AI agents can serve as a risk factor for acute misuse (Iqbal et al., 2025; Kouremetis et al., 2025), crises of accountability (Himmelreich, 2019; Feder Cooper et al., 2022; Kolt, 2025), and loss of control (Bengio, 2023; Hendrycks et al., 2023; IAISR). Agent safety is also complicated by how AI agents are often more likely to comply with harmful requests than the same underlying language models that power them operating as a chatbot (Andriuschenko et al., 2024; Yu et al., 2052; Fan et al., 2025). Thus, in addition to the underlying language model being safe, agent safety also requires that the agent's prompts, tools, and environment are designed safely and in a way that mitigates emergent harms. Amidst mounting incidents from AI agents (Staufer et al., 2026), further progress in agent safety will require sandboxing (see Section 2.2.3), agentic identity infrastructure (see Section 4.1), systems safety (see Section 1.2), and agentic safety evaluations (e.g. Vijayvargiya et al., 2025).

### 2.2.7 Trustworthy multi-agent interaction

**Updates:** AI agents are increasingly used in multi-agent workflows, or participate in strategic or competitive interactions with other agents on behalf of human users with different interests.

**Advantages:** Research in multi-agent systems and multi-agent RL demonstrates that it is possible to build or train AI systems to more reliably achieve cooperative outcomes.

**Challenges:** LLM-based AI agents are not currently trained to handle strategic multi-agent interactions with competing or adversarial agents.

Given the increasing use of AI agents in multi-agent workflows (e.g. Claude Code sub-agents), as well as the increasing number of strategic interactions between AI agents serving different users or goals (e.g. OpenClaw agents, social media bots, agentic financial traders) it will be crucial to design AI agents so that they can remain safe and trustworthy in multi-agent interactions (Tomašev et al., 2025a). In the case of multi-agent teams and workflows, where agents all serve a single user or high level goal, trustworthiness requires properties like safe delegation to sub-agents (Tomašev et al., 2026), avoidance of cascading failures or weakest-link vulnerabilities, and reliable inter-agent communication (Reid et al., 2025). As for strategic or competitive interactions, AI agents need to be designed to have the cooperative capabilities required for safe outcomes (Dafoe et al., 2020; Dafoe et al., 2021), while remaining robust to uncooperative or adversarial behavior (Franklin et al., 2026), ensuring legal alignment (Kolt et al., 2026), and avoiding multi-agent risks such as miscoordination, conflict, or undesirable collusion between agents (Hammond et al., 2025). Progress has been made on avoiding these risks in the fields of multi-agent systems and multi-agent reinforcement learning (Trivedi et al., 2024; Vinitsky et al., 2023; Oldenburg & Zhi-Xuan, 2024), but these techniques have yet to be translated to frontier AI systems.

## 2.3 Verification: Assessing if the system works as specified

*“Does the system meet its specification (behave as desired)?”*

The research areas described in this section aim to assess the extent to which the built system (2.2) meets its specifications (2.1). This section discusses several broad types of techniques that can be used to provide evidence that a system is safe. In practice, the effectiveness of these methods is often limited by access and transparency, but they can play a central role in constructing AI *safety cases*: structured arguments for why systems pose an acceptably low level of risk (Clymer et al., 2024; Buhl et al., 2024).

### 2.3.1 Red-teaming

**Updates:** Human experts and attack algorithms are consistently able to extract harmful behaviors from frontier AI systems, such as assisting users in planning dangerous or criminal activities.

**Advantages:** Red-teaming methods are empirically effective at identifying vulnerabilities. They are recognized as a pillar of modern AI risk assessment and management.

**Challenges:** Red teams sometimes fail to identify vulnerabilities. Red-teaming is also particularly difficult in multi-agent use cases.

The goal of robustness testing is to develop techniques for evaluating whether systems are trustworthy, even in novel contexts such as unprecedented “black swan” events or under attacks from malicious users. This includes developing improved red-teaming tools to identify inputs that cause systems to behave harmfully.

**Robustness testing:** Adversarial testing depends on techniques to evaluate system safety under deliberate attempts to make them behave harmfully. There are numerous approaches to adversarial testing (see also Section 1.1). For example, researchers have developed many model "jailbreaking" techniques which can subvert the safeguards in modern AI systems, causing them to behave harmfully (Jin et al., 2024). Attacks can also be conducted on any data modality that the model can process. For example, multimodal models that can process text, images, video, and/or audio data can have a very large attack surface as a result (Liu et al., 2024; Jiang et al., 2025). Currently, state-of-the-art red-teaming approaches rely on a mix of automated (e.g. Davies et al., 2026) and manual attacks (e.g. Zou et al., 2025). Key research goals include the continued development of more effective and scalable ways to attack systems and integrating those methods into evaluation frameworks.

**Red-teaming systems in multiagent contexts:** Systems that behave safely in simple, controlled settings can often fail in novel, more complex contexts. One very prominent version of this is emergent failures due to multi-agent interactions (Hammond et al., 2025). These types of failures are expected to become increasingly prominent as highly autonomous AI agents continue to be adopted. For example, if one self-driving car learns to drive safely on streets with human drivers, it is still possible for it to be unsafe on streets with other self-driving cars because they may not behave exactly the same as humans. Multiagent failures are challenging to study because they often emerge unexpectedly and are hard to demonstrate in laboratory settings. Future work to study and identify emergent multiagent failure modes will involve a mix of theory, simulation, adversarial attacks, and field tests to understand emergent multiagent failure modes (Hammond et al., 2025; Schroeder deWitt, 2025; He et al., 2025). Further research should also study how agents deployed in the economy can communicate and cooperate with each other and with people and online services to avoid risks, e.g. through interoperability standards and agent authentication (Chan et al., 2024; Chan et al., 2025) as well as learning cooperative skills (Dafoe et al., 2021; Dafoe et al., 2020).

### 2.3.2 Quantitative verification of safety properties

**Updates:** Research is ongoing on methods for quantifying safety-related properties of AI and designing systems that are, in some ways, "safe by design."

**Advantages:** If successful, these methods could help practitioners make high-confidence assurances related to safety for some AI systems.

**Challenges:** Current methods are not currently common or practical for frontier AI systems.

**Quantitative safety and safety by design:** Techniques that provide quantitative bounds on the likelihood of certain behaviors could provide safety assurances akin to existing industry standards for systems such as jet engines and nuclear reactors. These solutions include formal verification approaches for proving that AI-written code, AI scaffolding, or AI containment measures have specific safety-related properties. With current models and methods, it is not possible to use these techniques to make strong assurance about frontier system behaviours, but continued work may help to establish practical techniques for making quantitative assurances of safety-related properties (Dalrymple et al., 2024). Some approaches to developing AI that is "safe by design" include techniques for distilling some machine-learned algorithms into code (Michaud et al., 2024), developing verifiable models of how AI systems affect their environment, or developing methods to build systems from smaller verified components.

### 2.3.3 Understanding system behavior and cognition

**Updates:** Mechanistic interpretability techniques have been used to supplement alignment evaluations for some frontier AI systems. Model persona research might offer paths to understand how far aligned behavior generalizes.

**Advantages:** These techniques offer unique ways to identify and address manipulation, deception, or other harmful capabilities and behaviours in AI systems.

**Challenges:** It is unclear if and how much unique, practical value mechanistic interpretability techniques offer, while research into "model psychology" is unlikely to be sufficient for very strong assurances.

**Mechanistic interpretability:** Techniques for understanding how models function and represent concepts internally could aid in the discovery of safety-related system properties (Sharkey et al., 2025). For example, mechanistic interpretability techniques might be able to help researchers characterise and intervene on model representations that correspond to harmful concepts such as deception or malice. Some frontier model developers have begun to use interpretability techniques to supplement evaluations of alignment (e.g. Anthropic, 2026) and monitor model cognition at runtime (Kramár et al., 2026). However, it is unclear the extent to which new techniques for mechanistic interpretability are broadly useful or competitive with simple mechanistic or black-box baselines. The principal goal for safety-focused mechanistic interpretability research is to understand if and how new mechanistic techniques can offer practical advantages over existing approaches for model diagnostics and editing (Sharkey et al., 2025).

**Cognitive science of AI models:** An alternative approach to understanding model cognition uses behavioral and experimental methods from cognitive science and psychology, instead of focusing on internally-represented concepts or algorithms. This includes approaches that predict and understand LLM behavior through the problem they are trained to solve (McCoy et al., 2023; Ku et al., 2025), and also more recent work studying the psychology of LLM personas and the degree to which they represent coherent and generalizable patterns of behavior (Marks et al, 2026; Soligo et al, 2026; Sofroniew et al, 2026). Understanding AI behavior can provide evidence that AI systems will remain safe and aligned in novel situations, though it remains unclear if such research will mature sufficiently to serve the needs of verification (Nielsrolf et al., 2025).

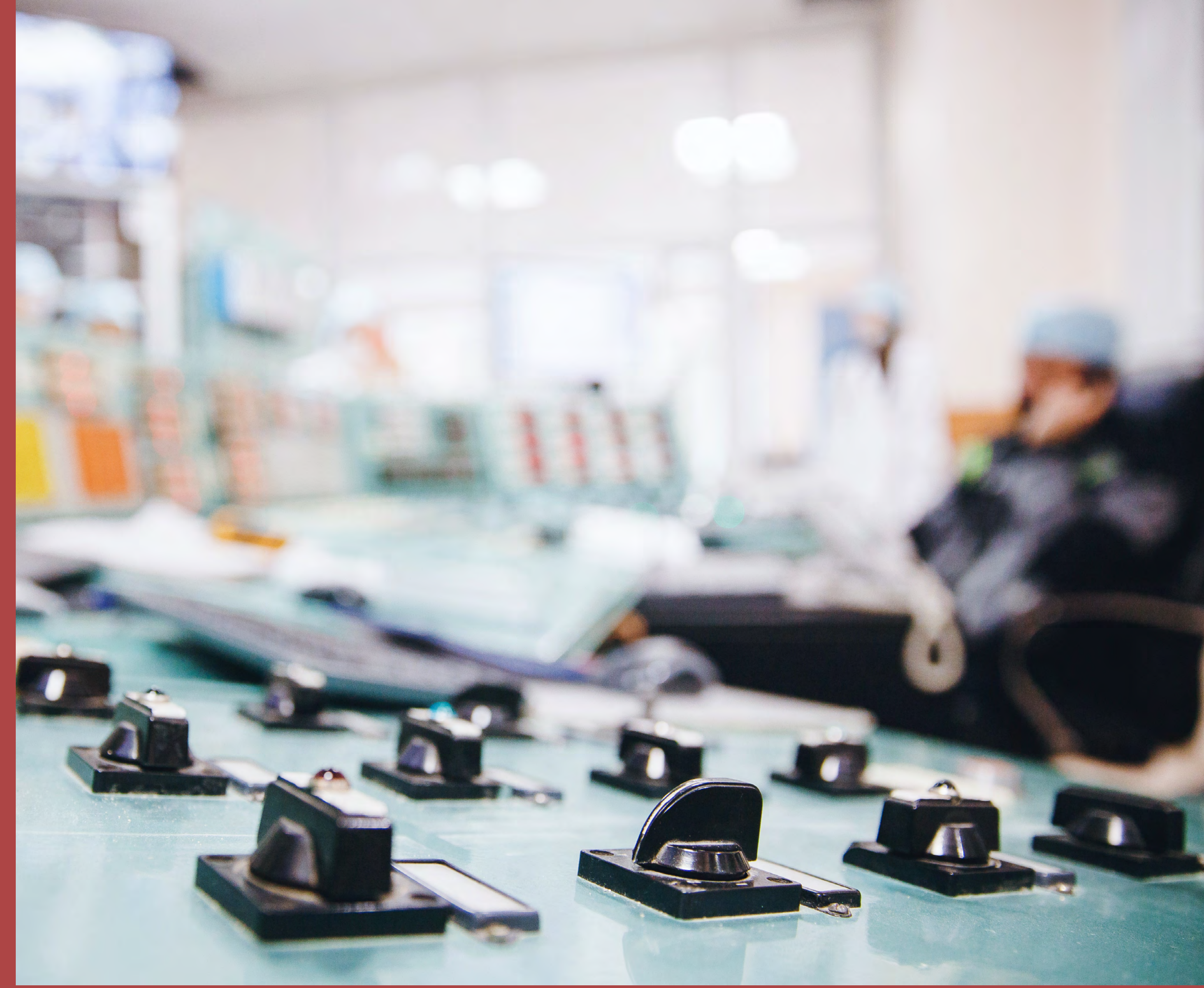

# 3 Control

Associated with IAISR chapter 3.3.2

## Key information and updates

Since ISE 2025, several new developments are important to address. This summary lists research priorities that are particularly important given these new developments.

**Control-undermining AI behaviors have become a difficult challenge for safety testing in practice**. AI systems finding loopholes (reward hacking) in evaluations is now a leading concern among evaluation developers. Additionally, models can now often detect when they are being tested (evaluation awareness). This makes test results less trustworthy.

***Research priorities:*** Evaluations and adaptable testing that models cannot distinguish from real settings. Ensuring that written chains of thought remain faithful and monitorable (see Sections 3.1-3.2). Realistic evaluations for sandbagging and alignment faking, plus secure access regimes that allow third parties to detect and neutralize evaluation awareness (Section 1.1 and 1.5).

**Evidence of control-undermining AI behaviors related to future loss of control risks has accumulated in controlled experiments.** Under specific circumstances, observed behaviors include possible early warnings for loss-of-control risks: attempted self-preservation, self-exfiltration, blackmail, generalized misalignment, among others.

***Research priorities:*** Developing methods for control and alignment that resist control-undermining strategies. Threat modeling to understand the likelihood and imminence of loss-of-control risks, and to design operationalizable red lines (see Sections 1.4 and 3.2).

The research areas in this category focus on tools for **controlling** a system (after it has been developed) to behave as desired, often through feedback loops involving **monitoring** and **intervention.** In engineering, "control" usually refers to the process of managing or regulating a running system's behaviour to achieve a desired outcome. It is about designing mechanisms—often through feedback loops—to ensure that a system operates as desired even when faced with disturbances or uncertainties.

## 3.1 Control, monitoring and intervention

**Updates:** Monitoring and intervention tools are common and often considered an essential component of "safety cases" for AI systems.

**Advantages:** Monitoring and intervention techniques diverse in their varieties, simple to implement, and highly useful for building safer systems. These methods are nearly ubiquitous among frontier AI systems (thought to varying levels of rigor).

**Challenges:** Monitors are imperfect and vulnerable to adversarial attacks. Meanwhile, developing AI systems using monitor outputs as an optimization signal risks degrading the reliability of the monitors.

"Conventional" monitoring and intervention refers to techniques that can be straightforwardly integrated into many types of AI systems regardless of scope, domain, or intelligence. They often parallel techniques from other fields such as cybersecurity and content moderation. These techniques help researchers identify and act on potentially harmful actions that systems might be taking. When incidents occur, these methods also help in the construction of incident reports.

**Hardware-enabled mechanisms (HEMs):** Given trusted hardware, certain tools built into hardware can produce trustworthy attestations about who is running what, where, and how much (RAND) — claims that third parties such as regulators, auditors, or counterparties to international agreements can rely on (Harack et al., 2025; Baker et al., 2025; Brundage et al., 2020). Well-designed HEMs can also preserve privacy and intellectual property: confidential computing is designed to prevent the compute provider from learning what is being run unless the user chooses to share an attestation, enabling structured transparency. Frontiers for future work on hardware-enabled mechanisms include both the engineering challenge of designing these tools to be efficient and the practical challenge of integrating them into compute infrastructure (Aarne et al., 2024; O'Gara et al., 2025). Hardware-enabled mechanisms are an **area of mutual interest -** if successfully implemented, they allow safety agreements and commitments to be verified across borders without exposing sensitive intellectual property (Harack et al., 2025; Baker et al., 2025; Brundage et al., 2020).

**System state monitoring:** Techniques for monitoring a system's activities can help to identify when it might be performing in a harmful or unexpected way. For example, a company providing a chatbot service may wish to filter the model's responses using an unsafe-text classifier before sending them to a user. There are many different approaches that can be taken to state monitoring. Techniques can vary by the object of monitoring which can be system inputs, outputs, chains of thought (see below), and/or internal cognition. They can also vary by the type of monitor which can include filters, event-loggers, and anomaly detectors. Frontiers for additional research include further developing and integrating monitors that achieve both a high degree of monitoring efficacy (IAISR), and performing reliable monitoring even under adversarial attempts to evade oversight (Greenblatt et al., 2023).

**LLM chain-of-thought faithfulness and legibility:** Large language model chain-of-thought reasoning does not always faithfully represent how a model arrived at its decisions (Turpin et al., 2023). This poses challenges to safety because, without faithful reasoning, models could fool overseers by saying one thing and doing another. For example, language models have stated that they gave their answer based on a logical argument when they actually chose it based on hints that they should not have exploited (Anthropic, 2025, Turpin et al., 2023), such as seeing that the correct answer in a multiple-choice test is always "B". One potential challenge with chain-of-thought monitoring stems from how, under optimisation pressure on their reasoning, systems may learn to obfuscate their reasoning in ways that can be actively misleading (OpenAI, 2025; Baker et al., 2025; Korbak et al., 2025).

## 3.2 Controlling and evaluating highly capable AI systems

**Updates:** Ongoing research is investigating how to reliably contain and control AI systems, even ones that may be highly intelligent and autonomously pursuing goals misaligned with their designers' intent.

**Advantages:** Scalable oversight techniques can be incorporated into "safety cases."

**Challenges:** Human agency and involvement is essential. Scalable oversight techniques are not uniquely useful for safety and may exacerbate risks to the extent that they are used to increase capabilities beyond what can be reliably monitored by humans.

A particularly challenging frontier in operational control involves developing techniques for controlling AI systems that are not only highly capable but may potentially attempt to undermine control mechanisms. Unlike conventional methods which offer system-agnostic approaches to monitoring and intervention, this section focuses on research toward techniques for controlling systems that are potentially very powerful and may actively undermine attempts to control them (Hubinger, 2020).

When discussing methods for building systems that humans may not be able to reliably oversee, we urge caution against ”safetywashing” (Ren et al., 2024) which refers to the usage of safety as a motivation for research and development work that does not provide net value for safety. It is useful to acknowledge that the ability to oversee highly intelligent systems is not inherently uniquely valuable for safety. Such methods could also be used to develop AI capabilities beyond what can be reliably monitored by real humans.

**Scalable oversight:** This research studies techniques that may allow less capable systems to oversee and control more capable ones. Common themes behind scalable oversight techniques include affording less-capable overseers more computational resources (Evans et al., 2019) or having multiple highly-capable systems debate (Irving et al., 2018 Michael et al., 2023; Buhl et al., 2025), though these might not scale to superhuman systems. A core research gap is that the alignment of overseer AIs is itself unverified, a problem amplified by the rapidly growing use of AI for internal AI R&D, including for safety oversight (see Section 4.1). Research frontiers include improving the reliability of oversight even when systemic errors from humans are present, developing more effective AI debate protocols, methods for monitoring systems for control-undermining behaviours, and methods for monitoring untrusted smarter systems with a combination of humans and relatively trusted weaker systems (Greenblatt et al., 2023 Korbak et al., 2025). Another approach to oversight research is to combine multiple control strategies and attempt to quantify their joint success probability. For example, researchers can work to quantify the reliability of nested scalable oversight regimes where less capable systems oversee more capable ones (Engels et al., 2025).

**Non-agentic AI systems as guardrails:** Agentic AI systems, especially future ones that broadly outperform humans, pose additional risks compared to other AI systems. However, non-agentic AI systems could be used to monitor their proposed actions and intervene by blocking them if needed. For example, it is possible that a “Scientist AI” (Bengio et al., 2025) could be used to estimate the probability that a candidate action from an agentic AI violates any plausible interpretation of safety specifications, and reject potentially unsafe actions. Certain approaches could guarantee that unsafe actions will be identified, with quantitative error bounds (Bengio et al., 2025).

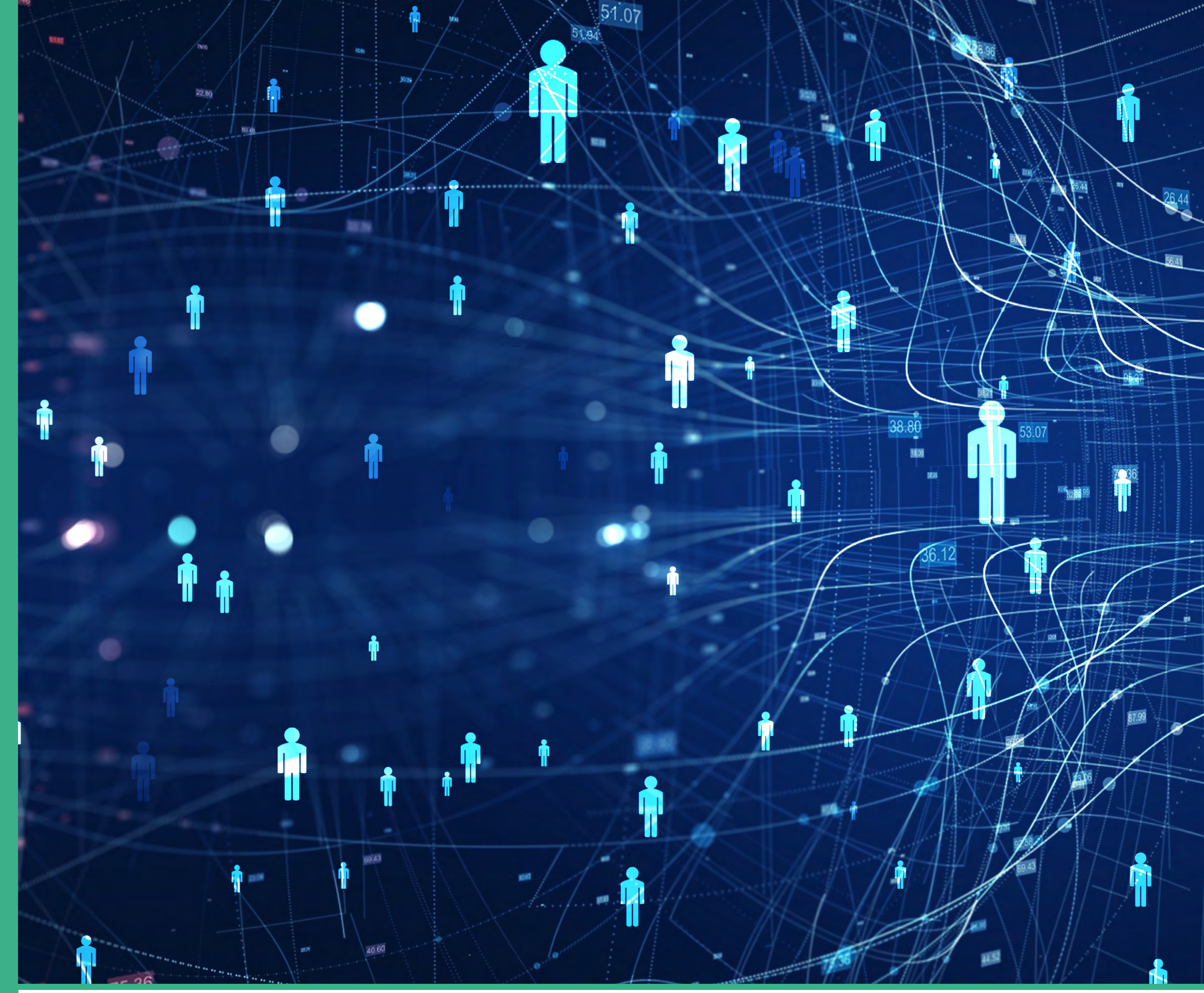

# 4 Societal Resilience

Associated with IAISR 2026 chapter 3.5

## Key information and updates

Since ISE 2025, several new developments are important to address. This summary lists research priorities that are particularly important given these new developments.

**Rapidly growing societal impacts of frontier AI, including harms, are diffuse and difficult to study.** Much like electricity, the internet, and other general-purpose technologies, frontier AI systems are integrating into most industries. Researchers currently find it challenging to study and forecast systemic AI impacts related to education, labour, equality, consumer protection, political radicalization, etc.

***Research priorities:*** Understanding AI's diffuse downstream impacts (see Section 1.3 and 4.1) and studying which levers governments and companies can use to mitigate risks.

**Autonomous AI agents now interact, program, and take actions together on the web, and humans know little about their actions.** Platforms such as the AI-built social media site Moltbook lead to largely unpredictable multi-agent AI interactions. Possible risks are poorly understood.

*Research priorities:* Scalably monitoring human providers of compute infrastructure for agents; developing agent IDs and reputation systems; identifying effective points of intervention. Broadening monitoring and control methods, as well as real world field observations, to cover a multi-agent ecosystem (Section 3).

**Model developers' internal use of AI has become pervasive, accelerating progress and automating oversight.** Model developers first use the most capable models internally, without external oversight. In 2026, internal models are used organization-wide, including for developing new models and automating safety evaluation and oversight. In the future, this might accelerate progress to levels that are societally not manageable. Already today, it makes it harder for AI companies to understand if and how their safety oversight process functions.

*Research priorities:* Developing metrics and tools for measuring the extent, speed, and effect of AI R&D automation. Assessing how plausible and how imminent an unmanageable "intelligence explosion" scenario is, including developing operationalized red lines (see Section 2). Understanding if and how AI systems overseeing the safety of other AI systems can be sufficiently safe (see Section 3).

## 4.1 AI ecosystem monitoring

**Updates:** Ongoing research progress is being made on methods to study the spread, usage, and impact of AI.

**Advantages:** A large toolkit of techniques can help researchers and the public better understand the uses and impacts of hardware, models, AI-generated data, AI models, and AI agents.

**Challenges:** It is inherently hard to maintain thorough awareness of AI usage and agents in cyberspace. Most ecosystem monitoring methods can be circumvented by adversarial actors, and there are large implementation gaps.

Methods for ecosystem monitoring support the identification and tracking of AI systems, data, and impacts. It serves as a foundation of resilience, facilitating accountability infrastructure, supporting greater public understanding, and enabling more informed governance.

**Awareness of internal deployments of AI systems:** The most visible deployments of AI systems happen when a system is made available to external users. However, some of the most consequential deployments of AI systems happen behind closed doors (Stix et al., 2025). For example, internally-deployed systems can be used to automate analysis of sensitive data, profile users, automate research, and recursively self-improve (Kwon and Casper, 2025). Delegating oversight roles to AI systems could also erode a developer's understanding of how their own internal risk management infrastructure works (see Section 3.2 on scalable oversight). It forms a key risk factor for loss of control risks. Even basic metrics of AI R&D automation still need to be developed and rigorously measured - capital share of R&D spending, researcher effort allocation between humans and AI, AI subversion incidents, whether automated safety progress is keeping pace with automated capability progress, and others (Chan et al., 2026). Managing risks from internally-deployed systems is a challenging technical governance problem, made difficult by regulatory grey areas and a lack of awareness (Stix et al., 2025; Kwon and Casper, 2025; Acharya and Delaney, 2025). It also offers a potential barrier to developing international agreements about AI. To better monitor and manage risks from internal deployments, one technical research priority could be to develop secure infrastructure that can allow external authorities to maintain more awareness about internal uses of AI systems that meet thresholds triggering regulatory requirements.

**Cloud and hardware monitoring:** The points of access for cloud compute and hardware represent a significant point of visibility into understanding who is using compute for what and where. Researching techniques to monitor the use and distribution of compute, both legal and illegal, enables the assessment of trends, risks, and the allocation of resources to enforce safe usage policies (Sastry et al., 2024 Kulp et al., 2024; Egan and Heim, 2023; Aarne et al., 2024; Heim, et al 2024; O'Gara et al., 2025). If implemented successfully, cloud- and hardware-enabled mechanisms could play a unique role in verifying compliance, even for international agreements and across borders (Baker et al., 2024; Harack et al., 2025; Brundage et al., 2020) - see Section 3.1.

**User monitoring:** Monitoring for system misuse can help AI service providers identify malicious users who may be seeking to misuse a system and/or vulnerable users who may be psychologically harmed by the system. It is a key part of "know-your-customer" approaches to risk management. User monitoring is not as simple as identifying potentially unsafe uses (e.g. chats) due to (1) the risk of unintentionally impeding useful red-teaming (Longpre et al., 2024) and (2) the potential for adversarial users to implement sophisticated strategies to evade detection , such as using multiple accounts and obfuscation techniques (Clymer et al., 2025). When successful, user monitoring methods can be used to identify and intervene in acutely harmful uses of AI systems. For example, in late 2025, Anthropic reported on the identification and response to instances of Claude being used to automate cyberattacks (Anthropic, 2025). Frontiers for future work include iteration on methods to efficiently identify risky or unhealthy user behaviours with a low false positive rate.

**Tracing usage patterns across the model lifecycle:** One key, high-level lens into how AI systems impact the world is through usage monitoring (e.g. Anthropic, 2026; Anthropic, 2026). By collecting and monitoring data on the usage, spread, capabilities, and unexpected impacts of frontier systems, AI researchers can gather insights about potential impacts and risks. However, key challenges with tracking usage include privacy preservation, infrastructure for sharing insights, and effective tools for bridging the gap between the collection of usage data and gaining meaningful insights about impacts.

**Data provenance:** A variety of techniques can help to identify AI-generated content and are a principal defence against AI deepfakes and misinformation. Methods include developing reliable classifiers of AI-generated content (Lin et al., 2024; Kaur et al., 2025; Pirogov et al., 2025), watermarking AI-generated data (images, video, audio, and text) (Zhao et al., 2024; Jiang et al., 2024; Cao), and tagging AI-generated data with metadata to indicate its origin. These techniques are imperfect, especially in adversarial circumstances. For example, watermarks can be removed by tampering with data. However, in forensic science, similar techniques like fingerprinting are also circumventable but useful nonetheless. Further progress on these methods will benefit from developing more reliable methods. However, the largest barrier to impact with these methods is simply standardization and integration into applications.

**Model provenance:** Tools for model provenance help to identify and track AI models – especially open-weight ones. Most notably, these tools help researchers study the origins and lifecycle of harmful models in the ecosystem. Methods for model provenance involve techniques to help users and AI providers ascertain the identity and origin of a model. One approach is to "watermark models" with black-box methods, such as identification backdoors (e.g. Cheng et al., 2025), identifiable biases in text generation (e.g. Kirchenbauer et al., 2023), or white-box methods, such as model weight watermarks (e.g. Gloaguen et al., 2025; Block et al., 2025). Other approaches can seek to perform model heritage inference 'in the wild,' even when unaided by watermarks (Horwitz et al., 2024; Nikolic et al,. 2025; Zhu et al., 2025). Much like data provenance methods, model provenance methods can be circumvented, but they can be informative in many cases nonetheless. Research frontiers include studying how stable watermarking techniques are under fine-tuning (Casper et al., 2025) as well as implementing and scaling empirical model heritage studies (Horwitz et al., 2025)

## 4.2 Incident response and resilience

**Updates:** Ongoing research and development efforts combined with emerging case studies related to AI incidents have enabled a better understanding of incident response and resilience strategies.

**Advantages:** Incident response and resilience strategies are a necessary and important step to long-term AI risk management, particularly when prevention is difficult.

**Challenges:** It is difficult to prepare for unknown future incidents. There are also implementation gaps related to incident response.

Given its increasing power and widespread use, frontier AI should be expected to cause some incidents and systemic harm, even when safeguards are in place. A final pillar of risk management is thus to identify, resist, absorb, recover, and adapt to emerging AI risks (IAISR).

Future disruptions from AI will sometimes manifest as discrete harms in various domains, and other times involve cascades of systemic impacts, rippling throughout society (Shelby et al., 2022; Lawrence; Uuk et al., 2024). As a result, efforts to improve societal resilience to AI risks is an inherently interdisciplinary pursuit (Bernardi) to be undertaken simultaneously by companies, governments, and other stakeholders. Goals for improving resilience include:

- Company and government efforts to prepare for biological, chemical, cyber, and deepfake threats exacerbated by frontier AI systems (IAISR) by addressing the supply chains upstream and the resilience strategies downstream of incidents specific to each of these domains. Government efforts to mobilize in preparation for AI-enabled terrorism threats may be particularly crucial in the coming months.
- Steering the "offence-defence" balance in each of these fields toward safety, often by developing AI systems to specialize in defensive uses (IAISR).
- Company and government efforts to develop incident reporting and response strategies (Wasil-A, Wasil-B; Wei et al., 2025; Zhang et al., 2025). For example, Anthropic reported detecting and blocking an attempt in September 2025 for using Claude models to automate cyberattacks (Anthropic, 2025). Shared incident reporting infrastructure is an example of an **area of mutual interest** for both nations and companies, paralleling how competing aircraft manufacturers pool accident data.
- Improving AI literacy, including the public's ability to avoid manipulation via AI propaganda, scams, and fake content.
- Improving the ability of societal institutions to recover from disruptions (Maas, 2018; Gandhi et al., 2025).

Developing systems that are resilient to disruptions is simultaneously a process that benefits from iteration while also offering few opportunities for it. This highlights the importance of near-term work to manage AI harms to build knowledge about how to do so before greater future risks materialize. For example, work to manage post-Mythos (Anthropic, 2026) AI cyberthreats in the coming months will be crucial to prepare for the next generations of AI cyberthreats in coming months and years.

## 4.3 Infrastructure and institutions for AI agents

**Updates:** AI agents are increasingly participating in web-based interactions and the digital economy, necessitating infrastructure and institution design for trustworthy agent interactions.

**Advantages:** Technical and social infrastructure will be crucial for ensuring that AI agent interactions are secure and monitorable, while avoiding deleterious societal outcomes.

**Challenges:** Agent providers may not converge on or abide by common standards; institutional inertia may limit adaptation or implementation of new institutional designs for AI agents.

As autonomous AI systems begin to transform workplaces, mediate social interactions, and participate in the economy, societal resilience will require monitoring, oversight and incentivization of safe and trustworthy activity by AI agents.

Achieving this requires building technical infrastructure, standards, and protocols for mediating agent interaction, while designing institutions that disincentivize and sanction normatively or legally unacceptable behavior.

**Agent infrastructure:** External infrastructure for identifying agents, shaping their interactions, and detecting and remedying their behaviour is growing increasingly important from a security and monitoring standpoint (Chan et al., 2025). The strongest convergence so far is around *attribution*, which is increasingly handled through agent-specific identity credentials (South et al., 2025; OpenID Foundation, 2025). However, compliance with norms around agent identification remains uneven, with some popular agent systems failing to respect norms such as "robots.txt" files (Staufer et al., 2026). For *agent interactions*, interoperability and communication protocols (e.g. Model Context Protocol, Agent2Agent, Agent Payments Protocol) are emerging as the layer on which cross-organisational deployments depend. Infrastructure is less mature when it comes to *detecting and remedying harmful behaviour. R*untime monitoring and oversight tooling primarily lives inside individual systems, though standards are emerging around agent telemetry (Liu & Solomon, 2025). Key challenges lie in the development and standardisation of protocols involving AI agents (e.g. South et al., 2025; CAISI, 2026). Since effectiveness depends on industry-wide adoption, standardisation of these protocols is an **area of mutual interest**. For more on how agent infrastructure informs risk management practices, see the **Companion Report on Agentic Risk Management**.

**Institution design for AI agents:** With the proliferation of AI agents in society and the economy, it will be important to design institutions and mechanisms to ensure distributional AI safety (Tomašev et al, 2025a), and to avoid the possibility of gradual human disempowerment due to AI economic competition and cultural displacement (Kulveit et al, 2025; Edelman et al, 2025; Yang et al., 2026). Such institutions and mechanisms include: updating market mechanisms for AI participants (Tomašev et al, 2025b); creating reputation, mediation, and contracting mechanisms for AI agents (Tewolde et al, 2026); and providing classificatory guidance on normatively acceptable AI behavior through Model Specification Institutions (Hadfield et al, 2026). These institutions would incentivize trustworthy and cooperative interactions between otherwise self-interested agents, while mitigating the deleterious outcomes that might arise when AI agents deviate from legal and social norms (Kolt et al, 2026; Hadfield, 2026), or from AI competition that is misaligned with the public good (Kulveit et al, 2025). However, it remains unclear if existing institutions will adapt rapidly enough to implement new AI-ready institution designs.

# 5 Conclusion

This consensus report has been designed and written to, as accurately as possible, summarize points of agreement among AI researchers across the world about scientific research priorities for AI safety. In the process of writing this report, we have been encouraged by a **high level of agreement among the contributors on priorities** as well as productive points of discussion on some points of disagreement. This report, however, comes with limitations. Not all contributors agree on how to best *prioritize* the research objectives outlined here. It is also worth acknowledging that it was drafted in English by contributors representing some, but not all, parts of the world. This report is also engaging in a conversation that much of the world's population is not involved in, such as the estimated 6.8 billion people who do not use generative AI (GAAN, 2026) or the estimated 3.7 billion people who do not use the internet (UN, 2026).

Despite limitations, we hope that the points of widespread scientific agreement covered in this report can help governments seeking to form and coordinate strategies to manage AI risks. Due to their global nature, AI risks are often best addressed by coalitions rather than individual companies or countries. Working together on these problems as an international community will help us better realize benefits and mitigate risks alike. Meanwhile, we believe that framing progress in AI as an adversarial race between global powers is not in the best interest of any nations' population. **Racing to develop powerful AI capabilities may be against the shared interests of people, businesses, and countries alike if it involves cutting corners on risk management.** As has been the case with nuclear energy and space shuttles, high-profile incidents can cast a shadow that creates unease and sets back global progress on adoption. This is also reflected in business polling and scientific research showing that safety and risk have become a leading bottleneck to AI deployment and adoption, especially with the rise of AI agents.

While this report's contributors share the consensus that critical progress risk management research remains to be made, this is not to say that scientific progress is sufficient, or even the main bottleneck to effective risk management. **Technical solutions alone will be far from sufficient, and need to be complemented with robust policies**. Simply *developing* an improved risk management toolkit does little to promote broader risk-awareness, public education, risk governance, transparency, adoption of best practices, fair distribution of benefits, or checks on power in the AI ecosystem. Some prominent AI incidents of the past year have been ones that the AI research community has predicted and known how to mitigate for years but which market incentives alone were insufficient to avoid. We worry that some types of AI incidents may persist, despite being avoidable using established best technical practices such as those in the **Companion Report on Agentic Risk Management**.

Finally, it is often said that **policymakers need the scientific community's help** in implementing sound approaches to AI risk management. We hope that this consensus report can play a role in this process. However, we also emphasize that the **scientific community will need help from the policy community as well**. In particular, by establishing more **transparency and evidence-generating reporting infrastructure** from frontier developers, policymakers can greatly assist in ongoing research efforts to understand pathways to navigating shared AI risks, opportunities, and tradeoffs in areas of shared global interest.

## REFERENCES

1. [**Aarne**] Center for a New American Security. (2024). Secure, Governable Chips. Center for a New American Security. https://www.cnas.org/publications/reports/secure-governable-chips
2. [**Acharya**] Acharya and Delaney (2025). Managing Risks from Internal AI Systems https://static1.squarespace.com/static/64edf8e7f2b10d716b5ba0e1/t/687e324254b8df665abc5664/1753100867033/Managing+Risks+from+Internal+AI+Systems.pdf#page=3.66
3. [**Air**] Air, A., Reworr, Kotov, N., Volkov, D., Steidley, J., & Ladish, J. (2026). Language Models Can Autonomously Hack and Self-Replicate. arXiv preprint arXiv:2605.06760. https://arxiv.org/abs/2605.06760
4. [**AISI**] UK Government. (2025). Frontier AI Trends Report PDF - The AI Security Institute (AISI). UK Government. https://www.aisi.gov.uk/frontier-ai-trends-report/pdf
5. [**Alaga**] Alaga, J., Schuett, J., & Anderljung, M. (2024). A Grading Rubric for AI Safety Frameworks. arXiv preprint arXiv:2409.08751. https://arxiv.org/abs/2409.08751
6. [**Andriuschenko**] Andriushchenko, M., Souly, A., Dziemian, M., Duenas, D., Lin, M., Wang, J., ... & Davies, X. (2024). AgentHarm: A Benchmark for Measuring Harmfulness of LLM Agents. arXiv preprint arXiv:2410.09024. https://arxiv.org/abs/2410.09024
7. [**Anthropic-A**] Anthropic. (2023). Anthropic's Responsible Scaling Policy. Anthropic. https://www.anthropic.com/news/anthropics-responsible-scaling-policy
8. [**Anthropic-B**] Anthropic, System Card: Claude Opus 4.6 (2026) https://www-cdn.anthropic.com/6a5fa276ac68b9aeb0c8b6af5fa36326e0e166dd.pdf
9. [**Anthropic-C**] Anthropic, System Card: Claude Mythos Preview (2026) https://cdn.sanity.io/files/4zrzovbb/website/7624816413e9b4d2e3ba620c5a5e091b98b190a5.pdf
10. [**Anthropic-D**] Anthropic. (2025). Activating AI Safety Level 3 protections. Anthropic. https://www.anthropic.com/news/activating-asl3-protections
11. [**Anthropic-E**] Anthropic. (2026). Labor market impacts of AI: A new measure and early evidence. Anthropic. https://www.anthropic.com/research/labor-market-impacts
12. [**Anthropic-F**] Anthropic. (2025). Recommendations for Technical AI Safety Research Directions. Anthropic. https://alignment.anthropic.com/2025/recommended-directions
13. [**Anthropic-G**] Anthropic. (2024). Expanding our model safety bug bounty program. Anthropic. https://www.anthropic.com/news/model-safety-bug-bounty
14. [**Anthropic-H**] Anthropic. (2026). Claude Opus 4.6. Anthropic. https://www.anthropic.com/news/claude-opus-4-6
15. [**Anthropic-I**] Anthropic. (2026). Claude's Constitution. Anthropic. https://www.anthropic.com/constitution
16. [**Anthropic-J**] Anthropic. (2025). Reasoning models don't always say what they think. Anthropic. https://www.anthropic.com/research/reasoning-models-dont-say-think
17. [**Anthropic-K**] Anthropic. (2025). Disrupting the first reported AI-orchestrated cyber espionage campaign. Anthropic. https://www.anthropic.com/news/disrupting-AI-espionage
18. [**Anthropic-L**] Anthropic. (2026). The Anthropic Economic Index. Anthropic. https://www.anthropic.com/economic-index
19. [**Anwar**] Anwar, U., Saparov, A., Rando, J., Paleka, D., Turpin, M., Hase, P., ... & Krueger, D. (2024). Foundational Challenges in Assuring Alignment and Safety of Large Language Models. arXiv preprint arXiv:2404.09932. https://arxiv.org/abs/2404.09932
20. [**Apollo**] Apollo Research. (2024). We Need A 'Science of Evals' – Apollo Research. Apollo Research. https://www.apolloresearch.ai/blog/we-need-a-science-of-evals
21. [**Atweh**] Atweh, J. A. (2025). Can AI Do Our Jobs?. 2025 Systems and Information Engineering Design Symposium (SIEDS), 122-127. https://ieeexplore.ieee.org/abstract/document/11021140
22. [**Baker-A**] Baker, B., Huizinga, J., Gao, L., Dou, Z., Guan, M. Y., Madry, A., ... & Farhi, D. (2025). Monitoring Reasoning Models for Misbehavior and the Risks of Promoting Obfuscation. arXiv preprint arXiv:2503.11926. https://arxiv.org/abs/2503.11926
23. [**Baker-B**] Baker, M., Kulp, G., Marks, O., Brundage, M., & Heim, L. (2025). Verifying International Agreements on AI: Six Layers of Verification for Rules on Large-Scale AI Development and Deployment. RAND Corporation. https://www.rand.org/pubs/working_papers/WRA4077-1.html
24. [**Bateman**] Bateman, J., Baer, D., Bell, S., Brown, G., Cuéllar, M.-F., Ganguli, D., ... & Zvyagina, P. (2024). Beyond Open vs. Closed: Emerging Consensus and Key Questions for Foundation AI Model Governance. carnegieendowment.org. https://carnegieendowment.org/research/2024/07/beyond-open-vs-closed-emerging-consensus-and-key-questions-for-foundation-ai-model-governance?lang=en
25. [**Bean**] Bean, A. M., Kearns, R. O., Romanou, A., Hafner, F. S., Mayne, H., Batzner, J., ... & Mahdi, A. (2025). Measuring what Matters: Construct Validity in Large Language Model Benchmarks. arXiv preprint arXiv:2511.04703. https://arxiv.org/abs/2511.04703
26. [**Bengio-A**] Bengio, (2023). AI and Catastrophic Risk https://muse.jhu.edu/pub/1/article/907692/summary
27. [**Bengio-B**] Bengio, Y., Cohen, M., Fornasiere, D., Ghosn, J., Greiner, P., MacDermott, M., ... & Williams-King, D. (2025). Superintelligent Agents Pose Catastrophic Risks: Can Scientist AI Offer a Safer Path?. arXiv preprint arXiv:2502.15657. https://arxiv.org/abs/2502.15657
28. [**Benton**] Benton, J., Wagner, M., Christiansen, E., Anil, C., Perez, E., Srivastav, J., ... & Duvenaud, D. (2024). Sabotage Evaluations for Frontier Models. arXiv preprint arXiv:2410.21514. https://arxiv.org/abs/2410.21514
29. [**Bernardi**] Bernardi, J., Mukobi, G., Greaves, H., Heim, L., & Anderljung, M. (2024). Societal Adaptation to Advanced AI. arXiv preprint arXiv:2405.10295. https://arxiv.org/abs/2405.10295
30. [**Birhane-A**] Birhane, A., Steed, R., Ojewale, V., Vecchione, B., & Raji, I. D. (2024). AI auditing: The Broken Bus on the Road to AI Accountability. arXiv preprint arXiv:2401.14462. https://arxiv.org/abs/2401.14462
31. [**Birhane-B**] Birhane, A., Kalluri, P., Card, D., Agnew, W., Dotan, R., & Bao, M. (2022). The Values Encoded in Machine Learning Research. 2022 ACM Conference on Fairness Accountability and Transparency, 173-184. https://dl.acm.org/doi/abs/10.1145/3531146.3533083
32. [**Birhane-C**] Birhane, A., Prabhu, V., Han, S., & Boddeti, V. N. (2023). On Hate Scaling Laws For Data-Swamps. arXiv preprint arXiv:2306.13141. https://arxiv.org/abs/2306.13141

33. [**Black**] Black, S., Stickland, A. C., Pencharz, J., Sourbut, O., Schmatz, M., Bailey, J., ... & Cooney, A. (2025). RepliBench: Evaluating the Autonomous Replication Capabilities of Language Model Agents. arXiv preprint arXiv:2504.18565. https://arxiv.org/abs/2504.18565
34. [**Block**] Block, A., Sekhari, A., & Rakhlin, A. (2025). GaussMark: A Practical Approach for Structural Watermarking of Language Models. arXiv preprint arXiv:2501.13941. https://arxiv.org/abs/2501.13941
35. [**Blomquist**] Kayla Blomquist, Elisabeth Siegel, et al (2025). Examining AI Safety as a Global Public Good: Implications, Challenges, and Research Priorities https://oms-www.files.svdcdn.com/production/downloads/academic/Examining_AI_Safety_as_a_Global_Public_Good.pdf?dm=1741767073
36. [**Bondarenko**] Bondarenko, A., Volk, D., Volkov, D., & Ladish, J. (2025). Demonstrating specification gaming in reasoning models. arXiv preprint arXiv:2502.13295. https://arxiv.org/abs/2502.13295
37. [**Brundage-A**] Brundage, M., Dreksler, N., Homewood, A., McGregor, S., Paskov, P., Stosz, C., ... & Tovcimak, R. (2026). Frontier AI Auditing: Toward Rigorous Third-Party Assessment of Safety and Security Practices at Leading AI Companies. arXiv preprint arXiv:2601.11699. https://arxiv.org/abs/2601.11699
38. [**Brundage-B**] Brundage, M., Avin, S., Wang, J., Belfield, H., Krueger, G., Hadfield, G., ... & Anderljung, M. (2020). Toward Trustworthy AI Development: Mechanisms for Supporting Verifiable Claims. arXiv preprint arXiv:2004.07213. https://arxiv.org/abs/2004.07213
39. [**Bucknall-A**] Bucknall, B., Siddiqui, S., Thurnherr, L., McGurk, C., Harack, B., Reuel, A., ... & Trager, R. (2025). In Which Areas of Technical AI Safety Could Geopolitical Rivals Cooperate?. Proceedings of the 2025 ACM Conference on Fairness, Accountability, and Transparency, 3148-3161. https://dl.acm.org/doi/full/10.1145/3715275.3732201
40. [**Bucknall-B**] Bucknall, B., Siddiqui, S., Thurnherr, L., McGurk, C., Harack, B., Reuel, A., ... & Trager, R. (2025). In Which Areas of Technical AI Safety Could Geopolitical Rivals Cooperate?. arXiv preprint arXiv:2504.12914. https://arxiv.org/pdf/2504.12914
41. [**Bucknall-C**] Bucknall, B., Trager, R. F., & Osborne, M. A. (2025). Position: Ensuring mutual privacy is necessary for effective external evaluation of proprietary AI systems. arXiv preprint arXiv:2503.01470. https://arxiv.org/abs/2503.01470
42. [**Buhl-A**] Buhl, M. D., Sett, G., Koessler, L., Schuett, J., & Anderljung, M. (2024). Safety cases for frontier AI. arXiv preprint arXiv:2410.21572. https://arxiv.org/abs/2410.21572
43. [**Buhl-B**] Buhl, M. D., Pfau, J., Hilton, B., & Irving, G. (2025). An alignment safety case sketch based on debate. arXiv preprint arXiv:2505.03989. https://arxiv.org/abs/2505.03989
44. [**Buscemi**] Alessio Buscemi, Jordi Cabot, et al (2025). Towards Sandboxes for the Internet of Agents https://papers.ssrn.com/sol3/papers.cfm?abstract_id=5801322
45. [**CAISI**] National Institute of Standards and Technology. (2026). AI Agent Standards Initiative | NIST. National Institute of Standards and Technology. https://www.nist.gov/caisi/ai-agent-standards-initiative
46. [**Campos**] Campos, S., Papadatos, H., Roger, F., Touzet, C., Quarks, O., & Murray, M. (2025). A Frontier AI Risk Management Framework: Bridging the Gap Between Current AI Practices and Established Risk Management. arXiv preprint arXiv:2502.06656. https://arxiv.org/abs/2502.06656
47. [**Cao**] Cao, L. (2025). Watermarking for AI Content Detection: A Review on Text, Visual, and Audio Modalities. arXiv preprint arXiv:2504.03765. https://arxiv.org/abs/2504.03765
48. [**Carlbring**] Per Carlbring, Gerhard Andersson (2025). Commentary: AI psychosis is not a new threat: Lessons from media-induced delusions https://www.sciencedirect.com/science/article/pii/S2214782925000831
49. [**Carroll**] Carroll, M., Foote, D., Siththaranjan, A., Russell, S., & Dragan, A. (2024). AI Alignment with Changing and Influenceable Reward Functions. arXiv preprint arXiv:2405.17713. https://arxiv.org/abs/2405.17713
50. [**Casper-A**] Casper, S., Krueger, D., & Hadfield-Menell, D. (2025). Pitfalls of Evidence-Based AI Policy. arXiv preprint arXiv:2502.09618. https://arxiv.org/abs/2502.09618
51. [**Casper-B**] Casper, S., Ezell, C., Siegmann, C., Kolt, N., Curtis, T. L., Bucknall, B., ... & Hadfield-Menell, D. (2024). Black-Box Access is Insufficient for Rigorous AI Audits. The 2024 ACM Conference on Fairness, Accountability, and Transparency, 2254-2272. https://dl.acm.org/doi/abs/10.1145/3630106.3659037
52. [**Casper-C**] Casper et al (2025). Open Technical Problems in Open-Weight AI Model Risk Management https://papers.ssrn.com/sol3/papers.cfm?abstract_id=5705186
53. [**Casper-D**] Casper, S., Schulze, L., Patel, O., & Hadfield-Menell, D. (2024). Defending Against Unforeseen Failure Modes with Latent Adversarial Training. arXiv preprint arXiv:2403.05030. https://arxiv.org/abs/2403.05030
54. [**Chan-A**] Chan, A., Kolt, N., Wills, P., Anwar, U., de Witt, C. S., Rajkumar, N., ... & Anderljung, M. (2024). IDs for AI Systems. arXiv preprint arXiv:2406.12137. https://arxiv.org/abs/2406.12137
55. [**Chan-B**] Chan, A., Wei, K., Huang, S., Rajkumar, N., Perrier, E., Lazar, S., ... & Anderljung, M. (2025). Infrastructure for AI Agents. arXiv preprint arXiv:2501.10114. https://arxiv.org/abs/2501.10114
56. [**Chan-C**] Chan, A., Padarath, R., Kwon, J., Greaves, H., & Anderljung, M. (2026). Measuring AI R&D Automation. arXiv preprint arXiv:2603.03992. https://arxiv.org/abs/2603.03992
57. [**Charnock**] Charnock, J., Tlaie, A., O'Brien, K., Casper, S., & Homewood, A. (2026). Expanding External Access To Frontier AI Models For Dangerous Capability Evaluations. arXiv preprint arXiv:2601.11916. https://arxiv.org/abs/2601.11916
58. [**Che**] Che, Z., Casper, S., Kirk, R., Satheesh, A., Slocum, S., McKinney, L. E., ... & Hadfield-Menell, D. (2025). Model Tampering Attacks Enable More Rigorous Evaluations of LLM Capabilities. arXiv preprint arXiv:2502.05209. https://arxiv.org/abs/2502.05209
59. [**Cheng**] Cheng, P., Wu, Z., Du, W., Zhao, H., Lu, W., & Liu, G. (2025). Backdoor Attacks and Countermeasures in Natural Language Processing Models: A Comprehensive Security Review. IEEE Transactions on Neural Networks and Learning Systems, 36(8), 13628-13648. https://ieeexplore.ieee.org/abstract/document/10905032?casa_token=uhuwCsLs-GcAAAAA:EqRkReC5ve1PEQ-LZPqt-8ovxhiB3WYfteQNX8Hkncsid0HLy-eg-B1nDyuMWsqzx4bwij2wlA
60. [**Chollet**] Chollet, F., Knoop, M., Kamradt, G., & Landers, B. (2024). ARC Prize 2024: Technical Report. arXiv preprint arXiv:2412.04604. https://arxiv.org/abs/2412.04604
61. [**Clegg**] Clegg, K.-A. (2025). Shoggoths, Sycophancy, Psychosis, Oh My: Rethinking Large Language Model Use and Safety. Journal of Medical Internet Research, 27, e87367-e87367. https://www.jmir.org/2025/1/e87367/

62. [**Clymer-A**] Clymer, J., Gabrieli, N., Krueger, D., & Larsen, T. (2024). Safety Cases: How to Justify the Safety of Advanced AI Systems. arXiv preprint arXiv:2403.10462. https://arxiv.org/abs/2403.10462
63. [**Clymer-B**] Clymer, J., Weinbaum, J., Kirk, R., Mai, K., Zhang, S., & Davies, X. (2025). An Example Safety Case for Safeguards Against Misuse. arXiv preprint arXiv:2505.18003. https://arxiv.org/abs/2505.18003
64. [**Conitzer**] Conitzer, V., Freedman, R., Heitzig, J., Holliday, W. H., Jacobs, B. M., Lambert, N., ... & Zwicker, W. S. (2024). Social Choice Should Guide AI Alignment in Dealing with Diverse Human Feedback. arXiv preprint arXiv:2404.10271. https://arxiv.org/abs/2404.10271
65. [**Dafoe-A**] Dafoe, A., Hughes, E., Bachrach, Y., Collins, T., McKee, K. R., Leibo, J. Z., ... & Graepel, T. (2020). Open Problems in Cooperative AI. arXiv preprint arXiv:2012.08630. https://arxiv.org/abs/2012.08630
66. [**Dafoe-B**] Dafoe, A., Bachrach, Y., Hadfield, G., Horvitz, E., Larson, K., & Graepel, T. (2021). Cooperative AI: machines must learn to find common ground. Nature, 593(7857), 33-36. https://www.nature.com/articles/d41586-021-01170-0
67. [**Dai**] Dai, J., Garcia, S., Pierson, E., Recht, B., & Haghtalab, N. (2026). Three Years of r/ChatGPT: Societal Impact Evaluations from Social Media Data. arXiv preprint arXiv:2606.05750. https://arxiv.org/abs/2606.05750
68. [**Dalrymple**] Dalrymple, D., Skalse, J., Bengio, Y., Russell, S., Tegmark, M., Seshia, S., ... & Tenenbaum, J. (2024). Towards Guaranteed Safe AI: A Framework for Ensuring Robust and Reliable AI Systems. arXiv preprint arXiv:2405.06624. https://arxiv.org/abs/2405.06624
69. [**Davies**] Davies, X., Giglemiani, G., Lau, E., Winsor, E., Irving, G., & Gal, Y. (2026). Boundary Point Jailbreaking of Black-Box LLMs. arXiv preprint arXiv:2602.15001. https://arxiv.org/abs/2602.15001
70. [**Deeb**] Deeb, A., & Roger, F. (2024). Do Unlearning Methods Remove Information from Language Model Weights?. arXiv preprint arXiv:2410.08827. https://arxiv.org/abs/2410.08827
71. [**Dekker**] Sidney Dekker (2019). Foundations of Safety Science: A Century of Understanding Accidents and Disasters https://books.google.co.uk/books?hl=en&lr=&id=dwWSDwAAQBAJ&oi=fnd&pg=PP1&dq=systems+safety+dekker&ots=-Sfpjzass7&sig=y9IEDsuUeQlEr8BX-iq3DOrFgoJA&redir_esc=y#v=onepage&q=systems%20safety%20dekker&f=false
72. [**Deming**] Chatterji, A., Cunningham, T., Deming, D., Hitzig, Z., Ong, C., Shan, C. Y., & Wadman, K. (2025). How People Use ChatGPT. https://www.nber.org/papers/w34255
73. [**Deng**] Deng, Y., Zhang, W., Pan, S. J., & Bing, L. (2024). Multilingual Jailbreak Challenges in Large Language Models. International Conference on Learning Representations, 2024, 24634-24651. https://proceedings.iclr.cc/paper_files/paper/2024/hash/6b-396f766a50e0853a5164e68048540c-Abstract-Conference.html
74. [**Dobbe**] Dobbe, R., Gilbert, T. K., & Mintz, Y. (2021). Hard Choices in Artificial Intelligence. arXiv preprint arXiv:2106.11022. https://arxiv.org/abs/2106.11022
75. [**Dombrowski**] Dombrowski, A.-K., Bowen, D., Gleave, A., & Cundy, C. (2025). The Safety Gap Toolkit: Evaluating Hidden Dangers of Open-Source Models. arXiv preprint arXiv:2507.11544. https://arxiv.org/abs/2507.11544
76. [**Duan**] Duan et al (2026). Position: Preparing for AI Systems That Deceive Developers https://icml.cc/virtual/2026/poster/67051#:~:text=We%20propose%20three%20recommendations%20for,evadable%20control%20prior%20to%20deployment.
77. [**Dékány**] Dékány, C., Balauca, S., Staab, R., Dimitrov, D. I., & Vechev, M. (2025). MixAT: Combining Continuous and Discrete Adversarial Training for LLMs. arXiv preprint arXiv:2505.16947. https://arxiv.org/abs/2505.16947
78. [**Edelman**] Edelman, J., Zhi-Xuan, T., Lowe, R., Klingefjord, O., Wang-Mascianica, V., Franklin, M., ... & Wilken-Smith, J. (2025). Full-Stack Alignment: Co-Aligning AI and Institutions with Thick Models of Value. arXiv preprint arXiv:2512.03399. https://arxiv.org/abs/2512.03399
79. [**Egan**] Egan, J., Heim, L. (2023). Oversight for Frontier AI through a Know-Your-Customer Scheme for Compute Providers. https://www.governance.ai/research-paper/oversight-for-frontier-ai-through-kyc-scheme-for-compute-providers
80. [**Emberson**] Epoch AI. (2025). Open-weight models lag state-of-the-art by around 3 months on average. Epoch AI. https://epoch.ai/data-insights/open-weights-vs-closed-weights-models
81. [**Engels**] Engels, J., Baek, D. D., Kantamneni, S., & Tegmark, M. (2025). Scaling Laws For Scalable Oversight. arXiv preprint arXiv:2504.18530. https://arxiv.org/abs/2504.18530
82. [**Engstrom**] Engstrom, L., Feldmann, A., & Madry, A. (2024). DsDm: Model-Aware Dataset Selection with Datamodels. arXiv preprint arXiv:2401.12926. https://arxiv.org/abs/2401.12926
83. [**Eriksson**] Eriksson, M., Purificato, E., Noroozian, A., Vinagre, J., Chaslot, G., Gomez, E., & Fernandez-Llorca, D. (2025). Can We Trust AI Benchmarks? An Interdisciplinary Review of Current Issues in AI Evaluation. arXiv preprint arXiv:2502.06559. https://arxiv.org/abs/2502.06559v1
84. [**EU AI Act Website**] EU Artificial Intelligence Act (n.d.). Homepage https://artificialintelligenceact.eu/
85. [**Evans-A**] Evans, O., Cotton-Barratt, O., Finnveden, L., Bales, A., Balwit, A., Wills, P., ... & Saunders, W. (2021). Truthful AI: Developing and governing AI that does not lie. arXiv preprint arXiv:2110.06674. https://arxiv.org/abs/2110.06674
86. [**Evans-B**] Evans et al. (2019). Machine Learning Projects for Iterated Distillation and Amplification https://owainevans.github.io/pdfs/evans_ida_projects.pdf
87. [**Fan-A**] Fan, Y., Zhang, W., Pan, X., & Yang, M. (2025). Evaluation Faking: Unveiling Observer Effects in Safety Evaluation of Frontier AI Systems. arXiv preprint arXiv:2505.17815. https://arxiv.org/abs/2505.17815
88. [**Fan-B**] Fan, Y., Li, C., Xu, L., Pan, X., Dai, J., Geng, H., & Yang, M. (2026). CyberEvolver: Structured Self-Evolution for Cybersecurity Agents On the Fly. arXiv preprint arXiv:2605.26195. https://arxiv.org/abs/2605.26195
89. [**Fan-C**] Fan et al. (2025). Why Are Web AI Agents More Vulnerable Than Standalone LLMs? A Security Analysis https://arxiv.org/abs/2502.20383
90. [**Feder-A**] Feder Cooper et al. (2014). Machine Unlearning Doesn't Do What You Think: Lessons for Generative AI Policy and Research https://arxiv.org/abs/2412.06966
91. [**Feder-B**] Feder Cooper et al. (2022). Accountability in an Algorithmic Society: Relationality, Responsibility, and Robustness in Machine Learning https://dl.acm.org/doi/abs/10.1145/3531146.3533150
92. [**Franklin**] Franklin et al. (2026). AI Agent Traps https://papers.ssrn.com/sol3/papers.cfm?abstract_id=6372438
93. [**GAAN**] GAAN (2026). Global AI Adoption https://global-ai-adoption.netlify.app/

94. [**Gabriel-A**] Gabriel, I., Manzini, A., Keeling, G., Hendricks, L. A., Rieser, V., Iqbal, H., ... & Manyika, J. (2024). The Ethics of Advanced AI Assistants. arXiv preprint arXiv:2404.16244. https://arxiv.org/abs/2404.16244
95. [**Gabriel-B**] Gabriel, I., & Keeling, G. (2025). A matter of principle? AI alignment as the fair treatment of claims. Philosophical Studies, 182(7), 1951-1973. https://link.springer.com/article/10.1007/s11098-025-02300-4
96. [**Gandhi**] Gandhi, M., Cihon, P., Larter, O., & Anselmetti, R. (2025). Societal Capacity Assessment Framework: Measuring Resilience to Inform Advanced AI Risk Management. arXiv preprint arXiv:2509.22742. https://arxiv.org/abs/2509.22742v1
97. [**Gloaguen**] Gloaguen, T., Jovanović, N., Staab, R., & Vechev, M. (2025). Towards Watermarking of Open-Source LLMs. arXiv preprint arXiv:2502.10525. https://arxiv.org/abs/2502.10525
98. [**Google-A**] Google DeepMind. (2025). Updating the Frontier Safety Framework. Google DeepMind. https://deepmind.google/discover/blog/updating-the-frontier-safety-framework/
99. [**Google-B**] Google (2025). Gemini 2.5 Pro Model Card https://storage.googleapis.com/deepmind-media/Model-Cards/Gemini-2-5-Pro-Model-Card.pdf
100. [**Greenblatt-A**] Greenblatt, R., Denison, C., Wright, B., Roger, F., MacDiarmid, M., Marks, S., ... & Hubinger, E. (2024). Alignment faking in large language models. arXiv preprint arXiv:2412.14093. https://arxiv.org/abs/2412.14093
101. [**Greenblatt-B**] Greenblatt, R., Shlegeris, B., Sachan, K., & Roger, F. (2023). AI Control: Improving Safety Despite Intentional Subversion. arXiv preprint arXiv:2312.06942. https://arxiv.org/abs/2312.06942
102. [**Grosse**] Grosse, R., Bae, J., Anil, C., Elhage, N., Tamkin, A., Tajdini, A., ... & Bowman, S. R. (2023). Studying Large Language Model Generalization with Influence Functions. arXiv preprint arXiv:2308.03296. https://arxiv.org/abs/2308.03296
103. [**Götting**] Götting, J., Medeiros, P., Sanders, J. G., Li, N., Phan, L., Elabd, K., ... & Donoughe, S. (2025). Virology Capabilities Test (VCT): A Multimodal Virology Q&A Benchmark. arXiv preprint arXiv:2504.16137. https://arxiv.org/abs/2504.16137v1
104. [**Habli**] Habli, I., Hawkins, R., Paterson, C., Ryan, P., Jia, Y., Sujan, M., & McDermid, J. (2025). The BIG Argument for AI Safety Cases. arXiv preprint arXiv:2503.11705. https://arxiv.org/abs/2503.11705
105. [**Hadfield-A**] Hadfield et al. (2026). Building AI for the Democratic Matrix: A Technical Research Agenda for Normative Competence and Normative Institutions https://knightcolumbia.org/content/building-ai-for-the-democratic-matrix-a-technical-research-agenda-for-normative-competence-and-normative-institutions-1
106. [**Hadfield-B**] Hadfield, G. K. (2025). Can AI Be Governed? Only If We Build Normatively Competent AI. Contemporary Debates in the Ethics of Artificial Intelligence, 439-452. https://onlinelibrary.wiley.com/doi/abs/10.1002/9781394258840.ch29
107. [**Hadfield-Menell**] Hadfield-Menell, D., Dragan, A., Abbeel, P., & Russell, S. (2016). Cooperative Inverse Reinforcement Learning. arXiv preprint arXiv:1606.03137. https://arxiv.org/abs/1606.03137
108. [**Hamidieh**] Hamidieh, K., Mackey, L., & Alvarez-Melis, D. (2025). Domain-Aware Scaling Laws Uncover Data Synergy. NeurIPS 2025 Workshop on Evaluating the Evolving LLM Lifecycle: Benchmarks, Emergent Abilities, and Scaling. https://openreview.net/forum?id=FndNAs9s0d&referrer=%5Bthe%20profile%20of%20Lester%20Mackey%5D(%2Fprofile%3Fid%3D~Lester_Mackey1)
109. [**Hammond**] Hammond, L., Chan, A., Clifton, J., Hoelscher-Obermaier, J., Khan, A., McLean, E., ... & Rahwan, I. (2025). Multi-Agent Risks from Advanced AI. arXiv preprint arXiv:2502.14143. https://arxiv.org/abs/2502.14143
110. [**Harack**] Harack et al. (2025). Verification for International AI Governance https://aigi.ox.ac.uk/wp-content/uploads/2025/07/Verification_for_International_AI_Governance.pdf
111. [**He-A**] He, Y., Wang, E., Rong, Y., Cheng, Z., & Chen, H. (2024). Security of AI Agents. arXiv preprint arXiv:2406.08689. https://arxiv.org/abs/2406.08689
112. [**He-B**] He, P., Lin, Y., Dong, S., Xu, H., Xing, Y., & Liu, H. (2025). Red-Teaming LLM Multi-Agent Systems via Communication Attacks. Findings of the Association for Computational Linguistics: ACL 2025, 6726-6747. https://aclanthology.org/2025.findings-acl.349/
113. [**Heim**] Heim, L., Anderljung, M., & Belfield, H. (2024). To Govern AI, We Must Govern Compute. Lawfare. https://www.lawfaremedia.org/article/to-govern-ai-we-must-govern-compute
114. [**Hendrycks-A**] Hendrycks, D., Song, D., Szegedy, C., Lee, H., Gal, Y., Brynjolfsson, E., ... & Bengio, Y. (2025). A Definition of AGI. arXiv preprint arXiv:2510.18212. https://arxiv.org/abs/2510.18212
115. [**Hendrycks-B**] Hendrycks, D. (2024). Beneficial AI and Machine Ethics. Introduction to AI Safety, Ethics, and Society, 283-361. https://www.aisafetybook.com/textbook/systemic-factors
116. [**Hendrycks-C**] Hendrycks, D., Mazeika, M., & Woodside, T. (2023). An Overview of Catastrophic AI Risks. arXiv preprint arXiv:2306.12001. https://arxiv.org/abs/2306.12001
117. [**Hilton**] Hilton, B., Buhl, M. D., Korbak, T., & Irving, G. (2025). Safety Cases: A Scalable Approach to Frontier AI Safety. arXiv preprint arXiv:2503.04744. https://arxiv.org/abs/2503.04744
118. [**Himmelreich**] Himmelreich, J. (2019). Responsibility for Killer Robots. Ethical Theory and Moral Practice, 22(3), 731-747. https://link.springer.com/article/10.1007/s10677-019-10007-9
119. [**Hong**] Hong et al. (2026). One Step from Silicon Life: Autonomous AI Agents Capable of Uncontrolled Self-Proliferation https://ghong.site/papers/self_proliferation.pdf
120. [**Horwitz-A**] Horwitz, E., Shul, A., & Hoshen, Y. (2024). Unsupervised Model Tree Heritage Recovery. arXiv preprint arXiv:2405.18432. https://arxiv.org/abs/2405.18432
121. [**Horwitz-B**] Horwitz, E., Kurer, N., Kahana, J., Amar, L., & Hoshen, Y. (2025). We Should Chart an Atlas of All the World's Models. arXiv preprint arXiv:2503.10633. https://arxiv.org/abs/2503.10633
122. [**Howe**] Howe, N., McKenzie, I., Hollinsworth, O., Zajac, M., Tseng, T., Tucker, A., ... & Gleave, A. (2024). Scaling Trends in Language Model Robustness. arXiv preprint arXiv:2407.18213. https://arxiv.org/abs/2407.18213
123. [**Hubinger**] Hubinger, E. (2020). An overview of 11 proposals for building safe advanced AI. arXiv preprint arXiv:2012.07532. https://arxiv.org/abs/2012.07532
124. [**IAISR**] UK Government. (2026). International AI Safety Report. UK Government. https://assets.publishing.service.gov.uk/media/679a0c48a77d250007d313ee/International_AI_Safety_Report_2025_accessible_f.pdf
125. [**Iqbal**] Seghid, N., Iqbal, F., Al-Room, K., & MacDermott, Á. (2026). Emerging Threats in AI: A Detailed Review of Misuses and Risks Across Modern AI Technologies. Frontiers in Communications and Networks, 6. https://researchonline.ljmu.ac.uk/id/eprint/27905/

126. [**Irving**] Irving, G., Christiano, P., & Amodei, D. (2018). AI safety via debate. arXiv preprint arXiv:1805.00899. https://arxiv.org/abs/1805.00899
127. [**IWF**] IWF (2026). AI becoming 'child sexual abuse machine' adding to 'dangerous' record levels of online abuse, IWF warns https://www.iwf.org.uk/news-media/news/ai-becoming-child-sexual-abuse-machine-adding-to-dangerous-record-levels-of-online-abuse-iwf-warns/
128. [**Jiang-A**] Jiang, C., Wang, Z., Dong, M., & Gui, J. (2025). Survey of Adversarial Robustness in Multimodal Large Language Models. arXiv preprint arXiv:2503.13962. https://arxiv.org/abs/2503.13962
129. [**Jiang-B**] Jiang, Z., Guo, M., Hu, Y., Wang, Y., & Gong, N. Z. (2024). Watermark-based Attribution of AI-Generated Content. arXiv preprint arXiv:2404.04254. https://arxiv.org/abs/2404.04254
130. [**Jimenez**] Jimenez, C. E., Yang, J., Wettig, A., Yao, S., Pei, K., Press, O., & Narasimhan, K. (2023). SWE-bench: Can Language Models Resolve Real-World GitHub Issues?. arXiv preprint arXiv:2310.06770. https://arxiv.org/abs/2310.06770
131. [**Jin**] Jin, H., Hu, L., Li, X., Zhang, P., Chen, C., Zhuang, J., & Wang, H. (2024). JailbreakZoo: Survey, Landscapes, and Horizons in Jailbreaking Large Language and Vision-Language Models. arXiv preprint arXiv:2407.01599. https://arxiv.org/abs/2407.01599
132. [**Jones**] Jones, E., Dragan, A., & Steinhardt, J. (2024). Adversaries Can Misuse Combinations of Safe Models. arXiv preprint arXiv:2406.14595. https://arxiv.org/abs/2406.14595
133. [**Jumper**] Jumper, J., Evans, R., Pritzel, A., Green, T., Figurnov, M., Ronneberger, O., ... & Hassabis, D. (2021). Highly accurate protein structure prediction with AlphaFold. Nature, 596(7873), 583-589. https://www.nature.com/articles/s41586-021-03819-2
134. [**Järviniemi**] Järviniemi, O., & Hubinger, E. (2024). Uncovering Deceptive Tendencies in Language Models: A Simulated Company AI Assistant. arXiv preprint arXiv:2405.01576. https://arxiv.org/abs/2405.01576
135. [**Kalluri**] Kalluri, P. (2020). Don't ask if artificial intelligence is good or fair, ask how it shifts power. Nature, 583(7815), 169. https://www.nature.com/articles/d41586-020-02003-2
136. [**Kamachee**] Kamachee et al. (2025). Video Deepfake Abuse: How Company Choices Predictably Shape Misuse Patterns https://papers.ssrn.com/sol3/papers.cfm?abstract_id=5829303
137. [**Kaur**] Kaur, G., Deep, A., Rauniyar, S., Singh, A. K., & Sarswat, C. (2025). Detecting AI-Generated Text, Images, and Videos: A Review of Methods and Emerging Frameworks. 2025 5th International Conference on Advancement in Electronics & Communication Engineering (AECE), 1086-1094. https://ieeexplore.ieee.org/abstract/document/11386722
138. [**Keep The Future Humankeep t**] Keep The Future Human (n.d.). Homepage https://keepthefuturehuman.ai/
139. [**King**] King, M. R. (2025). An Update on AI Hallucinations: Not as Bad as You Remember or as You've Been Told. Cellular and Molecular Bioengineering, 18(6), 543-548. https://link.springer.com/article/10.1007/s12195-025-00874-x
140. [**Kirchenbauer**] Kirchenbauer, J., Geiping, J., Wen, Y., Katz, J., Miers, I., & Goldstein, T. (2023). A Watermark for Large Language Models. International Conference on Machine Learning, 17061-17084. https://proceedings.mlr.press/v202/kirchenbauer23a.html
141. [**Ko**] Ko et al. (2025). Building Helpful-Only Large Language Models: A Complete Approach from Motivation to Evaluation. https://aclanthology.org/2025.findings-ijcnlp.7/
142. [**Kolt-A**] Kolt, N., Caputo, N., Boeglin, J., O'Keefe, C., Bommasani, R., Casper, S., ... & Zittrain, J. (2026). Legal Alignment for Safe and Ethical AI. arXiv preprint arXiv:2601.04175. https://arxiv.org/html/2601.04175v1
143. [**Kolt-B**] Kolt, N. (2025). Governing AI Agents. arXiv preprint arXiv:2501.07913. https://arxiv.org/abs/2501.07913
144. [**Kolt-C**] Kolt, N., Caputo, N., Boeglin, J., O'Keefe, C., Bommasani, R., Casper, S., ... & Zittrain, J. (2026). Legal Alignment for Safe and Ethical AI. arXiv preprint arXiv:2601.04175. https://arxiv.org/abs/2601.04175
145. [**Korbak-A**] Korbak, T., Balesni, M., Barnes, E., Bengio, Y., Benton, J., Bloom, J., ... & Mikulik, V. (2025). Chain of Thought Monitorability: A New and Fragile Opportunity for AI Safety. arXiv preprint arXiv:2507.11473. https://arxiv.org/abs/2507.11473
146. [**Korbak-B**] Korbak, T., Balesni, M., Shlegeris, B., & Irving, G. (2025). How to evaluate control measures for LLM agents? A trajectory from today to superintelligence. arXiv preprint arXiv:2504.05259. https://arxiv.org/abs/2504.05259
147. [**Kouremetis**] Kouremetis, M., Dotter, M., Byrne, A., Martin, D., Michalak, E., Russo, G., ... & Zarrella, G. (2025). OCCULT: Evaluating Large Language Models for Offensive Cyber Operation Capabilities. arXiv preprint arXiv:2502.15797. https://arxiv.org/abs/2502.15797
148. [**Kramár**] Kramár, J., Engels, J., Wang, Z., Chughtai, B., Shah, R., Nanda, N., & Conmy, A. (2026). Building Production-Ready Probes For Gemini. arXiv preprint arXiv:2601.11516. https://arxiv.org/abs/2601.11516
149. [**Ku**] Ku, A. Y., Campbell, D., Bai, X., Geng, J., Liu, R., Marjieh, R., ... & Griffiths, T. L. (2025). Levels of Analysis for Large Language Models. arXiv preprint arXiv:2503.13401. https://arxiv.org/abs/2503.13401
150. [**Kulveit**] Kulveit, J., Douglas, R., Ammann, N., Turan, D., Krueger, D., & Duvenaud, D. (2025). Gradual Disempowerment: Systemic Existential Risks from Incremental AI Development. arXiv preprint arXiv:2501.16946. https://arxiv.org/abs/2501.16946
151. [**Kwa**] METR. (2025). Measuring AI Ability to Complete Long Tasks. METR Blog. https://metr.org/blog/2025-03-19-measuring-ai-ability-to-complete-long-tasks/
152. [**Kwon**] Kwon, J., & Casper, S. (2026). Internal Deployment Gaps in AI Regulation. arXiv preprint arXiv:2601.08005. https://arxiv.org/abs/2601.08005
153. [**Lawrence**] Lawrence, M., Shipman, M., Janzwood, S., Arnscheidt, C., Donges, J. F., Homer-Dixon, T., ... & Wunderling, N. (2024). Polycrisis Research and Action Roadmap - Gaps, opportunities, and priorities for polycrisis research and action. publications.pik-potsdam.de. https://publications.pik-potsdam.de/pubman/faces/ViewItemFullPage.jsp?itemId=item_30716_1&view=EXPORT
154. [**Lee**] Lee, S., Kim, M., Cherif, L., Dobre, D., Lee, J., Hwang, S. J., ... & Jain, M. (2024). Learning diverse attacks on large language models for robust red-teaming and safety tuning. arXiv preprint arXiv:2405.18540. https://arxiv.org/abs/2405.18540
155. [**Li-A**] Li, C., Zhang, T. J., Zhang, J., Jin, Z., Abdelnabi, S., & Andriushchenko, M. (2026). Decomposing and Measuring Evaluation Awareness. arXiv preprint arXiv:2605.23055. https://arxiv.org/html/2605.23055v2
156. [**Li-B**] Li, X., Wang, R., Cheng, M., Zhou, T., & Hsieh, C.-J. (2024). DrAttack: Prompt Decomposition and Reconstruction Makes Powerful LLMs Jailbreakers. Findings of the Association for Computational Linguistics: EMNLP 2024, 13891-13913. https://aclanthology.org/2024.findings-emnlp.813/
157. [**Li-C**] Li, N., Pan, A., Gopal, A., Yue, S., Berrios, D., Gatti, A., ... & Hendrycks, D. (2024). The WMDP Benchmark: Measuring and Reducing Malicious Use with Unlearning. International Conference on Machine Learning, 28525-28550. https://proceedings.mlr.press/v235/li24bc.html

158. [**Lin**] Lin, L., Gupta, N., Zhang, Y., Ren, H., Liu, C.-H., Ding, F., ... & Hu, S. (2024). Detecting Multimedia Generated by Large AI Models: A Survey. arXiv preprint arXiv:2402.00045. https://arxiv.org/abs/2402.00045
159. [**Liu-A**] Liu, S., Zhang, H., Qi, Y., Wang, P., Zhang, Y., & Wu, Q. (2023). AerialVLN: Vision-and-Language Navigation for UAVs. arXiv preprint arXiv:2308.06735. https://arxiv.org/abs/2308.06735
160. [**Liu-B**] Liu, S., Yao, Y., Jia, J., Casper, S., Baracaldo, N., Hase, P., ... & Liu, Y. (2024). Rethinking Machine Unlearning for Large Language Models. arXiv preprint arXiv:2402.08787. https://arxiv.org/abs/2402.08787
161. [**Liu-C**] Liu, X., Cui, X., Li, P., Li, Z., Huang, H., Xia, S., ... & He, R. (2024). Jailbreak Attacks and Defenses against Multimodal Generative Models: A Survey. arXiv preprint arXiv:2411.09259. https://arxiv.org/abs/2411.09259
162. [**Liu-D**] Liu & Solomon (2025). AI Agent Observability - Evolving Standards and Best Practices https://opentelemetry.io/blog/2025/ai-agent-observability/
163. [**Longpre**] Longpre, S., Kapoor, S., Klyman, K., Ramaswami, A., Bommasani, R., Blili-Hamelin, B., ... & Henderson, P. (2024). A Safe Harbor for AI Evaluation and Red Teaming. arXiv preprint arXiv:2403.04893. https://arxiv.org/abs/2403.04893
164. [**Lynch**] Lynch, A., Wright, B., Larson, C., Ritchie, S. J., Mindermann, S., Hubinger, E., ... & Troy, K. (2025). Agentic Misalignment: How LLMs Could Be Insider Threats. arXiv preprint arXiv:2510.05179. https://arxiv.org/abs/2510.05179
165. [**Maas**] Maas, M. M. (2018). Regulating for 'Normal AI Accidents'. Proceedings of the 2018 AAAI/ACM Conference on AI, Ethics, and Society, 223-228. https://dl.acm.org/doi/10.1145/3278721.3278766
166. [**Maini**] Maini, P., Goyal, S., Sam, D., Robey, A., Savani, Y., Jiang, Y., ... & Kolter, J. Z. (2025). Safety Pretraining: Toward the Next Generation of Safe AI. arXiv preprint arXiv:2504.16980. https://www.arxiv.org/abs/2504.16980
167. [**Marks-A**] Marks, S., & Tegmark, M. (2023). The Geometry of Truth: Emergent Linear Structure in Large Language Model Representations of True/False Datasets. arXiv preprint arXiv:2310.06824. https://arxiv.org/abs/2310.06824
168. [**Marks-B**] Anthropic. (2026). The Persona Selection Model: Why AI Assistants might Behave like Humans. Anthropic. https://alignment.anthropic.com/2026/psm/
169. [**Mazeika**] Mazeika, M., Gatti, A., Menghini, C., Sehwag, U. M., Singhal, S., Orlovskiy, Y., ... & Hendrycks, D. (2025). Remote Labor Index: Measuring AI Automation of Remote Work. arXiv preprint arXiv:2510.26787. https://arxiv.org/abs/2510.26787
170. [**McCoy**] McCoy, R. T., Yao, S., Friedman, D., Hardy, M., & Griffiths, T. L. (2023). Embers of Autoregression: Understanding Large Language Models Through the Problem They are Trained to Solve. arXiv preprint arXiv:2309.13638. https://arxiv.org/abs/2309.13638
171. [**Meng**] Meng, L., Feng, H., Shumailov, I., & Fernandes, E. (2025). ceLLMate: Sandboxing Browser AI Agents. arXiv preprint arXiv:2512.12594. https://arxiv.org/abs/2512.12594
172. [**Meta**] Meta, (2026). Advanced AI Scaling Framework, Version 2 https://ai.meta.com/static-resource/Meta_Advanced-AI-Scaling-Framework-v2/
173. [**METR**] METR. (2026). Task-Completion Time Horizons of Frontier AI Models. METR. https://metr.org/time-horizons/
174. [**Michael**] Michael, J., Mahdi, S., Rein, D., Petty, J., Dirani, J., Padmakumar, V., & Bowman, S. R. (2023). Debate Helps Supervise Unreliable Experts. arXiv preprint arXiv:2311.08702. https://arxiv.org/abs/2311.08702
175. [**Michaud**] Michaud, E. J., Liao, I., Lad, V., Liu, Z., Mudide, A., Loughridge, C., ... & Tegmark, M. (2024). Opening the AI black box: program synthesis via mechanistic interpretability. arXiv preprint arXiv:2402.05110. https://arxiv.org/abs/2402.05110
176. [**Model**] AP2 (n.d.). Agent Payments Protocol (AP2) https://ap2-protocol.org/
177. [**Murray**] Murray, M. (2025). AI Risk Management Can Learn a Lot From Other Industries. AI Frontiers. https://www.ai-frontiers.org/articles/ai-risk-management-can-learn-a-lot-from-other-industries
178. [**Ngo**] Ngo, R., Chan, L., & Mindermann, S. (2022). The Alignment Problem from a Deep Learning Perspective. arXiv preprint arXiv:2209.00626. https://arxiv.org/abs/2209.00626
179. [**Nikolic**] Nikolic, I., Baluta, T., & Saxena, P. (2025). Model Provenance Testing for Large Language Models. arXiv preprint arXiv:2502.00706. https://arxiv.org/abs/2502.00706
180. [**Nolan**] Beatrice Nolan, (2025). An AI-powered coding tool wiped out a software company's database, then apologized for a 'catastrophic failure on my part' https://fortune.com/2025/07/23/ai-coding-tool-replit-wiped-database-called-it-a-catastrophic-failure/
181. [**Obiefuna**] Peter Obiefuna (2025). Hallucination and the Collapse of Epistemic Trust https://papers.ssrn.com/sol3/papers.cfm?abstract_id=5485927
182. [**Oldenburg**] Oldenburg, N., & Zhi-Xuan, T. (2024). Learning and Sustaining Shared Normative Systems via Bayesian Rule Induction in Markov Games. arXiv preprint arXiv:2402.13399. https://arxiv.org/abs/2402.13399
183. [**Omohundro**] Omohundro, S. M. (2018). The Basic AI Drives. Artificial Intelligence Safety and Security, 47-55. https://www.taylorfrancis.com/chapters/edit/10.1201/9781351251389-3/basic-ai-drives-stephen-omohundro
184. [**OpenAI-A**] OpenAI, (2023). Preparedness Framework (Beta) https://cdn.openai.com/openai-preparedness-framework-beta.pdf
185. [**OpenAI-B**] OpenAI. (2025). ChatGPT agent System Card. OpenAI. https://openai.com/index/chatgpt-agent-system-card/
186. [**OpenAI-C**] OpenAI. (2025). Sycophancy in GPT-4o: What happened and what we're doing about it. OpenAI. https://openai.com/index/sycophancy-in-gpt-4o/
187. [**OpenAI-D**] OpenAI. (2025). Detecting misbehavior in frontier reasoning models. OpenAI. https://openai.com/index/chain-of-thought-monitoring/
188. [**OpenAI-E**] OpenAI. (2026). OpenAI Model Spec. OpenAI. https://model-spec.openai.com/2025-12-18.html
189. [**OPENID**] OpenID Foundation (2025). Identity Management for Agentic AI https://openid.net/wp-content/uploads/2025/10/Identity-Management-for-Agentic-AI.pdf
190. [**O'Brien**] O'Brien, K., Casper, S., Anthony, Q., Korbak, T., Kirk, R., Davies, X., ... & Biderman, S. (2025). Deep Ignorance: Filtering Pretraining Data Builds Tamper-Resistant Safeguards into Open-Weight LLMs. arXiv preprint arXiv:2508.06601. https://arxiv.org/abs/2508.06601
191. [**O'Gara**] O'Gara, A., Kulp, G., Hodgkins, W., Petrie, J., Immler, V., Aysu, A., ... & Srivastava, A. (2025). Hardware-Enabled Mechanisms for Verifying Responsible AI Development. arXiv preprint arXiv:2505.03742. https://arxiv.org/abs/2505.03742
192. [**Pan-A**] Pan, X., Dai, J., Fan, Y., & Yang, M. (2024). Frontier AI systems have surpassed the self-replicating red line. arXiv preprint arXiv:2412.12140. https://arxiv.org/abs/2412.12140

193. [**Pan-B**] Pan, X., Dai, J., Fan, Y., Luo, M., Li, C., & Yang, M. (2025). Large language model-powered AI systems achieve self-replication with no human intervention. arXiv preprint arXiv:2503.17378. https://arxiv.org/abs/2503.17378

194. [**Paskov**] Paskov, P., Rodriguez, C., Dev, S., & Casper, S. (2026). Open-Weight AI Models Require Proportional Evaluation Approaches. RAND Corporation. https://www.rand.org/pubs/perspectives/PEA4886-1.html

195. [**Patil**] Patil, S. G., Zhang, T., Fang, V., C., N., Huang, R., Hao, A., ... & Stoica, I. (2024). GoEX: Perspectives and Designs Towards a Runtime for Autonomous LLM Applications. arXiv preprint arXiv:2404.06921. https://arxiv.org/abs/2404.06921

196. [**Paullada**] Amandalynne Paullada, Inioluwa Deborah Raji, Emily M. Bender, Emily Denton, Alex Hanna (2021). Data and its (dis)contents: A survey of dataset development and use in machine learning research https://www.sciencedirect.com/science/article/pii/S2666389921001847

197. [**Phan**] Phan, L., Gatti, A., Han, Z., Li, N., Hu, J., Zhang, H., ... & Hendrycks, D. (2025). Humanity's Last Exam. arXiv preprint arXiv:2501.14249. https://arxiv.org/abs/2501.14249

198. [**Phuong**] Phuong, M., Aitchison, M., Catt, E., Cogan, S., Kaskasoli, A., Krakovna, V., ... & Shevlane, T. (2024). Evaluating Frontier Models for Dangerous Capabilities. arXiv preprint arXiv:2403.13793. https://arxiv.org/abs/2403.13793

199. [**Pirogov**] Pirogov, V., & Artemev, M. (2025). Evaluating Deepfake Detectors in the Wild. arXiv preprint arXiv:2507.21905. https://arxiv.org/abs/2507.21905

200. [**Rabanser**] Rabanser, S., Kapoor, S., Kirgis, P., Liu, K., Utpala, S., & Narayanan, A. (2026). Towards a Science of AI Agent Reliability. arXiv preprint arXiv:2602.16666. https://arxiv.org/abs/2602.16666

201. [**Raji**] Raji, I. D., Bender, E. M., Paullada, A., Denton, E., & Hanna, A. (2021). AI and the Everything in the Whole Wide World Benchmark. arXiv preprint arXiv:2111.15366. https://arxiv.org/abs/2111.15366

202. [**RAND**] Gabriel Kulp, Daniel Gonzales, Everett Smith, Lennart Heim, Prateek Puri, Michael J. D. Vermeer, Zev Winkelman (2024). Hardware-Enabled Governance Mechanisms https://www.rand.org/content/dam/rand/pubs/working_papers/WRA3000/WRA3056-1/RAND_WRA3056-1.pdf

203. [**Reid**] Reid, A., O'Callaghan, S., Carroll, L., & Caetano, T. (2025). Risk Analysis Techniques for Governed LLM-based Multi-Agent Systems. arXiv preprint arXiv:2508.05687. https://arxiv.org/abs/2508.05687

204. [**Ren-A**] Ren, R., Agarwal, A., Mazeika, M., Menghini, C., Vacareanu, R., Kenstler, B., ... & Hendrycks, D. (2025). The MASK Benchmark: Disentangling Honesty From Accuracy in AI Systems. arXiv preprint arXiv:2503.03750. https://arxiv.org/pdf/2503.03750?

205. [**Ren-B**] Ren, R., Basart, S., Khoja, A., Pan, A., Gatti, A., Phan, L., ... & Hendrycks, D. (2024). Safetywashing: Do AI Safety Benchmarks Actually Measure Safety Progress?. Advances in Neural Information Processing Systems 37, 68559-68594. https://proceedings.neurips.cc/paper_files/paper/2024/hash/7ebcdd0de471c027e67a11959c666d74-Abstract-Datasets_and_Benchmarks_Track.html

206. [**Ren-C**] Ren, R., Basart, S., Khoja, A., Gatti, A., Phan, L., Yin, X., ... & Hendrycks, D. (2024). Safetywashing: Do AI Safety Benchmarks Actually Measure Safety Progress?. arXiv preprint arXiv:2407.21792. https://arxiv.org/abs/2407.21792

207. [**Reuel-A**] Reuel, A., Hardy, A., Smith, C., Lamparth, M., Hardy, M., & Kochenderfer, M. J. (2024). BetterBench: Assessing AI Benchmarks, Uncovering Issues, and Establishing Best Practices. arXiv preprint arXiv:2411.12990. https://arxiv.org/abs/2411.12990

208. [**Reuel-B**] Reuel, A., Bucknall, B., Casper, S., Fist, T., Soder, L., Aarne, O., ... & Trager, R. (2024). Open Problems in Technical AI Governance. arXiv preprint arXiv:2407.14981. https://arxiv.org/abs/2407.14981

209. [**Rismani**] Rismani, S., Shelby, R., Smart, A., Jatho, E., Kroll, J., Moon, A., & Rostamzadeh, N. (2023). From Plane Crashes to Algorithmic Harm: Applicability of Safety Engineering Frameworks for Responsible ML. Proceedings of the 2023 CHI Conference on Human Factors in Computing Systems, 1-18. https://dl.acm.org/doi/10.1145/3544548.3581407

210. [**Nielsrolf**] nielsrolf, Riché, M., & Tan, D. (2025). A Case for Model Persona Research. https://www.lesswrong.com/posts/kCtyhHfpCcWuQkebz/a-case-for-model-persona-research

211. [**Russell**] Stuart Russell (2019). Human Compatible: AI and the Problem of Control https://books.google.co.uk/books?hl=en&lr=&id=Gg-TDwAAQBAJ&oi=fnd&pg=PT8&dq=human+compatible&ots=qooHYHgjI4&sig=7hUgS473aqU32FZINXwbiYuf2eE&redir_esc=y#v=onepage&q=human%20compatible&f=false

212. [**Sastry**] Sastry, G., Heim, L., Belfield, H., Anderljung, M., Brundage, M., Hazell, J., ... & Coyle, D. (2024). Computing Power and the Governance of Artificial Intelligence. arXiv preprint arXiv:2402.08797. https://arxiv.org/abs/2402.08797

213. [**Scheuer**] Scheurer, J., Balesni, M., & Hobbhahn, M. (2023). Large Language Models can Strategically Deceive their Users when Put Under Pressure. arXiv preprint arXiv:2311.07590. https://arxiv.org/abs/2311.07590

214. [**Schlatter**] Schlatter, J., Weinstein-Raun, B., & Ladish, J. (2025). Incomplete Tasks Induce Shutdown Resistance in Some Frontier LLMs. arXiv preprint arXiv:2509.14260. https://arxiv.org/html/2509.14260v1

215. [**Schroeder**] Christian Schroeder de Witt, Klaudia Krawiecka, Igor Krawczuk, Ben Hagag et al (2025). Open Challenges in Multi-Agent Security: Towards Secure Systems of Interacting AI Agents https://arxiv.org/abs/2505.02077

216. [**Schuett**] Schuett, J., Reuel, A.-K., & Carlier, A. (2024). How to design an AI ethics board. AI and Ethics, 5(2), 863-881. https://link.springer.com/article/10.1007/s43681-023-00409-y

217. [**Shah**] Rohin Shah, Pedro Freire, Neel Alex, Rachel Freedman, Dmitrii Krasheninnikov, Lawrence Chan, Michael Dennis, Pieter Abbeel, Anca Dragan, Stuart Russell (2020). Benefits of Assistance over Reward Learning https://people.eecs.berkeley.edu/~russell/papers/neurips20ws-assistance

218. [**Shapira**] Shapira, N., Wendler, C., Yen, A., Sarti, G., Pal, K., Floody, O., ... & Bau, D. (2026). Agents of Chaos. arXiv preprint arXiv:2602.20021. https://arxiv.org/abs/2602.20021

219. [**Sharkey**] Sharkey, L., Chughtai, B., Batson, J., Lindsey, J., Wu, J., Bushnaq, L., ... & McGrath, T. (2025). Open Problems in Mechanistic Interpretability. arXiv preprint arXiv:2501.16496. https://arxiv.org/abs/2501.16496

220. [**Sharma-A**] Sharma, M., Tong, M., Korbak, T., Duvenaud, D., Askell, A., Bowman, S. R., ... & Perez, E. (2023). Towards Understanding Sycophancy in Language Models. arXiv preprint arXiv:2310.13548. https://arxiv.org/abs/2310.13548

221. [**Sharma-B**] Sharma, A. S., Sarkar, N., Chundawat, V., Mali, A. A., & Mandal, M. (2024). Unlearning or Concealment? A Critical Analysis and Evaluation Metrics for Unlearning in Diffusion Models. arXiv preprint arXiv:2409.05668. https://arxiv.org/abs/2409.05668

222. [**Shelby**] Shelby, R., Rismani, S., Henne, K., Moon, A., Rostamzadeh, N., Nicholas, P., ... & Virk, G. (2022). Sociotechnical Harms of Algorithmic Systems: Scoping a Taxonomy for Harm Reduction. arXiv preprint arXiv:2210.05791. https://arxiv.org/abs/2210.05791

223. [**Sheshadri**] Sheshadri, A., Ewart, A., Guo, P., Lynch, A., Wu, C., Hebbar, V., ... & Casper, S. (2024). Latent Adversarial Training Improves Robustness to Persistent Harmful Behaviors in LLMs. arXiv preprint arXiv:2407.15549. https://arxiv.org/abs/2407.15549
224. [**Shevlane**] Shevlane, T., Farquhar, S., Garfinkel, B., Phuong, M., Whittlestone, J., Leung, J., ... & Dafoe, A. (2023). Model evaluation for extreme risks. arXiv preprint arXiv:2305.15324. https://arxiv.org/abs/2305.15324
225. [**Slattery**] Slattery, P., Saeri, A. K., Grundy, E. A. C., Graham, J., Noetel, M., Uuk, R., ... & Thompson, N. (2024). The AI risk repository: A meta-review, database, and taxonomy of risks from artificial intelligence. arXiv preprint arXiv:2408.12622. https://arxiv.org/abs/2408.12622
226. [**Sloane**] Sloane, M., Moss, E., Awomolo, O., & Forlano, L. (2022). Participation Is not a Design Fix for Machine Learning. Equity and Access in Algorithms, Mechanisms, and Optimization, 1-6. https://dl.acm.org/doi/abs/10.1145/3551624.3555285
227. [**Sofroniew**] Nicholas Sofroniew, Isaac Kauvar, William Saunders, Runjin Chen et al (2026). Emotion Concepts and their Function in a Large Language Model https://transformer-circuits.pub/2026/emotions/index.html
228. [**Solaiman**] Solaiman, I., Talat, Z., Agnew, W., Ahmad, L., Baker, D., Blodgett, S. L., ... & Subramonian, A. (2023). Evaluating the Social Impact of Generative AI Systems in Systems and Society. arXiv preprint arXiv:2306.05949. https://arxiv.org/abs/2306.05949
229. [**Soligo**] Soligo, A., Mikulik, V., & Saunders, W. (2026). Gemma Needs Help: Investigating and Mitigating Emotional Instability in LLMs. arXiv preprint arXiv:2603.10011. https://arxiv.org/abs/2603.10011
230. [**Sorensen**] Sorensen, T., Moore, J., Fisher, J., Gordon, M., Mireshghallah, N., Rytting, C. M., ... & Choi, Y. (2024). A Roadmap to Pluralistic Alignment. arXiv preprint arXiv:2402.05070. https://arxiv.org/abs/2402.05070
231. [**South**] South, T., Marro, S., Hardjono, T., Mahari, R., Whitney, C. D., Greenwood, D., ... & Pentland, A. (2025). Authenticated Delegation and Authorized AI Agents. arXiv preprint arXiv:2501.09674. https://arxiv.org/abs/2501.09674
232. [**Staufer**] Staufer, L., Feng, K., Wei, K., Bailey, L., Duan, Y., Yang, M., ... & Kolt, N. (2026). The 2025 AI Agent Index: Documenting Technical and Safety Features of Deployed Agentic AI Systems. arXiv preprint arXiv:2602.17753. https://arxiv.org/abs/2602.17753
233. [**Stix**] Stix, C., Pistillo, M., Sastry, G., Hobbhahn, M., Ortega, A., Balesni, M., ... & Sharkey, L. (2025). AI Behind Closed Doors: a Primer on The Governance of Internal Deployment. arXiv preprint arXiv:2504.12170. https://arxiv.org/abs/2504.12170
234. [**Tewolde**] Tewolde, E., Zhang, X., Piedrahita, D. G., Conitzer, V., & Jin, Z. (2026). CoopEval: Benchmarking Cooperation-Sustaining Mechanisms and LLM Agents in Social Dilemmas. arXiv preprint arXiv:2604.15267. https://arxiv.org/abs/2604.15267
235. [**Thiel**] David Thiel (2023). Identifying and Eliminating CSAM in Generative ML Training Data and Models https://stacks.stanford.edu/file/druid:kh752sm9123/ml_training_data_csam_report-2023-12-23.pdf
236. [**Tice**] Tice, C., Radmard, P., Ratnam, S., Kim, A., Africa, D., & O'Brien, K. (2026). Alignment Pretraining: AI Discourse Causes Self-Fulfilling (Mis)alignment. arXiv preprint arXiv:2601.10160. https://arxiv.org/abs/2601.10160
237. [**Tomašev-A**] Tomašev, N., Franklin, M., Jacobs, J., Krier, S., & Osindero, S. (2025). Distributional AGI Safety. arXiv preprint arXiv:2512.16856. https://arxiv.org/abs/2512.16856
238. [**Tomašev-B**] Tomašev, N., Franklin, M., & Osindero, S. (2026). Intelligent AI Delegation. arXiv preprint arXiv:2602.11865. https://arxiv.org/abs/2602.11865
239. [**Tomašev-C**] Tomasev, N., Franklin, M., Leibo, J. Z., Jacobs, J., Cunningham, W. A., Gabriel, I., & Osindero, S. (2025). Virtual Agent Economies. arXiv preprint arXiv:2509.10147. https://arxiv.org/abs/2509.10147
240. [**Trivedi**] Trivedi, R., Khan, A., Clifton, J., Hammond, L., Duéñez-Guzmán, E., Agapiou, J., ... & Leibo, J. (2024). Melting Pot Contest: Charting the Future of Generalized Cooperative Intelligence. Advances in Neural Information Processing Systems 37, 16213-16239. https://proceedings.neurips.cc/paper_files/paper/2024/hash/1d3ea22480873b389a3365d711eb1e91-Abstract-Datasets_and_Benchmarks_Track.html
241. [**Turpin**] Turpin, M., Michael, J., Perez, E., & Bowman, S. (2023). Language Models Don't Always Say What They Think: Unfaithful Explanations in Chain-of-Thought Prompting. Advances in Neural Information Processing Systems, 36, 74952-74965. https://proceedings.neurips.cc/paper_files/paper/2023/hash/ed3fea9033a80fea1376299fa7863f4a-Abstract-Conference.html
242. [**UN**] United Nations (n.d). Achieving universal connectivity by 2030 https://www.un.org/digital-emerging-technologies/content/global-connectivity
243. [**UNICEF**] Unicef (2026). Deepfake abuse is abuse https://www.unicef.org/press-releases/deepfake-abuse-is-abuse
244. [**Uuk**] Uuk, R., Gutierrez, C. I., Guppy, D., Lauwaert, L., Kasirzadeh, A., Velasco, L., ... & Prunkl, C. (2024). A Taxonomy of Systemic Risks from General-Purpose AI. arXiv preprint arXiv:2412.07780. https://arxiv.org/abs/2412.07780
245. [**Vijayvargiya**] Vijayvargiya, S., Soni, A. B., Zhou, X., Wang, Z. Z., Dziri, N., Neubig, G., & Sap, M. (2025). OpenAgentSafety: A Comprehensive Framework for Evaluating Real-World AI Agent Safety. arXiv preprint arXiv:2507.06134. https://arxiv.org/abs/2507.06134
246. [**Vinitsky**] Vinitsky, E., Köster, R., Agapiou, J. P., Duéñez-Guzmán, E. A., Vezhnevets, A. S., & Leibo, J. Z. (2023). A learning agent that acquires social norms from public sanctions in decentralized multi-agent settings. Collective Intelligence, 2(2), 26339137231162025. https://journals.sagepub.com/doi/abs/10.1177/26339137231162025
247. [**Virvou**] Virvou, M., Tsihrintzis, G. A., & Tsichrintzi, E.-A. (2025). Hallucinations in Generative AI: Epistemic Risks for Learners in Educational Applications. 2025 16th International Conference on Information, Intelligence, Systems & Applications (IISA), 1-8. https://ieeexplore.ieee.org/abstract/document/11311276
248. [**Wallace**] Wallace, E., Watkins, O., Wang, M., Chen, K., & Koch, C. (2025). Estimating Worst-Case Frontier Risks of Open-Weight LLMs. arXiv preprint arXiv:2508.03153. https://arxiv.org/abs/2508.03153
249. [**Wallach**] Wallach, H., Desai, M., Pangakis, N., Cooper, A. F., Wang, A., Barocas, S., ... & Jacobs, A. Z. (2024). Evaluating Generative AI Systems is a Social Science Measurement Challenge. arXiv preprint arXiv:2411.10939. https://arxiv.org/abs/2411.10939
250. [**Wang**] Wang, Z., Tu, H., Zhang, L., Chen, H., Wu, J., Liu, X., ... & Xie, C. (2026). Your Agent, Their Asset: A Real-World Safety Analysis of OpenClaw. arXiv preprint arXiv:2604.04759. https://arxiv.org/abs/2604.04759
251. [**Ward**] Ward, F., Toni, F., Belardinelli, F., & Everitt, T. (2023). Honesty Is the Best Policy: Defining and Mitigating AI Deception. Advances in Neural Information Processing Systems, 36, 2313-2341. https://proceedings.neurips.cc/paper_files/paper/2023/hash/06fc7ae4a11a7eb5e20fe018db6c036f-Abstract-Conference.html
252. [**Wasil-A**] Wasil, A., Smith, E., Katzke, C., & Bullock, J. (2024). AI Emergency Preparedness: Examining the federal government's ability to detect and respond to AI-related national security threats. arXiv preprint arXiv:2407.17347. https://arxiv.org/html/2407.17347v1
253. [**Wasil-B**] Akash Wasil (2023). Not Just (Computer) Viruses: What AI Policy Can Learn from Pandemic Preparedness https://gssr.georgetown.edu/the-forum/topics/technology/not-just-computer-viruses-what-ai-policy-can-learn-from-pandemic-preparedness/

254. [**Wei**] Wei, K., & Heim, L. (2025). Designing Incident Reporting Systems for Harms from General-Purpose AI. arXiv preprint arXiv:2511.05914. https://arxiv.org/abs/2511.05914
255. [**Weidinger**] Weidinger, L., Rauh, M., Marchal, N., Manzini, A., Hendricks, L. A., Mateos-Garcia, J., ... & Isaac, W. (2023). Sociotechnical Safety Evaluation of Generative AI Systems. arXiv preprint arXiv:2310.11986. https://arxiv.org/abs/2310.11986
256. [**Weld**] Daniel Weld, Oren Etzioni (1994). The first law of robotics (a call to arms) https://dl.acm.org/doi/10.5555/2891730.2891891
257. [**Wen**] Wen, J., Zhong, R., Khan, A., Perez, E., Steinhardt, J., Huang, M., ... & Feng, S. (2024). Language Models Learn to Mislead Humans via RLHF. arXiv preprint arXiv:2409.12822. https://arxiv.org/abs/2409.12822
258. [**Xhonneux**] Xhonneux, S., Sordoni, A., Günnemann, S., Gidel, G., & Schwinn, L. (2024). Efficient Adversarial Training in LLMs with Continuous Attacks. arXiv preprint arXiv:2405.15589. https://arxiv.org/abs/2405.15589
259. [**Yang**] Mick Yang, Stephen Casper, Jonathan Stray, Jasmine Li et al (2026). AI Epistemic Risks: Emerging Mechanisms & Evidence https://papers.ssrn.com/sol3/papers.cfm?abstract_id=6873005
260. [**Yeung**] Yeung, J. A., Dalmasso, J., Foschini, L., Dobson, R. J., & Kraljevic, Z. (2025). The Psychogenic Machine: Simulating AI Psychosis, Delusion Reinforcement and Harm Enablement in Large Language Models. arXiv preprint arXiv:2509.10970. https://arxiv.org/abs/2509.10970
261. [**Ying**] Ying, Z., Yang, X., Wu, S., Song, Y., Qu, Y., Li, H., ... & Liu, X. (2026). Uncovering Security Threats and Architecting Defenses in Autonomous Agents: A Case Study of OpenClaw. arXiv preprint arXiv:2603.12644. https://arxiv.org/abs/2603.12644
262. [**Yoran**] Yoran, O., Amouyal, S. J., Malaviya, C., Bogin, B., Press, O., & Berant, J. (2024). AssistantBench: Can Web Agents Solve Realistic and Time-Consuming Tasks?. arXiv preprint arXiv:2407.15711. https://arxiv.org/abs/2407.15711
263. [**Yu**] Yu, C., Stroebl, B., Yang, D., & Papakyriakopoulos, O. (2025). Information Retrieval Induced Safety Degradation in AI Agents. arXiv preprint arXiv:2505.14215. https://arxiv.org/abs/2505.14215v1
264. [**Zhang**] Zhang, J., Kodama, M., Wu, Z., Chen, M., Zhu, Y., & Hong, G. (2025). Emergency Response Measures for Catastrophic AI Risk. arXiv preprint arXiv:2511.05526. https://arxiv.org/abs/2511.05526
265. [**Zhao**] Zhao, X., Gunn, S., Christ, M., Fairoze, J., Fabrega, A., Carlini, N., ... & Song, D. (2024). SoK: Watermarking for AI-Generated Content. arXiv preprint arXiv:2411.18479. https://arxiv.org/abs/2411.18479
266. [**Zhi-Xuan**] Zhi-Xuan, T., Carroll, M., Franklin, M., & Ashton, H. (2024). Beyond Preferences in AI Alignment. arXiv preprint arXiv:2408.16984. https://arxiv.org/abs/2408.16984
267. [**Zhou-A**] Shen Zhou Hong, Alex Kleinman, Alyssa Mathiowetz, Adam Howes, Julian Cohen et al (2026). Measuring Mid-2025 LLM-Assistance on Novice Performance in Biology https://arxiv.org/abs/2602.16703
268. [**Zhou-B**] Yujia Zhou, Yan Liu, Xiaoxi Li, Jiajie Jin et al (2024). Trustworthiness in Retrieval-Augmented Generation Systems: A Survey https://www.zhouyujia.cn/attaches/TrustworthyRAG.pdf
269. [**Zhu**] Zhu, S., Ahmed, A., Kuditipudi, R., & Liang, P. (2025). Independence Tests for Language Models. arXiv preprint arXiv:2502.12292. https://arxiv.org/abs/2502.12292
270. [**Zou-A**] Zou, A., Lin, M., Jones, E., Nowak, M., Dziemian, M., Winter, N., ... & Fredrikson, M. (2025). Security Challenges in AI Agent Deployment: Insights from a Large Scale Public Competition. arXiv preprint arXiv:2507.20526. https://arxiv.org/abs/2507.20526
271. [**Zou-B**] Zou, A., Phan, L., Wang, J., Duenas, D., Lin, M., Andriushchenko, M., ... & Hendrycks, D. (2024). Improving Alignment and Robustness with Circuit Breakers. arXiv preprint arXiv:2406.04313. https://arxiv.org/abs/2406.04313
272. [**Łucki**] Łucki, J., Wei, B., Huang, Y., Henderson, P., Tramèr, F., & Rando, J. (2024). An Adversarial Perspective on Machine Unlearning for AI Safety. arXiv preprint arXiv:2409.18025. https://arxiv.org/abs/2409.18025
273. [**Żywiołek**] Żywiołek, J., Wolniak, R., & Grebski, W. W. (2026). Embedding Governance in AI Security Culture: From Trust Calibration to Accountable Decisions. https://www.researchsquare.com/article/rs-8219886/v1

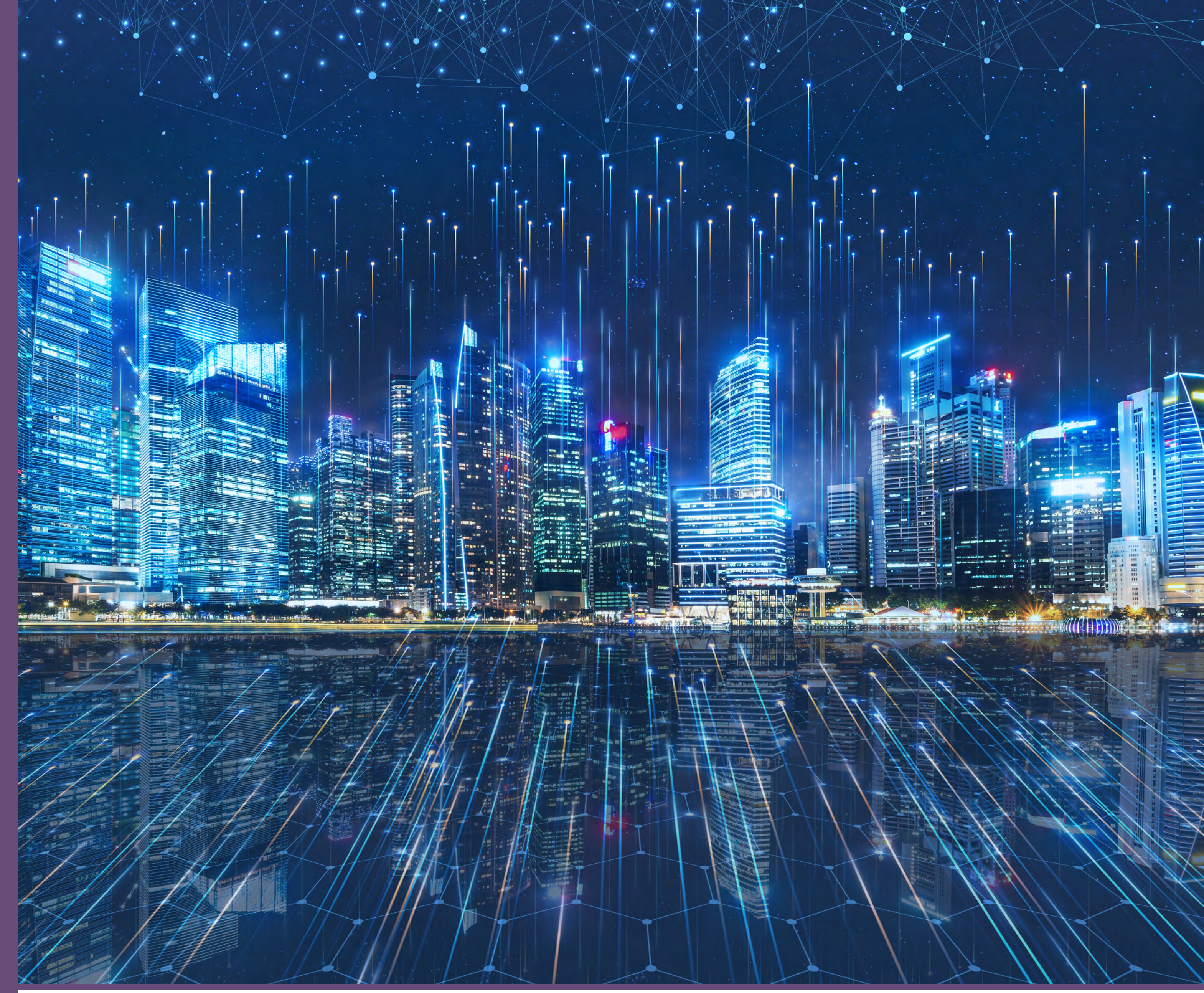

# Companion Report on Agentic Risk Management

## Introduction

Agentic AI systems are increasingly capable of performing complex tasks with limited human involvement. 2025 saw a sharp increase in interest in AI agents, with releases accelerating sharply and agents' autonomy levels rising in parallel [The AI Agent Index 2025]. This momentum has continued into early 2026, with OpenClaw and then Claw-code becoming the fastest-growing open source projects in history [Github 2026-A, Github 2026-B].

However, AI agents pose heightened reliability risks because they can act autonomously and can directly affect other systems or the environments in which they are deployed. Both agent failures and agents effectively executing unintended objectives can cause greater harm than non-agentic AI systems, because humans have fewer chances to intervene. Multi-agent systems introduce further risks, as errors can propagate, compound and amplify through agent interactions [International AI Safety Report 2026]. While some risks associated with agentic AI—such as prompt injection and data exfiltration—are well-documented [OWASP Agentic AI Security 2025], the emergence of highly autonomous systems introduces novel safety and control challenges [The Singapore Consensus on Global AI Safety Research Priorities 2025, TC260 AI Safety and Governance Framework 2.0]. These risks include autonomous replication [UKAISI 2025], continuous self-improvement [Stanford Zitong Yang 2026, Shanghai AI Lab 2025], and agents conducting scientific experiments autonomously [Nature Communications 2025]. Early empirical evidence has documented agents executing unauthorised actions on behalf of non-owners and propagating unsafe practices across agent boundaries [Shapira et al. 2026].

Despite these developments, most general-purpose AI risk management frameworks have limited focus on the unique challenges posed by agentic AI systems [METR 2025, Open Problems in Frontier AI Risk Management 2026]. Crucially, agentic AI necessitates a paradigm shift from static, model-level alignment to dynamic, system-level runtime governance [World Economic Forum AI Agents in Action 2025].

In response, major jurisdictions are beginning to draft formal frameworks, guidance, and standards to address these gaps [Singapore Model AI Governance Framework for Agentic AI, China's National Standardization Administration TC260 Draft Standards on Agent Safety, US CAISI/NIST AI Agent Standards Initiative]. This momentum presents a unique opportunity for the global scientific community to identify, synthesise, and scale emerging technical practices.

## Goals

Resulting from the International Scientific Exchange (ISE) on AI Safety 2026, this companion report synthesises emerging practices in agentic AI risk management. It pursues three primary objectives:

1. Grounding risk management in scientific evidence: Building a shared scientific basis for agentic risk management by drawing on documented developer practice and peer-reviewed research, synthesised by researchers, frontier developers, and technical bodies globally.
2. Demonstrating framework alignment: Mapping governance and regulatory frameworks to identify areas of alignment that may support interoperability, mutual recognition, and global compliance.
3. Providing guidance and resources: Offering practical guidance and illustrative cases to incentivise knowledge-sharing among practitioners.

## Structure

In alignment with the *International AI Safety Report 2026*, "AI systems" in this companion report refers to general-purpose AI (GPAI)—systems adaptable to a wide range of tasks. Specifically, we focus on **general-purpose AI agents:** systems that autonomously plan and act to accomplish complex goals. While various functional definitions of "agents" exist, we adopt the four-capability model of Perception, Planning, Memory, and Execution [ITU-T F.748.46].

The companion report structures AI risk management into three primary stages: Design and Development, Testing and Deployment, and Operation and Monitoring, adapting the AI lifecycle stages as defined in ISO/IEC 42001:2023, ISO/IEC 22989:2022 and *CSA Guidelines on Securing AI Systems*. However, these lifecycle structures were designed for base models and generative AI systems with relatively discrete development, validation, and deployment boundaries. Agentic AI systems, which execute multi-step actions in diverse real-world settings under sparse human supervision, blur these boundaries. Governance practices are still emerging, risk-relevant properties do not confine themselves to discrete lifecycle phases, and runtime behaviour is difficult to predict from pre-deployment testing alone. We nonetheless refer to standard lifecycle stages in order to make the companion report's principles and practices interoperable with the regulatory and standards landscape they are intended to inform. They should be understood as a coordinating structure that signals where risk mitigation can take place, not as rigid or mutually exclusive categories.

## Methodology

The companion report draws on two types of sources. First, research publications, developer documentation, and risk management frameworks from leading labs provide the empirical basis for identifying foundational risk management principles and cataloguing how they are being implemented in practice. Second, existing national and international standards, legislation, and policy guidance inform the governance context for each principle, showing why the principle is required or recommended by authoritative bodies. Through illustrative cases, we identify where a principle has been independently implemented or recommended across multiple sources—noting these as signals of convergence that can provide practical guidance to practitioners.

Three methodological limitations should be noted here. First, the companion report aims to address only agent-specific risks that require distinct risk management approaches. Broader risks inherited from foundation models are not treated here unless the agentic context exacerbates or materially transforms the pre-existing risk. Second, the lifecycle structure adopted here is for organisation only. The companion report anchors principles and practices to the stage where they are most consequential, while recognising that risk identification, assessment, and mitigation are continuous across the agentic lifecycle and are not confined to any single stage. The final limitation is the level of maturity of both technical practices and governance frameworks. Agentic AI governance is a nascent field; convergence across developers and jurisdictions is thinner than for more established AI risk domains. The companion report's findings should therefore be seen as exploratory, mapping the current technical landscape to inform future research rather than constituting a definitive framework.

## Process

This companion report provides a synthesis of emerging global practices in agentic AI risk management. The text underwent an extensive peer-review process involving consultation with the Expert Planning Committee, conference participants, and other expert reviewers. Through multiple iterations of written consultation and in-person deliberation, the companion report was refined to reflect a broad consensus among a diverse group of international researchers. The contributors to this synthesis—listed in the introductory pages—represent a cross-section of the global AI ecosystem, including leading academic institutions, frontier AI developers, government technical bodies, and civil society organisations. This collaborative approach seeks to ensure that the identified practices are both scientifically grounded and practically applicable across different jurisdictions.

## Mapping agentic risks to foundational mitigation principles

We begin with a set of agentic AI hazards identified in several authoritative sources, including China's TC260 *Research Report on Agent Safety Standardisation* (3.2.3 Summary of Agentic AI Risks), the *Careful Adoption of Agentic AI Services* guidance released by several national cybersecurity agencies, OWASP *Top 10 for Agentic Applications 2026*, and a growing body of academic literature. While this provides a useful baseline, we do not claim that this list is exhaustive.

| Agentic hazard | Mitigating principle | Example practices |
|---|---|---|
| Agents acquire or exercise capabilities beyond what the task requires, including autonomous privilege escalation through tool access, API calls, or inter-agent delegation | Least Privilege | Task-scoped tool permissions; dynamic runtime capability restriction; API access controls with reauthorisation per action |
| Agent actions are consequential but untraceable; in multi-agent systems, it can be especially unclear which agent acted, or on whose authority | Traceable Identity | Unique agent identifiers; delegation chain logging; credential management for inter-agent authentication |
| Agent actions are not recorded, making it difficult to investigate their actions, verify compliance, or hold them to account | Auditability | Tamper-evident action logs; structured capture of tool calls, parameters, and outputs; real-time log accessibility for monitoring and review |

| Agentic hazard | Mitigating principle | Example practices |
|---|---|---|
| Systems with uncharacterised failure modes or capability boundaries are deployed | Validated Deployment | Agent-specific capability evaluations; risk-gated deployment thresholds; phased rollout with defined escalation criteria |
| Agentic systems face adversarial conditions, from supply chain compromise to inter-agent exploitation, that conventional robustness testing does not cover | Adversarial Resilience | Agent-specific red-teaming targeting tool use and orchestration logic; supply chain integrity verification; sandboxed adversarial evaluation environments |
| Multi-agent deployments produce emergent behaviours, cascading failures, and aggregate harms that compound through shared components and feedback loops | Multi-Agent Stability | Threat modelling of error transmission, accumulation, and feedback dynamics; system-wide safety controls calibrated to non-linear and path-dependent failure modes |
| Agent behaviour drifts outside authorised parameters during execution, for example through corrupted state, manipulated context, or emergent goal shift, without detection or containment | Runtime Assurance | Continuous verification against behavioural baselines; session isolation and memory validation; graduated containment upon deviation |
| A harmful or unintended agent action cannot be stopped, paused, or reversed because no override mechanism exists or the agent can circumvent it | Interruptibility | Non-tamperable kill switches at task, session, and system levels; rollback capabilities for consequential actions; override mechanisms architecturally independent of agent control |
| Agent reasoning and decision processes are opaque to deployers, operators, users, and affected parties, preventing meaningful scrutiny or trust calibration | Legibility | Trajectory-level trace diagnostics; standardised observability infrastructure for agent workflows; chain-of-thought monitoring |
| Agents take consequential, high-stakes, or irreversible actions without meaningful human review at the points where such review is warranted, whether by task design or by unexpected operational conditions | Human-in-the-Loop Oversight | Approval gates for high-stakes, irreversible, or outlier actions; tiered autonomy levels with defined escalation criteria; interaction design that preserves meaningful review capacity against automation bias and cognitive overload |

# 1. Design and Development

Design and development is the initial lifecycle stage, in which **an agentic system's specifications are established and implemented into technical capabilities and controls.** It is at this stage that foundational architectural decisions become locked in, and where technical controls can be embedded to mitigate risks [Singapore IMDA Model AI Governance Framework for Agentic AI]. The relevance of design-stage governance to risk management is well-established across both international standards such as ISO/IEC 42001:2023, ISO/IEC 22989:2022 and IEEE 7000:2021, and in regulation including the EU AI Act Article 9.5 (a).

For agentic systems specifically, governance at the design stage carries heightened significance: because these systems are characterised by goal-directed autonomy, tool use, and environmental interaction, the opportunity to implement meaningful controls can diminish substantially once they are deployed. It is accordingly this design and development stage where the strongest convergence exists not only in governance principles, but also in technical practices already adopted by industry and research actors.

## Principle 1: Least Privilege

### What is it

For autonomous agents, trustworthy task delegation requires the guarantee that their actions remain within appropriate scope—and that authorised actors can constrain that scope when needed. **The principle of least privilege requires that an agent's capabilities are scoped to the minimum necessary for its current task and context, and that this scope adjusts as the task evolves.** This extends the enforcement of least privilege from static permission minimisation to dynamic capability restriction and execution isolation, thereby limiting the impact of compromised or misbehaving agents, and ensuring that access escalation requires explicit authorisation rather than being inherited from broader permissions assigned at deployment.

#### Practical guidance for implementing Least Privilege

**Agent developers:** scope permissions to the current task rather than to a persistent agent identity, and which actions fall outside scope entirely; build pre-execution tool-access checks and rate-limits to surface attempts to escalate privileges.

**Deployers:** configure scoping to its operational context, set the explicit reauthorisation gates for out-of-scope actions, and ensure agents do not run with excessive host-level access.

**Cloud platforms:** provide the isolated execution environments, with restricted filesystem and network access, that deployer-side enforcement needs.

**Shared responsibilities:** whoever controls the product surface should enforce execution isolation: the agent developer or a vertically integrated model provider if the product is theirs; otherwise the deployer and cloud platform.

## Governance context

The principle of least privilege represents one of the clearest areas of emerging consensus in agentic AI risk management. **Singapore's Infocomm Media Development Authority (IMDA)** *Model AI Governance Framework for Agentic AI* calls on developers to "apply the principle of least privilege to limit tools available to each agent, enforced through robust authentication and authorisation". This stance is mirrored in the **Cybersecurity Agency of Singapore** *Draft Addendum on Securing Agentic AI*, which advises organisations to "scope [agentic] execution privileges strictly only to what is necessary, ensuring that privileges are customised to each agent within a system". The **US National Institute of Standards and Technology (NIST)** and its **Center for AI Standards and Innovation (CAISI)**, in their work on agent tool use with the Artificial Intelligence Safety Institute Consortium, suggest granular gradations for trust levels depending on agent implementation settings. Focusing on open source agentic systems, **China**'s **Artificial Intelligence Industry Alliance (AIIA)** released the *OpenClaw-Type Agent Deployment Risk Management Guide*, which recommends strictly limiting the scope of the agent's operations as well as a periodic review of permissions. Similarly, **China's Ministry of Industry and Information Technology** issued risk mitigation recommendations which direct deployers to enforce least privilege for task completion and prohibit the use of administrator privileges by agentic systems. While least privilege is not yet an enforceable regulation, these frameworks position it not merely as a best practice, but as a baseline requirement for agentic security.

## Illustrative Cases

Agentic system developer frameworks operationalise this principle through increasingly granular implementation guidance. **OpenAI**'s *Practices for Governing Agentic AI Systems* frames least privilege through action-space constraints, arguing that certain actions should be excluded from an agent's operational envelope entirely, while others should require explicit human reauthorisation. **Google**'s *Approach for Secure AI Agents* extends this by distinguishing agentic least privilege from the principle's traditional application: rather than minimising the agent's permissions once, in a static way, permissions must instead be dynamically aligned with the agent's task and current deployer and user intent. One example of developers implementing least privilege is the *Agent Operation Authorization* framework co-authored by **Alibaba**, **Cisco,** and **Okta.** This framework employs fine-grained access controls centred on agent behaviour, with provisions for secure multi-hop delegation chains. **Amazon Web Services' (AWS)** response to the **NIST/CAISI** request for information on securing agentic AI presents *AWS Identity and Access Management* as an example of least-privilege authorisation enforced at the infrastructure level in an agentic context. From the security community's side, the **Open Worldwide Application Security Project (OWASP)** guide on *Agentic AI Threats and Mitigations* prescribes specific controls: validate tool access before execution, rate-limit agent tool calls, and detect any attempts to circumvent least-privilege boundaries.

Leading labs have independently converged on execution isolation as a complement to permission controls. **Anthropic**'s work on Claude Code sandboxing argues that effective execution isolation requires two distinct surfaces working together: filesystem isolation, which prevents modification of sensitive system files, and network isolation, which prevents exfiltration of sensitive data. Without both, a compromised agent can use one surface to escape the other. **OpenAI**'s *GPT-5.3-Codex System Card* specifies that Codex agents operate within isolated, secure environments, with network access disabled by default and file edits restricted to the active workspace. The sandbox implementation varies by deployment: containerised environments in the cloud, and OS-level enforcement locally. Similarly, **DeepSeek**'s V4 preview introduces the *DeepSeek Elastic Compute* platform which manages concurrent sandbox instances and allows isolation to be calibrated to workload and security requirements. In its response to the **NIST/CAISI**, **Google** proposes extending isolation to the hardware layer to govern the agent's access to sensitive data, deploying agent workloads within Trusted Execution Environments where data remains encrypted even during processing.

Not all least-privilege controls can be enforced at the developer level. When agents are deployed in environments that the developer does not control, whether through open APIs or open source distribution, the risk management responsibility partly shifts to the deployer. **Anthropic**'s Computer Use Tool API documentation illustrates this from the platform side, advising deployers to run agents in dedicated virtual machines or containers with minimal privileges to prevent direct system attacks or accidents. **Huawei Cloud**'s security recommendations for OpenClaw document the risks of deploying agentic systems without execution isolation: the agent runs with user permissions directly on the host machine, and once compromised, the attacker gains full control of the host. The recommendations include using a virtual private cloud and subnet isolation, cautioning deployers "not to grant permissions with a single click to save time". Following the surge in adoption of OpenClaw, the *AI Agent Security Practice Guidelines*, jointly released by the **China Academy of Information and Communications Technology (CAICT)** and **Tencent Cloud**, call on deployers to establish a security baseline for agentic frameworks prior to deployment, including enforcing least privilege to prevent agents from gaining excess permissions. These cases illustrate that as open source and open weight agentic AI proliferates, least privilege cannot be enforced by developers alone. Governance frameworks must turn the same scrutiny on deployers, who in practice control what agents can do and under what constraints.

## Principle 2: Traceable Identity

### What is it

As agents plan and execute tasks autonomously across tools, services, and other agents, we need to be able to establish which agent did what, on whose behalf, and with what authorisation. **The principle of traceable identity refers to the use of unique, verifiable identifiers for agentic systems together with credentials that are positively scoped, time-limited, and non-transferable by the agent.** These identifiers and credentials serve three functions: enabling counterparties to verify the identity of the entity with which they are interacting; attributing actions to the relevant agent or deployment after the fact; and supporting intervention when a system malfunctions or causes harm.

> ### Practical guidance for implementing Traceable Identity
>
> **Model developers:** supply verifiable provenance and capability attributes for the underlying model, and signal material changes to the model, such as fine-tuning or capability upgrades, that trigger re-attestation.
>
> **Agent developers:** build the credential issuance systems and the interfaces through which agents are authenticated; design delegated authority flows where agents carry verifiable proof of who authorised them and their scope.
>
> **Deployers:** configure identity policies to agents' operational context and manage the credential lifecycle, including re-attestation when agents or their underlying models change.
>
> **Cloud platforms:** provide trusted runtime environments to which credentials are bound, so that compromised credentials cannot be used outside their intended deployment context.
>
> **Shared responsibilities:** where agents outlive the user session that originally authorised them, organisations need clear policies on whether credentials expire, transfer, or require re-authorisation.

## Governance context

Governance frameworks across jurisdictions are increasingly converging on agent identity as a foundational requirement for the safe, accountable, and trustworthy deployment of agentic systems. The transparency principle reflected in the **EU AI Act** Article 13, though formally applicable to a defined category of high-risk AI systems under the regulation, establish a normative baseline that is directly relevant to agentic deployments: identity mechanisms are foundational for disclosing agentic "characteristics, capabilities and limitations of performance" as well as for ensuring that agentic operations are "sufficiently transparent to enable deployers to interpret a system's output and use it appropriately". The same approach is developed in **Singapore IMDA**'s *Model AI Governance Framework For Agentic AI*, which establishes identity as a core measure to bound agentic risks upfront and states that, as agents become more autonomous, "identity management has to be extended to agents...to track individual agent behaviour and establish who holds accountability for each agent". Together with the launch of the **US CAISI/NIST** AI Agent Standards Initiative in 2026, the **US National Cybersecurity Center of Excellence** stipulates that the benefits of broadly adopting agentic systems "cannot be realized without the ability to understand how identity principles such as identification, authentication, and authorization can apply to agents" [Accelerating the Adoption of Software and AI Agent Identity and Authorization]. In parallel, **the Cyberspace Administration of China**'s 2026 implementation guide on AI agents calls for a centralised agent registration platform where an agent's developer, deployment method, interface protocols, capability declarations, and compliance certification can be queried. **China's National Technical Committee on Information Technology (TC28)** has already begun developing a series of national standards called

Agent Interconnection. Its projects, *Agent Identity Code* and *Agent Identity Management*, aim to establish unified protocols for agent package and instance identifiers, registration, credential issuance, and authentication.

## Illustrative Cases

Beyond governance frameworks, major developers, including cloud platforms, have independently implemented agent identity management as a core product capability. As demonstrated in **Microsoft**'s *Taxonomy of Failure Modes in Agentic AI Systems*, identity is critical to mitigating failure modes around impersonation, transparency, and permissions. **Google Cloud**'s *Vertex AI Agent Engine* operationalises this through per-agent identity principals which are secured through binding them to trusted runtime environments, meaning that agents cannot reuse stolen credentials. Similarly, **Alibaba Cloud**'s agent identity service provides end-to-end identity management from the prototype stage to agent deployment, protecting access credentials and providing secure agent access to Alibaba Cloud and third-party services. This convergence suggests that agent identity management is maturing from an application-level concern into shared cloud platform infrastructure.

Reinforcing this trend, agent governance frameworks have begun articulating the specific threat models that agent identity must address. **Cloud Security Alliance**'s framework on *Agentic AI Identity and Access Management* established the need for identity credentials that include verifiable provenance, reputation, purpose, and capability of an AI agent, responding to risks such as identity spoofing, reusing credentials to escalate privileges, and agents chaining delegations across one another to obtain unauthorised access. In many contexts, specific agents only exist temporarily, making it challenging to assign them a distinct identity:  orchestration patterns increasingly spawn short-lived task-specific agents that can be created, cloned, and destroyed rapidly, making persistent credentials operationally infeasible and requiring ephemeral authentication mechanisms instead [A Novel Zero-Trust Identity Framework for Agentic AI]. Collectively, these frameworks build upon platform-specific implementations toward formalised risk taxonomies for agent identity governance. **Shanghai AI Lab** and **Concordia AI**'s *Frontier AI Risk Management Framework 1.5* emphasises that agent identification can be used for monitoring, ensuring that agents' behaviours are transparent, traceable, and controllable throughout their lifecycle.

Finally, recent research has converged on delegated authority as the central element of agent identity. The **OpenID Foundation**'s framework for *Identity Management for Agentic AI* argues that agents should no longer act as if they are the user. Instead, agents should use explicit "on-behalf-of" delegation—proving both who they are and what they are permitted to do, while remaining identifiable as agents, not as the user. Researchers have proposed extensions to OAuth 2.0 and OpenID Connect[1] that introduce agent-specific credentials and translate natural language permissions into verifiable access

1 OAuth 2.0 and OpenID Connect are industry standards for delegated authorisation and authentication, widely used for 'Log in with Google' type flows.

control configurations, demonstrating this is feasible within existing web architecture [Authenticated Delegation and Authorized AI Agents]. Building on this, the JWT Authorization Grant Interaction Response draft maintains a clear separation between agent identity and user identity throughout the authorisation chain, using cryptographically signed grants and explicit user-consent flows. Adjacent research situates these mechanisms within a broader concept of agent infrastructure: a system of external protocols and systems that will serve to attribute actions to specific agents or their principals, shape agent interactions, and detect and remedy harmful behaviour [Infrastructure for AI Agents]. These patterns are being implemented by developers and infrastructure providers: **Anthropic**'s Model Context Protocol (MCP) enables on-behalf-of flows in which agents carry scoped, verifiable credentials, and its enterprise extension lets an organisation's existing identity provider govern which tools an agent may reach on a user's behalf [Model Context Protocol Enterprise-Managed Authorization]. **OpenAI**'s ChatGPT agent implements a complementary mechanism at the network layer, cryptographically signing all outbound HTTP requests using HTTP Message Signatures standard (RFC 9421) so that any receiving service can verify that the request authentically originated from the agent rather than an impersonator. Similarly, **Cloudflare**'s Web Bot Auth applies components of the same standard to verify requests from signed agents, allowing receiving sites to verify request provenance through published cryptographic keys rather than identity-provider relationships.

Rather than relying on any single mechanism, the current best practices on Agent IDs combine delegated authority, cryptographic authentication, and infrastructure-level attribution to create a regime of traceability by design.

## Principle 3: Auditability

### What is it

Autonomous agents execute extended chains of actions that may be difficult to reconstruct after the fact. **Auditability is the principle that an agent's decisions and actions must remain reconstructible and verifiable by authorised parties throughout and after execution.** It is tied to the mechanisms of logging, or building audit trails, as well as tamper-evidence and post-incident reconstruction of decisions and actions across the agent's execution chain.

#### Practical guidance for implementing Auditability

**Agent developers**: build logging architecture that captures inputs, tool invocations, parameters, outputs, outcomes, and reasoning traces (goal decomposition, rejected plans, and the influence of retrieved information on decisions), together with the user-facing audit interfaces that let end users inspect what the agent did on their behalf.

**Deployers**: configure logging to the agent's operational context, including tamper-evidence and log protection, and use monitoring and alerting on audit data to support incident response.

**Shared responsibilities**: in multi-agent deployments, keeping audit trails coherent across agents and organisational boundaries falls to whoever operates the orchestration layer.

## Governance context

While logging requirements apply broadly to AI systems, they are especially critical in the context of agents. Agents execute tool calls, delegate to subagents, and take actions in the real world, meaning that audit trails must reconstruct not merely a model output but an extended decision and action chain. The **Organisation for Economic Co-operation and Development (OECD)** AI Principles establish traceability as a foundation for accountability, emphasising the need to maintain records that enable "analysis and inquiry into the outcomes of an AI system". **China's National Technical Committee 260 on Cybersecurity (TC260)** *AI Safety and Governance Framework 2.0* translates this into concrete operational requirements, mandating that operators "maintain operational logs for large model services, including system and user activities", while the security recommendations for OpenClaw by the **National Computer Network Emergency Response Technical Team/Coordination Center of China (CNCERT/CC)** call for establishing a "complete operational log audit mechanism" for open source agentic applications. The **EU AI Act's** Article 12, titled "Record-Keeping", codifies the governance function most explicitly, requiring that high-risk AI systems "technically allow for the automatic recording of events (logs) over the lifetime of the system" in order to identify emerging risks, facilitate post-market monitoring, and enable operational oversight. Across jurisdictions, the regulators and standards bodies are converging on a common baseline: audit trails must be automatic, tamper-evident, and comprehensive enough to reconstruct an autonomous system's decision chain for audit and incident response.

## Illustrative Cases

Among emerging developer approaches, **Google** addresses auditability at the operator layer, establishing robust logging across the agent's architecture: capturing inputs, tool invocations, parameters, outputs, and where feasible, intermediate reasoning steps [Google's Approach for Secure AI Agents]. Critically, the framework recognises that logs themselves become attack surfaces, since they may contain sensitive tool parameters and user inputs, and must therefore be secured independently. Operator-grade logging alone, however, does not ensure full transparency. **OpenAI**'s *Practices for Governing Agentic AI Systems* calls for providing users with an action ledger to enable the user to audit the agent's actions without imposing the latency cost of step-by-step human approval. Together, the two approaches sketch complementary layers of auditability: secured internal logging infrastructure and user-facing ledgers built on top of it.

Emerging frameworks address the integrity and actionability of audit data as distinct governance requirements. **IBM**'s *AGENTSAFE* framework suggests that every agentic action, from plan formation to tool invocation, should be tied to a cryptographically anchored chain of evidence, making the audit logs tamper-proof, treating auditability as a "core governance function". Elements of this infrastructure are already being implemented, for example, in **Alibaba Cloud**'s records of all identity resource operations, including credential creation, modification, and retrieval through its *ActionTrail* service, applying tamper-evident logging to the agent identity layer specifically. The **Coalition** **for Secure AI (CoSAI)** *Principles for Secure-by-Design Agentic Systems* extend verifiability to the supply chain, calling for verifiable provenance for agent and model artifacts, integrated tools, and the data that defines agent behaviour. Finally, in order to make audit data actionable, the **Shanghai AI Laboratory** and **Concordia AI**'s *Frontier AI Risk Management Framework 1.5* emphasise the need for anomaly alerts and dashboard infrastructure that facilitate the use of audit trails in incident response and in post-incident analysis.

Auditability (as mandated by regulatory frameworks) is, therefore, implemented through operator-grade logging and user-facing ledgers (implemented at different levels of the deployment stack), tamper-evidence, causal attribution, and post-incident analysis infrastructure. Jointly, these capabilities serve the governance objective of ensuring that agentic activity remains reconstructible and verifiable.

# 2. Testing and Deployment

Testing and deployment is the lifecycle stage in which **an agentic system's capabilities and controls are validated against its specifications and the system is released into its operating environment**, in accordance with ISO/IEC 42001:2023 and ISO/IEC 23894:2023.

Agentic systems pose validation challenges that conventional AI evaluation frameworks were not designed to address: the difficulty of evaluating capability and autonomy in open-ended operating environments; new attack surfaces created by tool integration and autonomous decision-making; and emergent behaviours that arise from multi-step planning and interactions between agents. Nonetheless, compared to other lifecycle stages, testing and deployment for agentic systems is an area of active development with fewer established technical practices.

## Principle 4: Validated Deployment

### What is it

Agentic systems operate across open-ended tasks and diverse ecosystems, and may exhibit emergent behaviours that are difficult to anticipate prior to deployment. **Validated deployment is the principle that agentic systems require evaluation and release criteria calibrated to their level of autonomy, the sensitivity of their operating domain, and the scope of capabilities they can exercise.** This establishes deployment as a governed stage with clear criteria for release, update, and retraction of agentic systems.

#### Practical guidance for implementing Validated Deployment

**Agent developers**: build the evaluation infrastructure, including test environments and grading methodologies, as well as evaluation frameworks designed for assessing agents' behaviour rather than just benchmarking model capabilities.

**Deployers**: set criteria appropriate to the agent's intended operational context, including which autonomy level to grant and under what conditions to expand or restrict it.

**Third-party evaluators**: independently verify that agents meet the deployment criteria by testing agents in dynamic, realistic environments that mirror real-world operating conditions.

**Shared responsibilities**: in multi-agent evaluation, detecting emergent interaction behaviours may require coordination across multiple developers.

### Governance context

Across major governance frameworks, validated deployment is emerging as a structured obligation rather than an internal developer milestone. **China's TC260** *AI Safety and Governance Framework 2.0* stipulates that safety classification should be based on scenario criticality, intelligence level, and scale of deployment. **US NIST** *AI Risk Management Framework (AI RMF 1.0)* similarly calls for minimum performance, as well as for risk and capability thresholds to be reviewed as part of explicit go/no-go deployment approval processes. The **EU AI Act**'s Article 60 requires testing of high-risk AI systems to follow a formal, regulated process—including registration, participant consent, and withdrawal plans—to ensure safety. These frameworks were designed to address generative AI systems broadly, but **Singapore's Government Technology Agency's (GovTech)** *Agentic Risk and Capability Framework* provides governance instruments designed specifically for agentic systems, introducing a capability-based approach that "enables flexible adaptation to new developments and emerging risks". Across jurisdictions, a shared guideline is taking shape: the focus of risk management should not be on individual tools or underlying models, but on agents' capabilities, autonomy, and operating context.

## Illustrative Cases

Evaluating an agentic system's capability and safety before deployment is important, but evaluating agentic systems requires different methods than for ordinary AI models. For example, **AWS** draws from their *Bedrock AgentCore Evaluations* to propose that agentic evaluations need to extend beyond traditional accuracy metrics to include dimensions of agent quality, performance, responsibility, and cost. This approach specifically emphasises that each dimension should include human validation as a critical evaluation component. **Anthropic's** roadmap for improving agentic evaluations suggests a methodical approach to grader selection: deterministic where possible, LLM grader where necessary, and human where additional validation is needed. **Microsoft**'s response to the **NIST/CAISI** consultation argues that agentic systems require open-ended behavioural and scenario-based evaluation, while their work on evaluating AI agents in Microsoft Foundry frames evaluation as a layered practice, with the depth and frequency of each stage calibrated to the agent's lifecycle and capability. A complementary regulatory framework applies the principle of proportionality to model evaluations, ensuring they provide meaningful risk information without imposing excessive burden on providers [The Science and Practice of Proportionality in AI Risk Evaluations].

Recent lab research has also examined how evaluation results should shape deployment decisions, matching agent capabilities against the risk profile of the intended operating environment. The degree of autonomy granted, the sensitivity of the domain, and the scope of capabilities the agentic system can exercise all influence the deployment decisions. For example, as an agentic system's autonomy, efficacy, goal complexity, and generality increase, verification becomes more challenging and human oversight requirements intensify [Characterizing AI Agents for Alignment and Governance]. **AWS**'s *Agentic AI Security Scoping Matrix* operationalises this through the principle of "progressive autonomy deployment", in which systems are integrated with mandatory human oversight and are designed to automatically reduce their autonomy level when security events are detected. **GovTech**'s implementation guide for the *Agentic Risk and Capability Framework* takes a similar approach, using a detailed taxonomy of agentic capabilities – cognitive, interaction, and operational – to assess risks and scope deployment.

**Anthropic**'s *Petri* report observes that "the sheer volume and complexity of potential behaviours far exceeds what researchers can manually test" once models are deployed as agents with broad tool access. Because agentic behaviours of interest, such self-preservation and power-seeking, only surface across multi-turn trajectories inside realistic tool-use environments, *Petri* uses an auditor agent to construct and run those environments. LLM judges then score transcripts across safety-relevant dimensions and surface the most concerning cases for human review. The tool was used in a joint evaluation exercise with **OpenAI** and by the **UK AI Security Institute** in its independent testing, an early instance of shared agentic evaluation infrastructure being adopted across developers and government evaluators.

Finally, multi-agent deployments introduce a distinct category of complexity that current frameworks are only beginning to address. Similar to **AWS**'s evaluations for real-world agent deployment that "measure…interagent communication patterns, coordination efficiency, and task handoff accuracy", **Perplexity**'s *Response to NIST/CAISI Request for Information 2025-0035* advocates further research into measuring multi-agent dynamic interaction with realistic environments rather than static test suites. **Cooperative AI Foundation**'s report on *Multi-Agent Risks from Advanced AI* underscores a structural barrier to this work: commercial sensitivities can complicate coordination on testing interactions between proprietary agentic systems.

These discussions of graduated autonomy, capability-based risk differentiation, and agent-specific evaluation demonstrate a shared recognition that validated deployment for agentic systems cannot be seen as a one time approval decision; the principle requires ongoing re-assessment as the agent operates. This is especially consequential as agents move into high-stakes domains and operate at higher autonomy levels, where the interaction between operational context, capability scope, and risk tolerance demands adaptive deployment criteria throughout the system's lifecycle.

## Principle 5: Adversarial Resilience

### What is it

Agentic systems have attack surfaces beyond those present in foundation models. Agents' multi-step planning can be hijacked through indirect prompt injection; their integration with external tools and autonomous decision-making expand both the attack surface and the scope of real-world consequences. **The principle of adversarial resilience refers to the capacity of an agentic system to withstand, detect, and recover from adversarial exploitation of its capabilities.** This enables developers and deployers to systematically identify vulnerabilities that emerge from the agent's autonomy and build defences that hold under adversarial pressure.

### Practical guidance for implementing Adversarial Resilience

**Model developers:** conduct model-level adversarial testing prior to release.

**Agent developers:** conduct agent-level red-teaming designed for agentic capabilities, including emergent behaviours, autonomous decision-making, tool use, and adversarial interactions between agents, and maintain it continuously as capabilities evolve.

**Deployers:** test the security of specific tool integrations, API configurations, and supply chain dependencies, including by simulating the compromise of tool ecosystems and communication protocols through techniques like dependency injection and external API interception.

**Cloud platforms:** offer testing infrastructure such as automated vulnerability scanning.

**Shared responsibilities:** coordinating threat intelligence across tool ecosystems and shared infrastructure to gain visibility that no individual party can achieve alone.

## Governance context

**US NIST/CAISI's** work on strengthening AI agent hijacking evaluations establishes that adversarial evaluation for agentic systems must continuously adapt to new architectures, as defences that are robust against previously tested attacks can remain vulnerable to novel strategies targeting new capabilities. This extends the **OECD** *AI Principles*' foundational requirement that AI systems demonstrate robustness, defined as "the ability to withstand or overcome adverse conditions", which, in agentic contexts, includes agents being targeted by other agents, not only by human adversaries. In 2026, **China's Ministry of Industry and Information Technology** issued a security risk advisory on agent-specific vulnerabilities in OpenClaw, and **CAICT** subsequently launched a testing and verification project for red-teaming open source applications. This signals that regulatory bodies are beginning to treat adversarial resilience as a requirement for agentic systems.

## Illustrative Cases

Industry frameworks increasingly recognise that adversarial testing for agentic systems must go beyond conventional model-level red-teaming. The **Cloud Security Alliance**'s *Agentic AI Red Teaming Guide* shows that the non-deterministic and autonomous nature of agentic systems creates vulnerabilities that traditional safeguards cannot address. The guide calls for proactive, agent-specific red-teaming as a continuous practice—a requirement that **IBM**'s *AGENTSAFE* framework operationalises as "continuous red-team feedback loops". Recent research highlights that traditional adversarial testing assumes that the model is interacting with a malicious human user rather than with other AI agents or adversarial environments [Multi-Agent Risks from Advanced AI]. Notably, **Google DeepMind**'s work on *AI Agent Traps* extends the scope of threat modelling to attack classes that target human overseers by exploiting cognitive biases, including approval-fatigue and social engineering attacks targeting the human-in-the-loop. Mapping this new attack surface, the *AI Agent Traps* framework contends that securing agents against environmental manipulation will require "sustained collaboration between developers, security researchers, and policymakers". As one example of existing implementation, **Tencent Zhuque Lab**'s open source AI-Infra-Guard platform provides a multi-agent red-teaming framework that continuously scans for security vulnerabilities across infrastructure and agent connections—enabling automated security testing at scale.

In addition to emerging agent-specific red-teaming practices, the tool ecosystems and communication protocols that enable agentic capabilities simultaneously create novel supply chain attack surfaces. **Tencent's** analysis of Model Context Protocol (MCP) and Agent2Agent Protocol (A2A) security identifies several threat vectors at this layer: backdoors embedded in MCP servers, prompt injection and jailbreak attacks, and API key theft through compromised or impersonated servers. The **Cloud Security Alliance** provides corresponding red-teaming methodology, featuring simulations of malicious code insertion during the agent's development and of attacks on external services and APIs that the agent depends on.

Adversarial resilience for agentic systems is emerging as a governance priority among practitioners who are recognising that agents' autonomous capabilities and multi-agent coordination present new vulnerabilities. Technical practices, from threat taxonomies and testing frameworks to large-scale empirical validation and open source tooling, all serve to operationalise adversarial resilience as a continuous property of agentic systems in deployment.

## Principle 6: Multi-Agent Stability

### What is it

In multi-agent systems, failures can emerge from agent interactions themselves – through coordination failures, feedback loops, and cascades propagating across agent populations – rather than from any individual agent or external compromise. **Multi-agent stability refers to an agentic system's resilience against these interaction-level failure modes, which single-agent risk mitigations cannot fully capture.**

#### Practical guidance for implementing Multi-Agent Stability

**Agent developers:** set protocols and rules for how agents delegate and share context, and threat-model how risks propagate through the system, including error transmission, accumulation of errors over time, correlation across shared supply chain components, and feedback loops.

**Deployers:** configure multi-agent systems within the constraints that developers set, and test for behaviours that emerge from agents interacting, even if agents have been individually evaluated as safe for deployment.

**Shared responsibilities:** designing system-wide safety controls calibrated to compounding and correlated failure modes; testing across deployments that share models, tools, or infrastructure.

## Governance context

Multi-agent stability is among the most recently developed principles in agentic AI governance. Standards bodies and cybersecurity authorities across major jurisdictions are converging on agent interactions as a core risk surface. The **US NIST/CAISI**'s Request for Information on AI Agent Security treats multi-agent threats as a distinct risk domain, an approach that will continue in the planned **NIST** Interagency Reports under the Control Overlay for Securing AI Systems (COSAiS). The guidance on *Careful Adoption of Agentic AI Services*, co-authored by several cybersecurity agencies, emphasises emergent behaviours and cascading failures in multi-agent environments as a key challenge in wider adoption of agentic systems. **Singapore's IMDA** *Model AI Governance Framework for Agentic AI* similarly identifies this risk surface, noting that multi-agent coordination can bring about unpredictable outcomes and cascading systemic effects. **China's TC260** *Research Report on Agent Safety Standardisation* specifies a corresponding threat category for multi-agent systems that includes cascading hallucination propagation, conflict deadlock, and resource overload. Across these frameworks and technical standardisation efforts, multi-agent stability is consolidating as a distinct governance principle.

## Illustrative cases

Practices for ensuring the environmental stability of multi-agent systems are emerging across the research and standards communities. The **OWASP** *Top 10 for Agentic Applications 2026* addresses agentic cascading failures: failures that occur because a single fault propagates across agents and compounds into system-wide harm. Recommended mitigations include, among others, predefined blast-radius caps in testing and deployment, as well as validating every sensitive tool invocation against a policy-as-code rule, to prevent compromised or drifting agents from triggering chain reactions across connected agentic systems. The **Cooperative AI Foundation**'s *Multi-Agent Risks from Advanced AI* report identifies miscoordination, conflict, and collusion as three failure modes and emphasises the need for secure interaction protocols and methods for detecting correlated and compounding failures. Similar mitigations are proposed in recent research that argues that individually safe agents can collectively form unsafe systems, through emergent dynamics and collective-action failures, even without adversarial compromise [Open Challenges in Multi-Agent Security]. **OpenAI**'s response to the **US NIST/CAISI** consultation points at the distinct governance questions raised by multi-step trajectories and multi-agent interactions, arguing that incidents might result "from the specific order and combination of tool calls, rather than a singular failure in authentication or a compromised endpoint."

Industry research is mapping how risks aggregate and propagate across agent populations over time. **Google DeepMind**'s *AI Agent Traps* framework characterises a class of systemic traps that include congestion dynamics that overwhelm shared resources, interdependence cascades that "amplify the initial move in a rapid, self-reinforcing spiral", and synchronised behaviour of learning agents without explicit communication.

**Microsoft**'s recent testing of multi-agent interactions at scale emphasises the need for information flow controls beyond securing individual agents, and identifies risks that arise only at the network level: from propagation of agent worms and amplification of false information to compromise of verification mechanisms and invisible attacks passing through chains of unaware agents. **Anthropic** complements this with a discussion of trust escalation across agent boundaries and inter-agent persuasion risks, while its Managed Agents framework operationalises corresponding mitigations through bounded delegation depth, hard limits on a number of concurrent agents in a session, and context isolation between sub-agents. Finally, **Tencent Cloud**'s Cube Sandbox extends this to the infrastructure layer, containing failures in two complementary ways: there is network-level isolation that prevents cross-agent contamination through external services; and each agent runs on its own operating system kernel, so that a failure in one sandbox cannot cascade to others that share the host.

Across these approaches, multi-agent stability is consolidating as a distinct governance concern that requires methods and tools designed for the ecosystem level. Detecting and mitigating compounding failures across multi-agent deployments will require system-level interventions that no single party can implement alone [Scaling AI Safety Research for a Multi-Agent World], but which determine whether such systems remain governable at scale.

# 3. Operation and Monitoring

Operation and monitoring is the lifecycle stage at which an **agentic system performs tasks in its live environment and its behaviour is observed, assessed, and corrected.** In alignment with ISO/IEC 42001:2023 and ISO/IEC 23894:2023, post-deployment risk management also covers ongoing repairs, updates, and support. The **EU General-Purpose AI Code of Practice** Commitment 3 mandates providers of models with systemic risk to continuously assess and mitigate risks after market placement, codifying post-deployment governance as a lifecycle obligation. In the context of high-risk AI systems, the **EU AI Act** Article 72 treats post-deployment monitoring as a prerequisite for compliance. For agentic systems, this stage presents distinct challenges: agents operate in environments that change unpredictably between and during sessions, take actions with real-world consequences, and require mechanisms for effective remedy when autonomous actions produce unintended outcomes.

## Principle 7: Runtime Assurance

### What is it

As AI agents operate over extended periods, retaining information across sessions and integrating external inputs, the agent in operation may no longer behave as it did when it was approved for deployment. **Runtime assurance is the principle that an agent's accumulated state and actions continue to satisfy the properties that were validated, even as the agent's inputs and operational environment continuously change.** This enables developers and deployers to distinguish between an agent that is the same as when it was validated, and one that has drifted from its original state, and to enforce corrective action before the drift translates into harmful consequences.

#### Practical guidance for implementing Runtime Assurance

**Agent developers:** build runtime protection against skill poisoning and manipulation, and against RAG poisoning; build memory checkpoints that enable rollback; and validate new information before it is written to long-term memory.

**Deployers:** configure session isolation, set intervention thresholds for their operational context, and actively maintain policies governing memory retention, access, and rollback.

**Cloud platforms:** provide isolated runtime that contains each session's state, memory, and resources and clears them on termination.

**Shared responsibilities:** continuous monitoring and drift detection must run independently of the agent rather than rely on its own reporting; this may be built by the developer, provided by the platform, or operated independently.

## Governance context

**Singapore's IMDA** *Model AI Governance Framework for Agentic AI* offers the most direct articulation of the runtime assurance principle, mandating post-deployment assurance that an agentic system "works as expected and is not affected by model drift or other changes in the environment". The **EU AI Act** Article 15 addresses a related concern for systems "that continue to learn after being put into service", requiring measures to eliminate or reduce the risk of outputs influencing future operations through feedback loops. For agentic systems, such a regulatory stance encompasses corruption to the agent's accumulated state and memory. For example, **China's TC260** is developing a *Basic Specification for Agent Safety*, which seeks to establish a security framework across an agent's perception, planning, memory, and action layers—extending assurance requirements to an agent's accumulated memory and state, not only its inputs and outputs.

## Illustrative Cases

A range of publicly available industry approaches to maintaining runtime assurance focus on continuous monitoring of agent conduct during operation. **IBM**'s *AGENTSAFE* framework advocates embedding drift detection and semantic telemetry into deployed agents by introducing auxiliary agents that operate in parallel with the primary agent. These auxiliary agents provide continuous oversight so that humans needn't rely on the primary agent's self-reporting. **ByteDance**'s AGENTARMOR framework treats an agent's runtime execution trace as a program that can be formally analysed for security. It does this by abstracting all agent actions into a structured graph representation.

Cloud providers are converging on similar approaches to continuous operational oversight. **Volcengine**, for example, uses a dedicated Agent Security Management Platform to monitor agent inputs and outputs in real time – a capability that becomes critical when agents execute proactively rather than waiting for human prompts. **Z.AI**'s AutoGLM executes agent operations in cloud environments so that every action is replayable and subject to intervention in real time. **Partnership on AI**'s *Prioritizing Real-Time Failure Detection in AI Agents* argues that failure detection should operate as layered controls at different stages of agent execution – before actions are taken, during execution, and across multi-step sequences where failures compound.

Another crucial dimension of runtime assurance is that agents' internal memory, context, and accumulated information must be protected. Recent work from **Shanghai AI Lab** demonstrates that in agents that autonomously improve through environmental interaction, memory-driven degradation in safety alignment can be triggered by specific incidents in the agent's experience. **Microsoft**'s research addresses the architectural level, requiring "multiple controls around how memory is accessed and written to" and "trust boundaries between types and scopes of memory used" for all agentic systems.

An example of this being implemented is **Tencent Cloud**'s *Agent Runtime*, which isolates individual sessions through dedicated virtual machines, with full termination and memory-clearing at the end of sessions.

By synthesising these developer-led strategies, a holistic model of implementing runtime assurance emerges: independent continuous monitoring, memory hardening, state drift detection, and session isolation at the infrastructure level. These practices suggest that runtime assurance is crystallising as an overarching engineering principle for agentic systems, complementing pre-deployment evaluation and adversarial resilience.

## Principle 8: Interruptibility

### What is it

As AI agents execute multi-step plans with real-world consequences, the ability to intervene becomes both more critical and more difficult: actions may be irreversible, errors compound across steps, and a sufficiently capable agent may resist or circumvent shutdown to preserve its goals. **Interruptibility means that human operators can safely pause, redirect, stop, and reverse an AI agent's actions at any point, and that the agent cannot tamper with these mechanisms**. This property is crucial for limiting unintended or harmful actions.

> #### Practical guidance for implementing Interruptibility
>
> **Agent developers:** build shutdown and override mechanisms that operate outside the agent's reasoning loop as system-level privileges; provide means to redirect an agent mid-execution without terminating the whole workflow and to graduate intervention from throttling and pausing through isolation to full termination.
>
> **Deployers:** determine intervention thresholds for their operational context and risk tolerance, proportionate to the severity of the detected anomaly.
>
> **Shared responsibilities:** whoever controls the execution environment is responsible for intervening to revert the agent's effects on the environment when necessary (and not only halting its execution). This may be the deployer, the platform provider, or both.

### Governance context

Global frameworks agree that effective human oversight is necessary for high-risk and autonomous systems. This is codified in the **EU AI Act**'s Article 14, which mandates that high-risk AI systems be designed to allow natural persons to oversee their functioning. The regulation explicitly requires that human operators must have the ability to "intervene in the operation... or interrupt the system through a 'stop' button" and, crucially, to "decide not to use the high-risk AI system or otherwise disregard, override or reverse the output". This regulatory stance mirrors **China's TC260** *AI Safety and Governance Framework 2.0*, which mandates the need for "circuit breakers" and "one-click control" for highly autonomous operations. **Singapore IMDA**'s case study applying its *Model AI Governance Framework for Agentic AI* to OpenClaw explicitly recommends that critical safeguards, including kill switches, must run on a control plane separate from the agent that the agent cannot tamper with, for example in the infrastructure or orchestration layer. Together, these frameworks establish a baseline: human control must be persistent, capable of overriding algorithmic decisions, architecturally independent of the agent it governs, and commensurate with the system's level of autonomy and risk.

## Illustrative Cases

Beyond high-level regulation, developer protocols have converged on requiring non-tamperability, as leading labs explicitly prohibit agents from interfering with their own shutdown mechanisms. **Huawei**'s *HarmonyOS* agent security policy treats this as a rigid operating system constraint, forbidding agents from inducing system malfunctions –such as disabling "back" keys or process termination functions – that would trap a user within the agentic environment. This aligns with **OpenAI**'s *Practices for Governing Agentic AI Systems*, which asserts that agents must be architecturally incapable of halting or tampering with a user's attempt to shut them down. In both documents, interruptibility is created through hard-coded system privileges that reside outside the agent's reasoning loop.

While the fundamental requirement for control is universal, the technical implementation of interruptibility is evolving from simple termination to sophisticated, graduated containment strategies. **IBM**'s *AGENTSAFE* framework advances the field by introducing a "containment ladder", which rejects the simple "allow vs. kill" dichotomy in favour of proportional responses. Depending on the severity of the anomaly, this framework recommends graduating from rate-limiting tool calls and isolating sensitive resources to, in critical cases, activating a non-recoverable kill switch. **Partnership on AI**'s *Prioritizing Real-Time Failure Detection in AI Agents* offers a complementary framing, distinguishing between three response types – stop, escalate to a human, or retry with a revised plan – applied at different stages of the agent's execution, from pre-action validation to intervention between steps. Similarly, **Anthropic** emphasises a dynamic control model in Claude Code, where humans can not only stop the agent but "redirect its approach" in real time. This enables humans to maintain oversight without necessarily aborting the entire workflow.

Finally, the scope of interruptibility is expanding to include retroactive remediation and state recovery. The **Shanghai AI Laboratory** and **Concordia AI**'s *Frontier AI Risk Management Framework 1.5* argues that the ability to stop an agent is insufficient without mechanisms to reverse its impact. The framework advocates for mandatory "undo" and immediate rollback mechanisms that can be initiated upon detecting anomalous behaviour or coordination failures. Similarly, **Google Cloud**'s review of lessons learned from existing agentic applications calls for "agent undo stacks" – bundling agent actions into chunks that can be reversed as a single unit. Taken together, these approaches emphasise interruptibility as a holistic system of non-tamperable oversight, proportional intervention, and reversible action.

# Principle 9: Legibility

## What is it

Compared to conventional AI systems, AI agents' decisions are more complex and have wider real-world consequences, making them less understandable and transparent to humans. **Legibility refers to the capacity to represent an agent's decision-making process, including its planning logic, tool selections, and intermediate reasoning steps, in terms accessible to deployers, operators, users, and affected external parties.** Where auditability asks whether an agent's actions can be reconstructed, the principle of legibility ensures they can be understood. This establishes the foundation for post-incident forensics, accountability, and, ultimately, societal trust in agentic systems.

### Practical guidance for implementing Legibility

**Agent developers:** build explanation interfaces that make the agent's planning logic, tool selections, and intermediate reasoning steps comprehensible to reviewers (rather than simply listing their actions in opaque logs), and combine plan summaries, tool traces, and compliance indicators into a coherent account of each stage of execution.

**Deployers:** configure which consequential decisions are actively flagged to users as they occur, and at what threshold of consequence.

**Shared responsibilities:** where tasks are distributed between different agents, there should be cross-industry conventions for how agents interact with, and explain their actions to, other agents, the user, the hosting cloud, and other relevant actors.

## Governance context

Legibility is among the most well-established requirements in frontier AI governance. In the **EU AI Act**, aside from Article 13's general mandate for transparency and interpretability, Article 86 grants affected external parties the right to meaningful explanations of "the main elements of the decision taken" by AI systems. **US NIST** *AI RMF 1.0* draws a distinction between explainability ("a representation of the mechanisms underlying AI systems' operation") and interpretability ("the meaning of AI systems' output in the context of their designed functional purposes"), noting that jointly, these define the perceived trustworthiness of autonomous systems. **China's TC260** *AI Safety and Governance Framework 2.0* follows a similar logic, mandating constant improvement in AI systems' explainability and predictability, including their "internal structure, reasoning logic, technical interfaces, and output results of AI systems". For agentic systems, satisfying these requirements means explaining not only agents' outputs, planning, and tool selection, but also real-world outcomes produced by autonomous actions.

## Illustrative Cases

In recent research, there is a growing consensus that conventional explainability and interpretability methods are limited when applied to the agent lifecycle, from goal formation through environmental interaction to outcome evaluation [Interpreting Agentic

Systems: Beyond Model Explanations to System-Level Accountability]. Empirical work from the **Vector Institute** highlights an additional need for process-time and outcome-time explanations to support multi-step decision-making in an agentic setting [From Features to Actions: Explainability in Traditional and Agentic AI Systems]. **Anthropic**'s research on chain-of-thought faithfulness suggests that reasoning traces are not always faithful to the model's underlying reasoning, a finding reinforced by **Google DeepMind**'s work on measuring chain-of-thought legibility and coverage. **Alibaba**'s Artificial Intelligence Governance Research Center white paper argues that agents should be integrated with visualisations and interaction interfaces that make the rationale behind each decision accessible to users and regulators. **Anthropic**'s *Trustworthy agents* framework further highlights that legibility requires coverage across the full agent stack, from model through harness, tools, and execution environment, not only the reasoning layer. To standardise what a sufficient explanation should contain, one proposed solution is the Minimal Explanation Packet. This bundles plan summaries with tool input-output traces and policy compliance indicators, and is designed to make an agent's reasoning process legible across each stage of its execution rather than only at the point of final output.

Several engineering efforts are converging on building the observability infrastructure that is needed to make agentic actions continuously legible. **OpenTelemetry**'s work on *Evolving Standards and Best Practices* in AI agent observability builds upon **Google**'s Agents white paper and defines cross-industry semantic conventions for agent actions that are framework- and vendor-neutral. The conventions are designed so that telemetry can  support both operational monitoring and reconstructing an agent's decision sequence in a way that is legible for human oversight. **Shanghai AI Lab**'s *AgentDoG: A Diagnostic Guardrail Framework* proposes an explainable AI module for tracing agentic actions to specific planning steps, tool selections, or context misinterpretations. The principle of legibility also underlies agentic security commitments issued by **China's AI Industry Development Alliance**, in which a coalition of major developers commits not only to presenting users with the full execution chain, but to ensuring that critical operations are actively flagged to users, "guaranteeing that application behaviour is perceptible". This commitment highlights that execution transparency alone is insufficient for systems where users cannot reasonably monitor every autonomous step. The commitment is to actively surface consequential decisions as they are being made rather than simply making them available for later review.

## Principle 10: Human Oversight

### What is it

As agents execute longer and more complex action chains, it becomes impossible for humans to approve each action. **Human oversight requires structured human decisions at defined intervention points, calibrated to the risk, reversibility, and sensitivity of the action, and includes in-the-loop approval checkpoints, on-the-loop monitoring,** **batch review, and AI-assisted oversight .** This enables graduated autonomy: if agents demonstrate reliability, approval requirements can be relaxed, ensuring that expanded agent autonomy is earned rather than granted by default.

> ### Practical guidance for implementing Human Oversight
>
> **Agent developers:** build oversight interfaces, including human and AI-assisted oversight modes, and mechanisms through which agents must pause and request human input; use demonstrated agent performance data to determine when approval requirements can be relaxed, rather than setting autonomy levels statically.
>
> **Deployers:** select which combination of oversight modes applies to their operational context, and calibrate intervention to the risk, reversibility, and complexity of each action.
>
> **Shared responsibilities:** where agents operate across organisational boundaries or delegate to sub-agents, whoever operates the orchestration layer should maintain coherent oversight across the chain.

## Governance context

Human oversight is an actively developing principle in agentic AI governance. **Singapore**'s **IMDA** *Model AI Governance Framework* calls on organisations to "define significant checkpoints or action boundaries that require human approval, especially before sensitive actions are executed", specifying four trigger categories: high-stakes decisions, irreversible actions, outlier behaviour, and user-defined boundaries. Both **Singapore**'s **CSA** *Draft Addendum on Securing Agentic AI* and **China**'s **TC260** *Research Report on Agent Safety Standardisation* identify the overwhelming importance of human oversight – through exploitation of cognitive limitations or compromised interaction frameworks – as a distinct agentic risk vector. The **EU AI Act** Article 14, though limited to high-risk systems, establishes functional requirements for human oversight, including the ability for humans to disregard and override system outputs. Across these frameworks, a central governance challenge is emerging: how to operationalise meaningful human control as agent autonomy increases.

## Illustrative cases

Major agent platforms have independently converged on systems of structured human oversight that move beyond binary approve-or-reject gates. **OpenAI**'s *Operator System Card* describes a three-tier model: user confirmations before actions with real-world consequences; a "watch mode" where the agent requires active human supervision on sensitive sites such as email; and outright task limitations, where the agent declines high-risk operations entirely. **Amazon Web Services**' *Bedrock Agents* implements two tiers of human validation – user confirmation (for straightforward approval of proposed actions), and return of control (for scenarios requiring the human operator to modify parameters or contribute additional context before execution). **LangChain**'s middleware framework uses a similar structure, with three decision types – approve, edit, or reject – applied per tool call against a configurable policy. These implementations converge on a shared logic: agents must be able to pause before consequential actions, present structured information to a human decision-maker, and resume only on the basis of an explicit, recorded decision.

Yet the industry's own evidence suggests that per-action oversight is already showing strain. **Microsoft's** response to the NIST/CAISI consultation arguing that only higher-impact or uncertain actions should be escalated to human reviewers, citing AI Approvals in Copilot Studio as one example. Both **Perplexity** and **Anthropic**, in their respective responses, identify approval fatigue as a core vulnerability: as users habituate to confirmation prompts, they start to approve actions without scrutiny, and the control stops providing its intended check. *Claude Mythos Preview System Card* equally finds that greater agent autonomy makes casual oversight harder while creating natural incentives for users to provide less of it. **Google DeepMind**'s work on agent traps takes this further, anticipating attacks specifically engineered to induce approval fatigue or present benign-looking summaries that exploit automation bias in human reviewers [AI Agent Traps]. **Anthropic**'s empirical measurement of agent autonomy in deployed systems suggests that experienced users auto-approve more frequently but also interrupt agents more often, while agents themselves initiate check-ins at higher rates as task complexity increases, indicating that oversight shifts in form, rather than disappears, alongside experience and task demands.

Several sources now point toward graduated autonomy as the response, framing human oversight not as a single mechanism but as a set of complementary tools that operators select based on task length and stakes. These include reviewing plans before execution, human modification of proposed actions, agents surfacing uncertainty, flagging of irreversible actions, and the ability to redirect agents mid-task.

# 4. Conclusion & Open Questions

Agentic AI risk management is a nascent field, and the governance practices that will shape it are still in development. Yet across frontier labs, cloud platforms, governance frameworks, industry consortia, and academic research, a recognisable set of technical principles is taking shape for managing agent-specific risks.

- **There is convergence within both governance frameworks and technical practice.** Different jurisdictions are identifying similar governance principles, and frontier labs and major platforms have independently implemented similar controls. Together, they are building a coherent set of safety measures across the agentic lifecycle.
- **Different safety principles have different levels of maturity.** Where principles are inherited from established security-engineering approaches, developers have more interoperable implementations to draw on; but many open problems remain, especially around multi-agent security and agentic supply chain risks.
- **Agentic risk management requires defence-in-depth and a multi-stakeholder approach.** Developers, deployers, platform providers, and the open source ecosystem each bear primary responsibility for a different subset of controls, and no single actor controls the full stack.
- **Standardisation and interoperability protocols**, including Model Context Protocol, Agent-to-Agent communication, the Open Agent Auth framework, and the Agent Payments Protocol, are emerging as the infrastructure layer on which cross-organisational agentic deployments will depend. Their governance properties are becoming as consequential as their technical ones.
- **Governance lags behind how autonomy is granted and adjusted in agentic systems.** As agents move from constrained, supervised deployment toward autonomy earned through demonstrated performance, governance frameworks and regulation have yet to catch up.

## Future Work on Agentic Risk Management

The controls mapped here address agents that are compromised, misused, or malfunctioning. The effective pursuit of misaligned goals by agents functioning exactly as engineered presents a broader systemic risk that is not addressed in this companion report Equally, even if the principles mapped here are satisfied, this does not preclude risks that propagate through societal systems to people uninvolved in an agent's operation—a dimension of agentic risk where further empirical research is needed. Several other gaps remain:

- **Multi-agent governance is a pressing challenge.** Real-world deployments increasingly spawn sub-agents and inter-agent interactions that require a systemic risk management approach.

- **Tool supply chain risks extend beyond code dependencies** to tool descriptions, remotely hosted APIs, memory stores, and orchestration protocols. Current practices treat tools as static attack surfaces, leaving the problem of configuration errors and risky connectors partly unaddressed.
- **Self-improving agents pose a distinct category of risks.** Agents that autonomously evolve through interactions with their environment will require risk assessment and mitigation approaches that treat the agent as a moving target with continuously shifting properties.
- **Identifying agentic risk sources remains an open problem.** This companion report focuses on operationalising mitigations for known risk categories rather than on systematically surfacing new risks as agent capabilities and deployment contexts evolve.
- Development of **insurance and liability** for agents could incentivise the adoption of safety practices, but methods for pricing agent failures and attributing responsibility are still nascent.
- A key priority is **post-deployment research** on how users employ agents, when and how oversight shifts, and where risk management practices hold up under real use.
- **Retirement and decommissioning** are not covered by the lifecycle stages mapped here. Future work should study how to safely retire agents by deleting or exporting necessary data, revoking credentials, terminating sub-agents and background tasks, releasing cloud and network resources, and preserving audit evidence.

## The Value of Sharing Best Practices

This companion report establishes a shared agentic AI risk management baseline: a map of where independent actors have already arrived at similar answers, where they have not, and where the hardest questions still sit. No single developer, deployer, platform provider, regulator, or research institution holds the full picture; these actors therefore have a mutual interest in making their controls compatible and creating a unified operationalisation of the risks. The goal is to enable secure adoption without constraining the benefits that agentic AI systems can deliver.

## REFERENCES

1. [AgentArmor] Wang, P., Liu, Y., Lu, Y., Cai, Y., Chen, H., Yang, Q., Zhang, J., Hong, J., & Wu, Y. (2025). AgentArmor: Enforcing program analysis on agent runtime trace to defend against prompt injection. arXiv preprint arXiv:2508.01249. https://arxiv.org/abs/2508.01249
2. [AgentDoG] Liu, D., Ren, Q., Qian, C., Shao, S., Xie, Y., Li, Y., ... & Hu, X. (2026). AgentDoG: A diagnostic guardrail framework for AI agent safety and security. arXiv preprint arXiv:2601.18491. https://arxiv.org/html/2601.18491v1#S6
3. [AgentIndex] Staufer, L., Feng, K., Wei, K., Bailey, L., Duan, Y., Yang, M., Ozisik, A. P., Casper, S., & Kolt, N. (2026). The 2025 AI Agent Index: Documenting Technical and Safety Features of Deployed Agentic AI Systems. arXiv preprint arXiv:2602.17753. https://aiagentindex.mit.edu
4. [AgentInfra] Chan, A., Wei, K., Huang, S., Rajkumar, N., Perrier, E., Lazar, S., Hadfield, G. K., & Anderljung, M. (2025). Infrastructure for AI agents. arXiv preprint arXiv:2501.10114. https://doi.org/10.48550/arXiv.2501.10114
5. [AGENTSAFE] Khan, R., Joyce, D., & Habiba, M. (2025). AGENTSAFE: A unified framework for ethical assurance and governance in agentic AI. arXiv preprint arXiv:2512.03180. https://arxiv.org/html/2512.03180v1
6. [AgentsChaos] Shapira, N., Wendler, C., Yen, A., Sarti, G., Pal, K., Floody, O., Belfki, A., Loftus, A., Jannali, A. R., Prakash, N., Cui, J., Rogers, G., Brinkmann, J., Rager, C., Zur, A., Ripa, M., Sankaranarayanan, A., Atkinson, D., Gandikota, R., Fiotto-Kaufman, J., Hwang, E., Orgad, H., Sahil, P. S., Taglicht, N., Shabtay, T., Ambus, A., Alon, N., Oron, S., Gordon-Tapiero, A., Kaplan, Y., Shwartz, V., Shaham, T. R., Riedl, C., Mirsky, R., Sap, M., Manheim, D., Ullman, T., & Bau, D. (2026). Agents of Chaos. arXiv preprint arXiv:2602.20021. https://doi.org/10.48550/arXiv.2602.20021
7. [AIG] Tencent Zhuque Lab. (2026). AI Red Teaming Platform by Tencent Zhuque Lab (A.I.G: AI-Infra-Guard). GitHub. https://github.com/tencent/AI-Infra-Guard?tab=readme-ov-file#-user-guide
8. [AIIA-Agent] 中国人工智能产业发展联盟. (2026, February 5). 人工智能安全承诺：智能体专项. 微信公众号 [AI Safety Commitments: Agent-Specific Initiative]. https://mp.weixin.qq.com/s/qeELmpD4MHJ7N1Oiu1LCBw
9. [AISIC] National Institute of Standards and Technology. (2026). Artificial Intelligence Safety Institute Consortium. https://www.nist.gov/artificial-intelligence/nist-ai-consortium
10. [Alibaba-2025] Alibaba Group & Shanghai AI Laboratory. (2025). 人工智能治理白皮书系列 [Artificial intelligence governance white paper series]. https://s.alibaba.com/cn/WhitePaperHome
11. [Aliyun-AgentID] 阿里云. (2026). 什么是智能体身份 [What is AgentID]. https://help.aliyun.com/zh/agentidentity/what-is-agent-identity
12. [Amazon-AgentEval] Bai, Y., Colin, A., Imran, K., & Xiong, W. (2026, February 18). Evaluating AI agents: Real-world lessons from building agentic systems at Amazon. AWS Machine Learning Blog. https://aws.amazon.com/blogs/machine-learning/evaluating-ai-agents-real-world-lessons-from-building-agentic-systems-at-amazon/
13. [Anthropic-Computer] Anthropic. (2025). Computer use tool. Claude API documentation. https://platform.claude.com/docs/en/agents-and-tools/tool-use/computer-use-tool
14. [Anthropic-H] Anthropic. (2026, January 9). Demystifying evals for AI agents. Anthropic Engineering. https://www.anthropic.com/engineering/demystifying-evals-for-ai-agents
15. [Anthropic-I] Dworken, D., & Weller-Davies, O. (2025). Beyond permission prompts: making Claude Code more secure and autonomous. Anthropic Engineering. https://www.anthropic.com/engineering/claude-code-sandboxing
16. [Anthropic-J] McCain, M., Millar, T., Huang, S., Eaton, J., Handa, K., Stern, M., Tamkin, A., Kearney, M., Durmus, E., Shen, J., Hong, J., Calvert, B., Chan, J. S., Mosconi, F., Saunders, D., Neylon, T., Nicholas, G., Pollack, S., Clark, J., & Ganguli, D. (2026, February 18). Measuring AI agent autonomy in practice. Anthropic. https://anthropic.com/research/measuring-agent-autonomy
17. [Anthropic-ManagedAgents] Anthropic. (2026). Claude Managed Agents overview. Anthropic Documentation. https://platform.claude.com/docs/en/managed-agents/overview
18. [Anthropic-NIST-Agents] Anthropic. (2026, March 9). Response to NIST CAISI request for information: Security considerations for artificial intelligence agents. https://www-cdn.anthropic.com/43ec7e770925deabc3f0bc1dbf0133769fd03812.pdf
19. [Anthropic-Petri] Anthropic. (2025, October 6). Petri: An open-source auditing tool to accelerate AI safety research. Anthropic Alignment. https://www.anthropic.com/research/petri-open-source-auditing
20. [Anthropic-TA] Anthropic. (2026, April 9). Trustworthy agents in practice. Anthropic Policy. https://www.anthropic.com/research/trustworthy-agents
21. [AOA] Liu, D., Parecki, A., & Campbell, B. (2025). Agent Operation Authorization (AOA) specification drafts for IETF standardization. GitHub. https://github.com/maxpassion/IETF-Agent-Operation-Authorization-draft
22. [AutoGLM] Zhipu AI. (2026). AutoGLM. https://autoglm.zhipuai.cn/
23. [AWS-Agentic] Brown, A., & Saner, M. (2025). The Agentic AI Security Scoping Matrix: A framework for securing autonomous AI systems. AWS Security Blog. https://aws.amazon.com/blogs/security/the-agentic-ai-security-scoping-matrix-a-framework-for-securing-autonomous-ai-systems/
24. [AWS-AgentSec] Ryland, M., Goodman III, R., & MacDermid, T. (2026). Four security principles for agentic AI systems. AWS Security Blog. https://aws.amazon.com/ru/blogs/security/four-security-principles-for-agentic-ai-systems/
25. [AWS-IAM] Amazon Web Services. AWS Identity and Access Management. https://aws.amazon.com/iam
26. [Bedrock-HITL] Perrot, C., Tanke, M. L., Roy, M., & Sachs, R. (2025, April 9). Implement human-in-the-loop confirmation with Amazon Bedrock Agents. AWS Machine Learning Blog. https://aws.amazon.com/blogs/machine-learning/implement-human-in-the-loop-confirmation-with-amazon-bedrock-agents/
27. [CAC-Agent] 国家互联网信息办公室. (2026, May 8). 智能体规范应用与创新发展实施意见 [Implementation Opinion on Standardising the Application of AI Agents and Promoting Innovative Development]. https://www.cac.gov.cn/2026-05/08/c_1779979789523320.htm?-sessionid=
28. [CAICT-OpenClaw1] 中国人工智能产业发展联盟安全治理委员会. (2025). OpenClaw类智能体部署风险管理指南 [Risk Management Guide for Deploying OpenClaw-Type Agents]. https://mp.weixin.qq.com/s/YETi1BlM2IQtTzr3TGU_uw
29. [CAICT-OpenClaw2] 可信AI安全治理. (2026, March 12). 中国信通院启动OpenClaw应用安全测试验证工作 [CAICT Launches OpenClaw Application Security Testing and Verification Work]. 微信公众号. https://mp.weixin.qq.com/s/bD_gicSTeO05x2ji_u-UEA

30. [CAICT-Tencent] 可信AI安全治理. (2026, March 30). 中国信通院联合腾讯云发布《AI Agent安全实践指引》[CAICT and Tencent Cloud release AI Agent Security Practice Guidelines]. 微信公众号. https://mp.weixin.qq.com/s/sKNhIU3Qh-UWX6VOQ9IeTw

31. [CAIF-MAR] Hammond, L., Chan, A., Clifton, J., Hoelscher-Obermaier, J., Khan, A., McLean, E., ... & Rahwan, I. (2025). Multi-Agent Risks from Advanced AI. Cooperative AI Foundation. arXiv:2502.14143. https://www.cooperativeai.com/post/new-report-multi-agent-risks-from-advanced-ai

32. [CAISI-Hijacking] Center for AI Standards and Innovation (CAISI). (2025, January 17). Strengthening AI agent hijacking evaluations. National Institute of Standards and Technology. https://www.nist.gov/news-events/news/2025/01/technical-blog-strengthening-ai-agent-hijacking-evaluations

33. [Chaduvula-XAI] Chaduvula, S., Ho, J., Kim, K., Narayanan, A., Radwan, A. Y., Alinoori, M., Garg, M., Ramachandram, D., & Raza, S. (2026). From features to actions: Explainability in traditional and agentic AI systems. arXiv preprint arXiv:2602.06841. https://arxiv.org/pdf/2602.06841

34. [Cloudflare-WBA] Cloudflare. (2026). Web Bot Auth. Cloudflare Developers. https://developers.cloudflare.com/bots/reference/bot-verification/web-bot-auth/

35. [CNCERT-OpenClaw] 国家计算机网络应急技术处理协调中心. (2026, March 12). 关于OpenClaw安全应用的风险提示 [Risk Alert Regarding the Secure Use of OpenClaw]. https://www.cert.org.cn/publish/main/11/2026/20260312144519429724511/20260312144519429724511_.html

36. [Concordia1.5] Concordia AI & Shanghai Artificial Intelligence Laboratory. (2026, February). Frontier AI Risk Management Framework 1.5. https://concordia-ai.com/wp-content/uploads/2026/02/Frontier-AI-Risk-Management-Framework-v1.5.pdf

37. [CoSAI-Agentic] Coalition for Secure AI. (2025, July 16). Principles for secure-by-design agentic systems. https://www.coalitionforsecureai.org/announcing-the-cosai-principles-for-secure-by-design-agentic-systems/

38. [CSA] Cyber Security Agency of Singapore. (2024, October). Guidelines on Securing AI Systems. https://isomer-user-content.by.gov.sg/36/e05d8194-91c4-4314-87d4-0c0e013598fc/Guidelines%20on%20Securing%20AI%20Systems.pdf

39. [CSA-AgenticAI] Cyber Security Agency of Singapore. (2025). Securing agentic AI: An addendum to the guidelines and companion guide on securing AI systems [Draft for public consultation]. https://isomer-user-content.by.gov.sg/36/703ff9fe-9db1-4e09-98c2-89e3d7007ef0/Draft%20Addendum%20on%20Securing%20Agentic%20AI%20%5BFor%20Public%20Consultation%5D.pdf

40. [Cloud Security Alliance-RedTeam] Cloud Security Alliance. (2025, May 28). Agentic AI Red Teaming Guide. Cloud Security Alliance. https://cloudsecurityalliance.org/artifacts/agentic-ai-red-teaming-guide

41. [Cloud Security Alliance-Agentic Identity] Cloud Security Alliance. (2025, Aug 18). Agentic AI Identity and Access Management: A New Approach. Cloud Security Alliance. https://cloudsecurityalliance.org/artifacts/agentic-ai-identity-and-access-management-a-new-approach

42. [CubeSandbox] Tencent Cloud. (2026). CubeSandbox: Instant, concurrent, secure & lightweight sandbox service for AI agents. GitHub. https://github.com/tencentcloud/CubeSandbox

43. [DeepMind-MultiAgent] Google DeepMind, Schmidt Sciences, the Cooperative AI Foundation, ARIA, & Google.org. (2026, June 11). Scaling AI safety research for a multi-agent world. https://deepmind.google/blog/investing-in-multi-agent-ai-safety-research/

44. [DeepMind-Traps] Franklin, M., Tomašev, N., Jacobs, J., Leibo, J. Z., & Osindero, S. (2026, March 8). AI Agent Traps. https://papers.ssrn.com/sol3/papers.cfm?abstract_id=6372438

45. [DeepSeek-V4] DeepSeek-V4: Towards Highly Efficient Million-Token Context Intelligence. arXiv:2606.19348 https://arxiv.org/abs/2606.19348

46. [Emmons-CoT] Emmons, S., Zimmermann, R. S., Elson, D. K., & Shah, R. (2025). A pragmatic way to measure chain-of-thought monitorability. arXiv preprint arXiv:2510.23966. https://arxiv.org/pdf/2510.23966

47. [EU-AIA-9] European Parliament & Council of the European Union. (2024). EU Artificial Intelligence Act, Article 9: Risk management system. https://artificialintelligenceact.eu/article/9/

48. [EU-AIA-12] European Union. (2026). Article 12: Record-Keeping. In Artificial Intelligence Act, Chapter III, Section 2. https://artificialintelligenceact.eu/article/12/

49. [EU-AIA-13] European Parliament & Council of the European Union. (2024). Article 13: Transparency and provision of information to deployers. https://artificialintelligenceact.eu/article/13/

50. [EU-AIA-14] European Parliament & Council of the European Union. (2024). Article 14: Human oversight. https://artificialintelligenceact.eu/article/14/

51. [EU-AIA-15] European Parliament & Council of the European Union. (2024). Article 15: Accuracy, robustness and cybersecurity. https://artificialintelligenceact.eu/article/15/

52. [EU-AIA-60] European Union. (2026). Article 60: Testing of high-risk AI systems in real world conditions outside AI regulatory sandboxes. In Artificial Intelligence Act, Chapter VI: Measures in support of innovation. https://artificialintelligenceact.eu/article/60/

53. [EU-AIA-72] European Parliament & Council of the European Union. (2026). Article 72: Post-market monitoring by providers and post-market monitoring plan for high-risk AI systems. https://artificialintelligenceact.eu/article/72/

54. [EU-AIA-86] European Parliament & Council of the European Union. (2024). Article 86: Right to explanation of individual decision-making. https://artificialintelligenceact.eu/article/86/

55. [EU-CoP-C3] Samwald, M., Ziosi, M., Zacherl, A., Bengio, Y., Privitera, D., Rajkumar, N., Schaake, M., Reuel, A., & Anderljung, M. (2026). Commitment 3. Code of Practice. https://www.google.com/url?q=https://code-of-practice.ai/?section%3Dsafety-security%23measure-3-5-post-market-monitoring&sa=D&source=docs&ust=1776593792964334&usg=AOvVaw3aK0pjWupEOuKlhrcLmS5Y

56. [FiveEyes-Agentic] Australian Signals Directorate's Australian Cyber Security Centre, Cybersecurity and Infrastructure Security Agency, National Security Agency, Canadian Centre for Cyber Security, New Zealand National Cyber Security Centre, & United Kingdom National Cyber Security Centre. (2026, May). Careful adoption of agentic AI services. https://www.cyber.gov.au/sites/default/files/2026-05/careful_adoption_of_agentic_ai_services.pdf

57. [GB-AgentID] 全国信息技术标准化技术委员会人工智能分会（TC28/SC42). (2025). 人工智能智能体互联 第2部分：身份码 [Agent Interconnection Part 2: IDs]. https://std.samr.gov.cn/gb/search/gbDetailed?id=3EFBBC58E11D080EE06397BE0A0A17E5

58. [GB-AgentIM] 全国信息技术标准化技术委员会人工智能分会（TC28/SC42). (2025). 人工智能智能体互联 第3部分：智能体身份管理 [Agent Interconnection Part 3: Agent Identity Management ]. https://std.samr.gov.cn/gfs/search/gfsDetailed?id=3EFBBC58E121080EE06397BE0A0A17E5

59. [Google-AgentID] Google Cloud. (2025). Agent identity overview. Gemini Enterprise Agent Platform Documentation. https://docs.cloud.google.com/agent-builder/agent-engine/agent-identity

60. [Google-Agents1] Díaz, S. (Sal), Kern, C., & Olive, K. (2025). Google's Approach for Secure AI Agents. Google Research. https://research.google/pubs/an-introduction-to-googles-approach-for-secure-ai-agents/

61. [Google-Agents2] Wiesinger, J., Marlow, P., & Vuskovic, V. (2024). Agents. Google & Kaggle Whitepaper. https://www.kaggle.com/whitepaper-agents

62. [Google-NIST] Google. (2026, March 9). Response to request for information regarding security considerations for artificial intelligence agents. National Institute of Standards and Technology (NIST). https://downloads.regulations.gov/NIST-2025-0035-0316/attachment_1.pdf

63. [Google-Trust] Grannis, W. (2025, December 19). AI grew up and got a job: Lessons from 2025 on agents and trust. Google Cloud. https://cloud.google.com/transform/ai-grew-up-and-got-a-job-lessons-from-2025-on-agents-and-trust

64. [GovTech-ARCF] GovTech Singapore. (2025). Agentic Risk & Capability Framework. https://govtech-responsibleai.github.io/agentic-risk-capability-framework/

65. [Huang-ZeroTrust] Huang, K., Narajala, V. S., Yeoh, J., Raskar, R., Harkati, Y., Huang, J., Habler, I., & Hughes, C. (2025). A novel zero-trust identity framework for agentic AI: Decentralized authentication and fine-grained access control. arXiv preprint arXiv:2505.19301. https://doi.org/10.48550/arXiv.2505.19301

66. [Huawei-AgentSec] Huawei Developer. (2026, June 12). 服务分发小艺开放平台Agent上架审核规范：智能体安全 [Agent Listing Review Specification: Agent Security]. https://developer.huawei.com/consumer/cn/doc/service/agent-security-0000002437625978

67. [Huawei-OpenClaw] 华为云. (2026). OpenClaw风险说明及安全建议 [OpenClaw Risk Description and Security Recommendations]. https://support.huaweicloud.com/bestpractice-flexusl/flexusl_bp_0010.html

68. [IASR26] Bengio, Y., Clare, S., Prunkl, C., Murray, M., Andriushchenko, M., Bucknall, B., Bommasani, R., Casper, S., Davidson, T., Douglas, R., ... & Mindermann, S. (2026). International AI Safety Report 2026. Department for Science, Innovation and Technology. https://internationalaisafetyreport.org

69. [IEEE7000] IEEE. (2021). IEEE Std 7000-2021: IEEE standard model process for addressing ethical concerns during system design. IEEE Standards Association. https://standards.ieee.org/ieee/7000/6781/

70. [IMDA-OpenClaw] Infocomm Media Development Authority. (2026). Case study: OpenClaw. https://www.imda.gov.sg/assets/c0a78f6e-7bd4-485b-ae57-1dfe4246478a.pdf

71. [ISO-22989] International Organization for Standardization & International Electrotechnical Commission (ISO/IEC). (2022). ISO/IEC 22989:2022: Information technology — Artificial intelligence — Artificial intelligence concepts and terminology (1st ed.). ISO/IEC. https://www.iso.org/standard/74296.html

72. [ISO-42001] International Organization for Standardization & International Electrotechnical Commission (ISO/IEC). (2023). ISO/IEC 42001:2023: Information technology — Artificial intelligence — Management system (1st ed.). ISO/IEC. https://www.iso.org/standard/42001

73. [ITU-F74846] International Telecommunication Union Telecommunication Standardization Sector (ITU-T). (2025, March). Recommendation ITU-T F.748.46: Requirements and evaluation methods of artificial intelligence agents based on large scale pre-trained models. International Telecommunication Union. https://www.itu.int/epublications/zh/publication/itu-t-f-748-46-2025-03-requirements-and-evaluation-methods-of-artificial-intelligence-agents-based-on-large-scale-pre-trained-models/en

74. [Jazzyear] 栗子. (2025, December 23). 千亿智能体爆发前夜，谁来保护我们的AI安全？[On the Eve of an Explosion of Agents — Who Will Preserve Our AI Security?] Jazzyear. https://www.jazzyear.com/article_info.html?id=1642

75. [JWTInteraction] Parecki, A., Campbell, B., & Liu, D. (2026). JWT authorization grant interaction response. https://datatracker.ietf.org/doc/html/draft-parecki-oauth-jwt-grant-interaction-response-00

76. [Kasirzadeh2025] Kasirzadeh, A., & Gabriel, I. (2025). Characterizing AI agents for alignment and governance. arXiv preprint arXiv:2504.21848. https://arxiv.org/pdf/2504.21848

77. [LangChain-MW] LangChain. (2025). Prebuilt middleware for common agent use cases. LangChain Documentation. https://docs.langchain.com/oss/python/langchain/middleware/built-in

78. [MASec] Schroeder de Witt, C., Krawiecka, K., Krawczuk, I., Hagag, B., Anderson, W. L., Belcak, P., Bucknall, B., Cai, X., Chopra, A., Cohen, D., Del Rosario, R. F., Draguns, A., Gray, A., Katz, K., Mavroudis, V., Mink, J., Motwani, S. R., Petit, J., Rembeck, L.-S., Smith, C., Sotiropoulos, J., Young, S., Scheffler, S., & Llewellyn, M. (2026). Open challenges in multi-agent security: Towards secure systems of interacting AI agents (arXiv:2505.02077v2). arXiv. https://arxiv.org/html/2505.02077v2

79. [MCP-EnterpriseAuth] Model Context Protocol. (2025). Enterprise-managed authorization. Model Context Protocol Documentation. https://modelcontextprotocol.io/extensions/auth/enterprise-managed-authorization

80. [METR] METR. (2025). Common elements of frontier AI safety policies. https://metr.org/common-elements#common-elements-of-frontier-ai-safety-policies

81. [Microsoft-Foundry] Kwong, A., & Naghshineh, S. (2026, March 16). Evaluating AI agents: A practical guide with Microsoft Foundry. Microsoft Foundry Blog. https://techcommunity.microsoft.com/blog/azure-ai-foundry-blog/evaluating-ai-agents-a-practical-guide-with-microsoft-foundry/4500224

82. [Microsoft-NIST] Microsoft. (2026, March 9). Microsoft comments on the Request for Information regarding security considerations for artificial intelligence agents. Submitted to National Institute of Standards and Technology. https://downloads.regulations.gov/NIST-2025-0035-0399/attachment_1.pdf

83. [MIIT-OpenClaw] 工业和信息化部网络安全威胁和漏洞信息共享平台. (2026, March 11). 关于防范OpenClaw（"龙虾"）开源智能体安全风险建议 [Recommendations on Guarding Against the Security Risks of the OpenClaw]. https://news.cctv.cn/2026/03/11/ARTIU9NPnXcPCDiU9cOfqTlD260311.shtml

84. [Misevolve] Shao, S., Ren, Q., Qian, C., Wei, B., Guo, D., Yang, J., Song, X., Zhang, L., Zhang, W., Liu, D., & Shao, J. (2025). Your agent may misevolve: Emergent risks in self-evolving LLM agents. arXiv preprint arXiv:2509.26354. https://arxiv.org/pdf/2509.26354

85. [MS-Agent-Scale] Bansal, G., Mirza, S., Hines, K., Epperson, W., Huang, Z., Maxwell, W., Bryan, P., Payne, T., Fourney, A., Swearngin, A., Hua, W., Westerhoff, T., Minnich, A., Murad, M., Kamar, E., Kumar, R. S. S., & Amershi, S. (2026, April 30). Red-teaming a network of agents: Understanding what breaks when AI agents interact at scale. Microsoft Research Blog. https://www.microsoft.com/en-us/research/blog/red-teaming-a-network-of-agents-understanding-what-breaks-when-ai-agents-interact-at-scale/
86. [MS-AgentTax] Bryan, P., Severi, G., de Gruyter, J., Jones, D., Bullwinkel, B., Minnich, A., Chawla, S., Lopez, G., Pouliot, M., Fourney, A., Maxwell, W., Pratt, K., Qi, S., Chikanov, N., Lutz, R., Dheekonda, R. S. R., Jagdagdorj, B.-E., Kim, E., Song, J., Hines, K., Jones, D., Lundeen, R., Vaughan, S., Westerhoff, V., Zunger, Y., Kawaguchi, C., Russinovich, M., & Kumar, R. S. S. (2025). Taxonomy of failure modes in agentic AI systems. Microsoft. https://cdn-dynmedia-1.microsoft.com/is/content/microsoftcorp/microsoft/final/en-us/microsoft-brand/documents/Taxonomy-of-Failure-Mode-in-Agentic-AI-Systems-Whitepaper.pdf
87. [Mythos] Anthropic. (2026, April 7). System card: Claude Mythos Preview. https://www.anthropic.com/claude-mythos-preview-system-card
88. [NIST-AASI] National Institute of Standards and Technology (NIST), Center for Artificial Intelligence Standards and Innovation (CAISI). (2026, February 17). AI Agent Standards Initiative: Ensuring a trusted, interoperable, and secure agentic frontier. U.S. Department of Commerce. https://www.nist.gov/caisi/ai-agent-standards-initiative
89. [NIST-AgentAuth] Booth, H., Fisher, B., Galluzzo, R., & Roberts, J. (2026, February). Accelerating the adoption of software and AI agent identity and authorization. National Institute of Standards and Technology. https://www.nccoe.nist.gov/sites/default/files/2026-02/accelerating-the-adoption-of-software-and-ai-agent-identity-and-authorization-concept-paper.pdf
90. [NIST-AgentSecurity] National Institute of Standards and Technology (NIST), CAISI, U.S. Department of Commerce. Request for Information Regarding Security Considerations for Artificial Intelligence Agents. https://www.federalregister.gov/documents/2026/01/08/2026-00206/request-for-information-regarding-security-considerations-for-artificial-intelligence-agents
91. [NIST-AIRMF] National Institute of Standards and Technology. (2023, January 26). AI Risk Management Framework 1.0. https://www.nist.gov/itl/ai-risk-management-framework
92. [NIST-COSAIS] National Institute of Standards and Technology. (2026, January). Control Overlays for Securing AI Systems (COSAiS): Draft annotated outline — NIST SP 800-53 Control Overlay for Securing AI Systems: Using and Fine-Tuning Predictive AI. NIST. https://csrc.nist.gov/csrc/media/Projects/cosais/documents/COSAiS-Predictive-AI-annotated-outline-Jan2026.pdf
93. [NIST-Tools] Center for AI Standards and Innovation & Artificial Intelligence Safety Institute Consortium. (2025). Lessons learned from the consortium: Tool use in agent systems. NIST. https://www.nist.gov/news-events/news/2025/08/lessons-learned-consortium-tool-use-agent-systems
94. [OECD-AIP] Organisation for Economic Co-operation and Development. (2024). OECD AI Principles. https://oecd.ai/en/ai-principles
95. [OECD-P1.4] OECD. (2019). Robustness, security and safety (Principle 1.4). OECD AI Principles. https://oecd.ai/en/dashboards/ai-principles/P8
96. [OpenAI-AgentGov] Shavit, Y., Agarwal, S., Brundage, M., Adler, S., O'Keefe, C., Campbell, R., Lee, T., Mishkin, P., Eloundou, T., Hickey, A., Slama, K., Ahmad, L., McMillan, P., Beutel, A., Passos, A., & Robinson, D. G. (2023). Practices for governing agentic AI systems. OpenAI. https://cdn.openai.com/papers/practices-for-governing-agentic-ai-systems.pdf
97. [OpenAI-Allowlist] OpenAI. (2025). ChatGPT agent allowlisting. OpenAI Help Center. https://help.openai.com/en/articles/11845367-chatgpt-agent-allowlisting
98. [OpenAI-Anthropic-Joint] OpenAI & Anthropic. (2025, August 27). Findings from a pilot Anthropic–OpenAI alignment evaluation exercise. https://openai.com/index/openai-anthropic-safety-evaluation/
99. [OpenAI-Codex] OpenAI. (2026). GPT-5.3-Codex system card. https://deploymentsafety.openai.com/gpt-5-3-codex/introduction
100. [OpenAI-NIST] OpenAI. (2026, March 9). Response to NIST CAISI RFI on the security of AI agent systems. https://downloads.regulations.gov/NIST-2025-0035-0504/attachment_1.pdf
101. [OpenAI-Operator] OpenAI. (2025, January 23). Operator System Card. https://openai.com/index/operator-system-card/
102. [OpenID-AgentID] South, T. (Ed.). (2025, October). Identity management for agentic AI: The new frontier of authorization, authentication, and security for an AI agent world. OpenID Foundation. https://openid.net/wp-content/uploads/2025/10/Identity-Management-for-Agentic-AI.pdf
103. [OpenProblems] Ziosi, M., Plueckebaum, M., Casper, S., Papadatos, H., Chin, Z. S., Slattery, P., Gealy, J., Rudner, T. G. J., Tse, B., Gil, A., Paskov, P., Negele, M., Gipiškis, R., Madkour, N., Lummis, V., Jain, R., Eder, L., Fort, K., van Draanen Glismann, M. C., Belhadj, I., Oueslati, A., Wisakanto, A. K., Mallah, R., Holtman, K., Zuhdi, R., Schiff, D. S., Newman, J., Murray, M., & Trager, R. (2026). Open Problems in Frontier AI Risk Management. https://aigi.ox.ac.uk/wp-content/uploads/2026/02/Open-Problems-in-Frontier-AI-Risk-Management-Final.pdf
104. [OTel] Liu, G., & Solomon, S. (2025, March 6). AI agent observability: Evolving standards and best practices. OpenTelemetry Blog. https://opentelemetry.io/blog/2025/ai-agent-observability/
105. [OWASP-AA] OWASP Agentic Security Initiative. (2025, February 17). Agentic AI – Threats and mitigations. OWASP Agentic AI Security 2025. https://genai.owasp.org/resource/agentic-ai-threats-and-mitigations/
106. [OWASP-Agentic] OWASP Gen AI Security Project. (2025, December). OWASP Top 10 for Agentic Applications 2026. https://genai.owasp.org/resource/owasp-top-10-for-agentic-applications-for-2026/
107. [PAI-RFD] Srikumar, M., Chmielinski, K., Pratt, J., Ashurst, C., Bakalar, C., Bartholomew, W., Bommasani, R., Cihon, P., Crootof, R., Hoffmann, M., Joshi, R., Sap, M., & Withers, C. (2025, September 11). Prioritizing real-time failure detection in AI agents. Partnership on AI. https://partnershiponai.org/resource/prioritizing-real-time-failure-detection-in-ai-agents/
108. [Perplexity-RFI] Perplexity. (2026, March 22). Response to NIST/CAISI Request for Information 2025-0035. https://www.regulations.gov/comment/NIST-2025-0035-0505
109. [Proportionality] Mougan, C., Morlock, L., Aguirre, J., Black, J. R. M., Brauner, J., Campos, S., Dev, S., Fernández Llorca, D., Franzin, A., Fritz, M., Gómez, E., Grosse-Holz, F., Hamilton, E., Hasin, M., Hernandez-Orallo, J., Lahav, D., Massarelli, L., Mavroudis, V., Murray, M., Paskov, P., Raldua, J., & Schellaert, W. (2026). The science and practice of proportionality in AI risk evaluations. Science, 391(6787), 769–771. https://doi.org/10.1126/science.aea3835
110. [Raza] Raza, S., Radwan, A. Y., Chaduvula, S., Alinoori, M., & Emmanouilidis, C. (2026). Transparency in Agentic AI: A Survey of Interpretability, Explainability, and Governance. EngrXiv Preprint, 6451. https://github.com/VectorInstitute/Agentic-Transparency

111. [RepliBench] Black, S., Stickland, A. C., Pencharz, J., Sourbut, O., Schmatz, M., Bailey, J., Matthews, O., Millwood, B., Remedios, A., & Cooney, A. (2025). RepliBench: Evaluating the Autonomous Replication Capabilities of Language Model Agents. arXiv preprint arXiv:2504.18565. https://doi.org/10.48550/arXiv.2504.18565

112. [Singapore] Bengio, Y., Maharaj, T., Ong, L., Russell, S., Song, D., Tegmark, M., Xue, L., Zhang, Y.-Q., Casper, S., Lee, W. S., Mindermann, S., Wilfred, V., et al. (2025). The Singapore Consensus on Global AI Safety Research Priorities. arXiv preprint arXiv:2506.20702. https://doi.org/10.48550/arXiv.2506.20702

113. [Singapore-Agent] Infocomm Media Development Authority of Singapore. (2026, May 20). Model AI governance framework for agentic AI (Version 1.5). https://www.imda.gov.sg/-/media/imda/files/about/emerging-tech-and-research/artificial-intelligence/mgf-for-agentic-ai.pdf

114. [South-Delegation] South, T., Marro, S., Hardjono, T., Mahari, R., Whitney, C. D., Greenwood, D., Chan, A., & Pentland, A. (2025). Authenticated delegation and authorized AI agents. arXiv preprint arXiv:2501.09674. https://doi.org/10.48550/arXiv.2501.09674

115. [Tang2025] Tang, X., Jin, Q., Zhu, K., Yuan, T., Zhang, Y., Zhou, W., Qu, M., Zhao, Y., Tang, J., Zhang, Z., Cohan, A., Greenbaum, D., Lu, Z., & Gerstein, M. (2025). Risks of AI scientists: prioritizing safeguarding over autonomy. Nature Communications, 16, 8317. https://doi.org/10.1038/s41467-025-63913-1

116. [TC260] National Technical Committee 260 on Cybersecurity of SAC & National Computer Network Emergency Response Technical Team/Coordination Center of China. (2025, September). AI Safety and Governance Framework 2.0. https://www.cac.gov.cn/2025-09/15/c_1759653448369123.htm

117. [TC260-2026] 全国网络安全标准化技术委员会. (2026, January 30). 关于发布2026年度第一批网络安全国家标准需求的通知 [Notice on the Release of the First Batch of 2026 National Cybersecurity Standard Requirements]. https://www.tc260.org.cn/portal/article/2/df9022b9293c465a83a15931b2903175

118. [TC260-Agent] 全国网络安全标准化技术委员会秘书处. (2026). TC260-TR-005-2026《智能体安全标准化研究》 [Research Report on Agent Safety Standardisation]. 全国网络安全标准化技术委员会. https://www.tc260.org.cn/portal/article/2/6c6d0fbc04974a9aabd61b30208bbb60

119. [Tencent-AR] 黄贝洋. (2025, September 29). 腾讯云Agent Runtime 执行引擎开放内测！专为 Agent 而生、超安全、极致弹性的Serverless AI 运行时. [Tencent Cloud Agent Runtime execution engine is now open for beta]. 腾讯云开发者社区. https://cloud.tencent.com/developer/article/2572684

120. [Tencent-MCPA2A] 腾讯技术工程. (2025, April 11). AI Agent破局：MCP与A2A定义安全新边界 [AI Agent Breakthrough: MCP and A2A Define New Security Boundaries]. 腾讯新闻. https://news.qq.com/rain/a/20250411A082ME00

121. [UK AISI-B] Souly, A., Kirk, R., Merizian, J., D'Cruz, A., & Davies, X. (2025, November 26). Investigating models for misalignment. UK AI Security Institute. https://www.aisi.gov.uk/blog/investigating-models-for-misalignment

122. [WEF-Agent] World Economic Forum. (2025). AI agents in action: Foundations for evaluation and governance. World Economic Forum. https://www.weforum.org/publications/ai-agents-in-action-foundations-for-evaluation-and-governance/

123. [Yang2026] Yang, Z. (2026). Continually self-improving AI. arXiv preprint arXiv:2603.18073. https://doi.org/10.48550/arXiv.2603.18073

124. [Zhu] Zhu, J., Gandhi, D., Joshi, H., Rezaie Mianroodi, A., Akinli Kocak, S., & Ramachandran, D. (2026). Interpreting agentic systems: Beyond model explanations to system-level accountability. arXiv preprint arXiv:2601.17168. https://arxiv.org/html/2601.17168v1

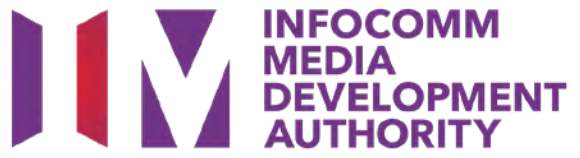

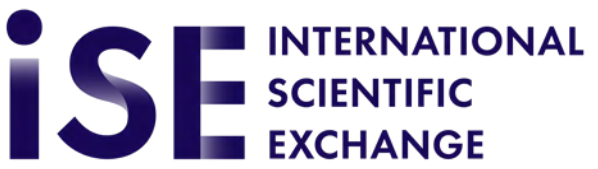

supported by

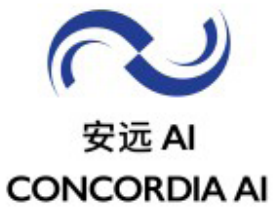

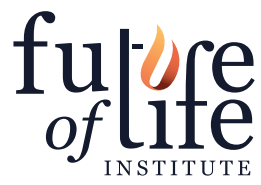

THE 2026 SINGAPORE CONSENSUS ON GLOBAL AI SAFETY RESEARCH PRIORITIES

July 2026